\documentclass[fleqn, usenatbib]{mnras}
\usepackage{newtxtext,newtxmath}
\usepackage[T1]{fontenc}

\usepackage{graphicx}	
\usepackage{amsmath}	
\usepackage{float}
\usepackage{mathrsfs}
\usepackage{color}
\usepackage{setspace}
\usepackage{comment}
\usepackage{stfloats}

\newcommand{\zhong}[1]{{\color{black}{#1}}}

\title[GaSNet-2]{Galaxy Spectra neural Network (GaSNet). II. Using Deep Learning for Spectral Classification and Redshift Predictions}
\author[F. Zhong et al.]{Fucheng Zhong$^{1}$, 
Nicola R. Napolitano$^{1,2}$\thanks{E-mail: napolitano@mail.sysu.edu.cn}, 
Caroline Heneka$^{3}$, 
Rui Li$^{4}$, 
Franz Erik Bauer$^{5,6,7}$, 
Nicolas Bouche$^{8}$, 
\newauthor{Johan Comparat$^{9}$}, 
Young-Lo Kim$^{10}$, 
Jens-Kristian Krogager$^{11}$, 
Marcella Longhetti$^{12}$, 
Jonathan Loveday$^{13}$, 
\newauthor{Boudewijn F. Roukema$^{14,15}$}, 
Benedict L. Rouse$^{5}$, 
Mara Salvato$^{16}$, 
Crescenzo Tortora$^{17}$, 
Roberto J. Assef$^{18}$, 
\newauthor{Letizia P. Cassarà$^{19}$}, 
Luca Costantin$^{20}$, 
Scott Croom$^{21}$, 
Luke J M Davies$^{22}$, 
Alexander Fritz$^{23}$, 
\newauthor{Guillaume Guiglion$^{24,25,26}$}, 
Andrew Humphrey$^{27,28,29}$, 
Emanuela Pompei$^{30}$, 
Claudio Ricci$^{31}$, 
Cristóbal Sifón$^{32}$, 
\newauthor{Elmo Tempel$^{33,34}$}, 
Tayyaba Zafar$^{35}$
\\
\\
$^{1}$School of Physics and Astronomy, Sun Yat-sen University, Zhuhai Campus, 2 Daxue Road, Xiangzhou District, 519082, Zhuhai, P. R. China \\ 
$^{2}$Department of Physics E. Pancini, University Federico II, Via Cinthia 6, 80126-I, Naples, Italy \\ 
$^{3}$Institute of Theoretical Physics, University of Heidelberg, Philosophenweg 12, 69120, Heidelberg, Germany \\ 
$^{4}$National Astronomical Observatories, Chinese Academy of Sciences, 20A Datun Road, Chaoyang District, 100101, Beijing, China \\ 
$^{5}$Instituto de Astrof{\'{\i}}sica and Centro de Astroingenier{\'{\i}}a, Facultad de F{\'{i}}sica, Pontificia Universidad Cat{\'{o}}lica de Chile, Campus San Joaquín, \\ Av. Vicuña Mackenna 4860, Macul Santiago, Chile, 7820436 \\ 
$^{6}$Millennium Institute of Astrophysics, Nuncio Monse{\~{n}}or S{\'{o}}tero Sanz 100, Of 104, Providencia, Santiago, Chile \\ 
$^{7}$Space Science Institute, 4750 Walnut Street, Suite 205, Boulder, Colorado 80301 \\ 
$^{8}$Centre National de la Recherche Scientifique (CNRS), Centre of Research in Astrophysics of Lyon (CRAL), 9 av Charles André, F-69230 Saint Genis Laval \\ 
$^{9}$Max-Planck-Institut für Extraterrestrische Physik (MPE), Giessenbachstrasse 1, 85748 Garching bei München, Germany \\ 
$^{10}$Department of Physics, Lancaster University, Lancs LA1 4YB, UK \\ 
$^{11}$Centre de Recherche Astrophysique de Lyon, 9 Avenue Charles André, 69230 Saint-Genis-Laval, France \\ 
$^{12}$INAF – Osservatorio Astronomico di Brera, via Brera 28, 20121, Milano, Italy \\ 
$^{13}$Astronomy Centre, University of Sussex, Falmer, Brighton, BN1 9QH, UK \\ 
$^{14}$Institute of Astronomy, Faculty of Physics, Astronomy and Informatics, Nicolaus Copernicus, University, Grudziadzka 5, 87-100 Toru\'n, Poland \\ 
$^{15}$Univ Lyon, Ens de Lyon, Univ Lyon1, CNRS, Centre de, Recherche Astrophysique de Lyon UMR5574, F--69007, Lyon, France \\ 
$^{16}$Max-Planck-Institut für extraterrestrische Physik (MPE), Gießenbachstraße 1, D-85748 Garching bei München, Germany \\ 
$^{17}$INAF - Osservatorio Astronomico di Capodimonte, Salita Moiariello 16, 80131, Napoli (Italy) \\ 
$^{18}$Instituto de Estudios Astrofísicos, Facultad de Ingeniería y Ciencias, Universidad Diego Portales, Av. Ejército 441, Santiago, Chile \\ 
$^{19}$INAF – IASF  Milano, via A. Corti 12, 20133, Milano, Italy \\ 
$^{20}$Centro de Astrobiolog\'{\i}a (CAB), CSIC-INTA, Ctra. de Ajalvir km 4, Torrej\'on de Ardoz, E-28850, Madrid, Spain \\ 
$^{21}$Sydney Institute for Astronomy, School of Physics | Faculty of Science, The University of Sydney, Rm 351, School of Physics, A28, Australia \\ 
$^{22}$ICRAR, The University of Western Australia, 35 Stirling Highway, Crawley, WA 6009, Australia \\ 
$^{23}$OmegaLambdaTec GmbH, Parkring 6 , 85748 Garching, Germany \\ 
$^{24}$Zentrum f\"ur Astronomie der Universit\"at Heidelberg, Landessternwarte, K\"onigstuhl 12, 69117 Heidelberg, Germany \\ 
$^{25}$Max Planck Institute for Astronomy, K\"onigstuhl 17, 69117, Heidelberg, Germany \\ 
$^{26}$Leibniz-Institut f{\"u}r Astrophysik Potsdam (AIP), An der Sternwarte 16, 14482 Potsdam, Germany \\ 
$^{27}$Instituto de Astrof\'{i}sica e Ci\^encias do Espa\c{c}o, Universidade do Porto, CAUP, Rua das Estrelas, Porto, 4150-762, Portugal \\ 
$^{28}$DTx -- Digital Transformation CoLab, Building 1, Azur\'em Campus, University of Minho, 4800-058 Guimar\~aes, Portugal \\ 
$^{29}$Instituto de Astrof\'isica e Ci\^encias do Espa\c{c}o, Universidade do Porto, CAUP, Rua das Estrelas, PT4150-762 Porto, Portugal \\ 
$^{30}$European Southern Observatory, Science Operations, Alonso de Cordova 3107, Vitacura, 19001 Santiago, Chile \\ 
$^{31}$Instituto de Estudios Astrofísicos, Facultad de Ingeniería y Ciencias, Universidad Diego Portales (UDP), Santiago de Chile \\ 
$^{32}$Instituto de Física, Pontificia Universidad Católica de Valparaíso \\ 
$^{33}$Tartu Observatory, University of Tartu, Observatooriumi 1, 61602 Tõravere, Estonia \\ 
$^{34}$ Estonian Academy of Sciences, Kohtu 6, 10130 Tallinn, Estonia \\
$^{35}$School of Mathematical and Physical Sciences, Macquarie University, NSW 2109, Australia \\ 
}
\date{Accepted XXX. Received YYY; in original form ZZZ }
\pubyear{2023}
\begin{document}
\label{firstpage}
\pagerange{\pageref{firstpage}--\pageref{lastpage}}
\maketitle
\clearpage

\begin{abstract}
\zhong{The size and complexity reached by the large sky spectroscopic surveys}
require efficient, accurate, and flexible automated tools for data analysis and science exploitation. We present the Galaxy Spectra Network/GaSNet-II, a supervised multi-network deep learning tool for spectra classification and redshift prediction. GaSNet-II can be trained to identify a customized number of classes and optimize the redshift predictions.
\zhong{Redshift errors are determined via an ensemble/pseudo-Montecarlo test obtained by randomizing the weights of the network-of-networks structure.}
As a demonstration of the capability of 
GaSNet-II, we use 260k Sloan Digital Sky Survey spectra from Data Release 16, separated into 13 classes including 140k galactic, and 120k extragalactic objects. 
GaSNet-II achieves 92.4\% average classification accuracy over the 13 classes
and mean redshift errors of approximately 0.23\% for galaxies and 2.1\% for quasars. 
We further train/test the pipeline 
on a sample of 200k 4MOST mock spectra and 21k publicly released DESI spectra. On 4MOST mock data, we reach 93.4\% accuracy in 10-class classification and mean redshift error of 0.55\% for galaxies and 0.3\% for active galactic nuclei. On DESI data, we reach 96\% accuracy in (star/galaxy/quasar only) classification and mean redshift error of 2.8\% for galaxies and 4.8\% for quasars, despite the small sample size available. GaSNet-II can process $\sim40$k spectra in less than one minute, on a normal Desktop GPU. This makes the pipeline particularly suitable for real-time analyses \zhong{and feedback loops for optimization} of Stage-IV survey observations.
\end{abstract}

\section{Introduction}
\label{sec:intro}
With the upcoming 
all-sky spectroscopic survey infrastructures, including the Dark Energy Spectroscopic Instrument (DESI; \citealt{2022AJ....164..207D}), 4-metre Multi-Object Spectroscopic Telescope (4MOST; \citealt{2019Msngr.175....3D}), Multi-Object Optical and Near-infrared Spectrograph (MOONS; \citealt{2020Msngr.180...10C}), and considering also the slitless spectroscopic capabilities of the space-based missions like Chinese Space Station Telescope (CSST; \citealt{2011SSPMA..41.1441Z}) and Euclid \citep{2011arXiv1110.3193L}, hundreds of millions of spectra will be acquired in the next half-decade. The first samples from DESI are already publicly available \citep{2023arXiv230606308D}. To optimize the scientific outcome of these huge datasets, strategies to perform fast, efficient, and, most of all, accurate automated analyses have become mandatory. Machine learning (ML)  provides a large variety of efficient solutions to achieve this goal. We have already demonstrated that Convolutional Neural Network (CNN) models can be very effective in classifying spectra for specific tasks like the search for strong galaxy-galaxy lenses (GaSNet; \citealt{2022RAA....22f5014Z}), showing superior efficiency and flexibility compared to traditional methods (e.g., principal component analysis (PCA) eigenspectra fitting; see \citealt{2021MNRAS.502.4617T}). 

Object classification and redshift prediction are the first steps to be performed by standard pipelines of spectroscopy observations. They provide basic information to be used for science applications. 
For instance, the separation of quiescent early-type galaxies, from the starburst emitting systems is fundamental for galaxy formation \citep{1996ApJ...472..546L}, while the classification of active galactic nuclei (AGN) is crucial to understanding the role of supermassive black holes \citep{2017A&A...601A.143F}, and the identification of quasars (quasi-stellar objects, QSOs) is important for cosmological studies \citep{2021ApJ...908L..51S}. 
ML can be an efficient and practical alternative to traditionally automatic methods \citep{2012AJ....144..144B, 2016AJ....152..205H} to build entire ML-based parallel pipelines, similar to what is already done in astronomical imaging, where there have been enormous advances in recent years. 
Some examples of these latter applications are the galaxy morphology pipelines, like the one developed by \citet{2022MNRAS.509.4024D} for SDSS-DR17, and the pipeline developed by \citet{2020MNRAS.491.2481B} for Euclid. ML can offer huge decreases in computational time and resources \citep{2014MNRAS.441.1741G}, while providing close to human-level classification results, e.g., in the star/quasar separation \citep{2018arXiv180809955B}. This provides the chance to overcome the limits typically plaguing traditional classification methods in terms of computational resources, human intervention, limited real-time applications, scalability, etc. \citep{alzubaidi2021review}, thus giving us the opportunity to develop automatized ML-based tools
\citep{2018A&A...609A.111D, 2018MNRAS.476.1151P, 2019A&C....2900313M}.

With respect to spectroscopy, a variety of automatic redshift prediction tools and pipelines have been developed using traditional methods, but relatively little has been done in terms of ML applications.
Traditional codes, such as {\it spectro1d} \citep{2002SPIE.4847..452S} and {\it redmonster} \citep{2016AJ....152..205H}, based on cross-correlation methods \citep{1979AJ.....84.1511T}, or 
{\it redrock} \citep{2023ApJ...943...68L}, based on template fitting using a set of different PCA components \citep{2023arXiv230606308D}, are some examples of such automated tools.
They have been tested or successfully applied to larger-scale spectroscopy surveys, generally requiring
minimal human intervention. However, they are 
often time-consuming, e.g., if the number of templates increases, or require an optimization of the first guess redshifts to maximize the accuracy. 
Furthermore, in low signal-to-noise ratio (SNR) situations, the performance of some of these tools can highly be degraded (e.g., because of an increasing failure rate, \citealt{2012AJ....144..144B}).

Deep learning (DL) based methods, instead, have the advantage of efficiency, scalability, and flexibility. Here, the applications to spectroscopy are yet at the pioneer level 
and limited to the search for strong gravitational lenses, \citealt{2019MNRAS.482..313L}), with only a  deep learning tool previously tested to classify spectra and measure redshift (i.e. GaSNet, \citealt{2022RAA....22f5014Z})
yet with the specific goal of finding hidden strong lensing emissions in galaxy spectra. However, the first GaSNet is versatile enough to be adapted to answer most of the typical problems large sky surveys might need to face. In particular, it can easily perform tasks like real-time analysis for the detection of transients/peculiar objects, and still give a prediction of their redshift.

In this paper, we present a new DL tool that expands the capabilities of the former GaSNet to respond to the needs for upcoming spectroscopic surveys like 4MOST and DESI.
DESI is expected to observe 30 million galaxies/AGN and 10 million stars. On the other hand, 4MOST will cover approximately 15,000 square degrees and observe more than 25 million targets.
In particular, we design and test a full real-time pipeline based on deep learning that uses reduced 1-D spectra as input to 1) classify spectra in a given number of subclasses; 2) predict the redshift; 3) assign an error to the redshift. 
GaSNet-II is a deep-learning-based tool for spectroscopy classification and redshift prediction which provides the probability of the type of spectrum and the object redshift with uncertainty. To train and test the pipeline we start from a catalog from Sloan Digital Sky Survey Data Release 16 (SDSS-DR16, \citealt{2020AJ....160..120J}) which provides a large number of classified spectra grouped into about 180 classes. This allows us to randomly select 13 subclass spectra from the SDSS Data Release 16, each with more than 20,000 spectra. The 4MOST mock spectra (10 subclasses) and DESI early data release spectra (3 classes) are also randomly selected as additional datasets, to examine the flexibility and generality of the pipeline. In particular, the different properties of these three datasets will allow us to cover a large variety of classification situations from very specialized classifications for SDSS and 4MOST samples to a coarse-grained classification using DESI data.


The paper is organized as follows: in Section \ref{sec: Data}, SDSS data sets used for our analysis are introduced. In Section \ref{sec: Method}, we describe the ML models and our novel idea of building an ML pipeline. In Section \ref{sec: Results}, we present the training and testing results. In Section \ref{sec: Discussion}, we discuss the ML predicted results, including further improvements and perspectives for further ML pipelines. In the final section \ref{sec: Conclusion}, we draw some conclusions.

\section{Data}
\label{sec: Data}

The main purpose of this paper is to find a DL-based method, to classify and predict the redshift of 1-D spectra. As introduced above, we are interested in applying ``supervised'' networks, based on labeled data. For the scope of this work, the main labels we need to start with are a ``class'' and a ``redshift''. The generality of the tool depends on the number of classes we can separate from their spectral properties. While a basic separation can rely on a very coarse classification aiming to distinguish only stars, galaxies, and AGN/QSO \citep{2017A&A...597A..79P}, for many science applications, one might be interested in a more detailed classification that distinguishes various star, galaxy and AGN/QSO subclasses \citep{2015ApJ...798....7B, 2019ApJ...883..175Y}.
In this case, to best train any supervised tool we need datasets that can provide such kind of information. The ideal dataset would be an observed sample of objects for which a qualitative/quantitative classification has been performed \citep{2019ApJS..243...21L, 2020ApJS..250....8L}. However, as an alternative, one can use mock datasets, where physically motivated templates of different galactic and extragalactic objects in different instrumental conditions (resolution, seeing, etc.) and covering a realistic range of intrinsic object properties (e.g., luminosity, colors, redshifts, kinematics, etc.), can mimic the data one is expected to collect for a given science program (e.g. via spectral synthesis; \citealt{2005MNRAS.358..363C}). 

\begin{table}
    \caption{Some definitions and statistics of our reference dataset from SDSS. Col. 1: the name of the different subclass, constituted by the class name and subclass name. The subclass name `nan' denotes classes with no specific subclass. Col. 2: the label we used afterward. Col. 3: the mean redshift of the subset. Col. 4: the redshift range. Col. 5: mean median signal-to-noise, $\overline{SNR}$. 
    }
    \centering
    \begin{tabular}{|l||l|l|l|l|}
    \hline
     Col. 1 & 2 & 3 & 4 & 5 \\
    \hline \hline
     class\_subclass & label & $\bar z$ & $[z_{min},z_{max}]$ & $\overline{SNR}$   \\
    \hline \hline 
     STAR\_A0            & 0 & -- & -- & 26.2 \\
    \hline 
     STAR\_F5            & 1 & -- & -- & 30.5 \\
    \hline 
     STAR\_F9            & 2 & -- & -- & 34.9 \\
    \hline 
     STAR\_G2            & 3 & -- & -- & 33.7 \\
    \hline 
     STAR\_K1            & 4 & -- & -- & 32.8 \\
    \hline 
     STAR\_K3            & 5 & -- & -- & 31.1 \\
    \hline 
     STAR\_K5            & 6 & -- & -- & 31.0 \\
    \hline 
     GALAXY\_nan         & 7 & 0.46 & [0.00, 1.86] & 5.82 \\
    \hline
     GALAXY\_AGN         & 8 & 0.21 & [0.00, 0.57] & 14.3 \\
    \hline 
     GALAXY\_STARBURST   & 9 & 0.15 & [0.00, 0.57] & 9.78 \\
    \hline
     GALAXY\_STARFORMING & 10& 0.11 & [0.00, 0.56] & 12.4 \\
    \hline
     QSO\_nan            & 11& 1.68 & [0.01, 7.04] & 2.64 \\
    \hline
     QSO\_BROADLINE      & 12& 1.78 & [0.03, 5.29] & 6.54 \\
     \hline
    \end{tabular}
    \label{Table:1}
\end{table}

Below we describe the data we will use throughout the paper, covering the two typologies of training/test samples discussed above. In particular, as the observation-based dataset, we use the SDSS-DR16 dataset, which contains the most detailed classified subclass sample of sources available to date. As such, this will represent the reference dataset around which we want to construct and benchmark our pipeline. Furthermore, to explore the possible application of GaSNet-II to upcoming stage-IV surveys, we use a customized mock catalog, closely reproducing 4MOST observations \citep{2019Msngr.175....3D, 2019Msngr.175...23H, 2019Msngr.175...42M, 2019Msngr.175...46D, 2019Msngr.175...58S}.
Furthermore, we take advantage of the early data release of DESI \citep{2023arXiv230606308D},
to perform a first test of the novel GaSNet-II version performances on a first Stage-IV survey dataset. 
Notable for SDSS and DESI, the redshifts and classifications are not 100\% reliable (see, e.g., \citealt{2020ApJS..250....8L} and \citealt{2023AJ....165..124A}), which can potentially lead to deviations between the DL predictions and the pipeline results.

\begin{figure}
\centering
        \includegraphics[width=0.45\textwidth]{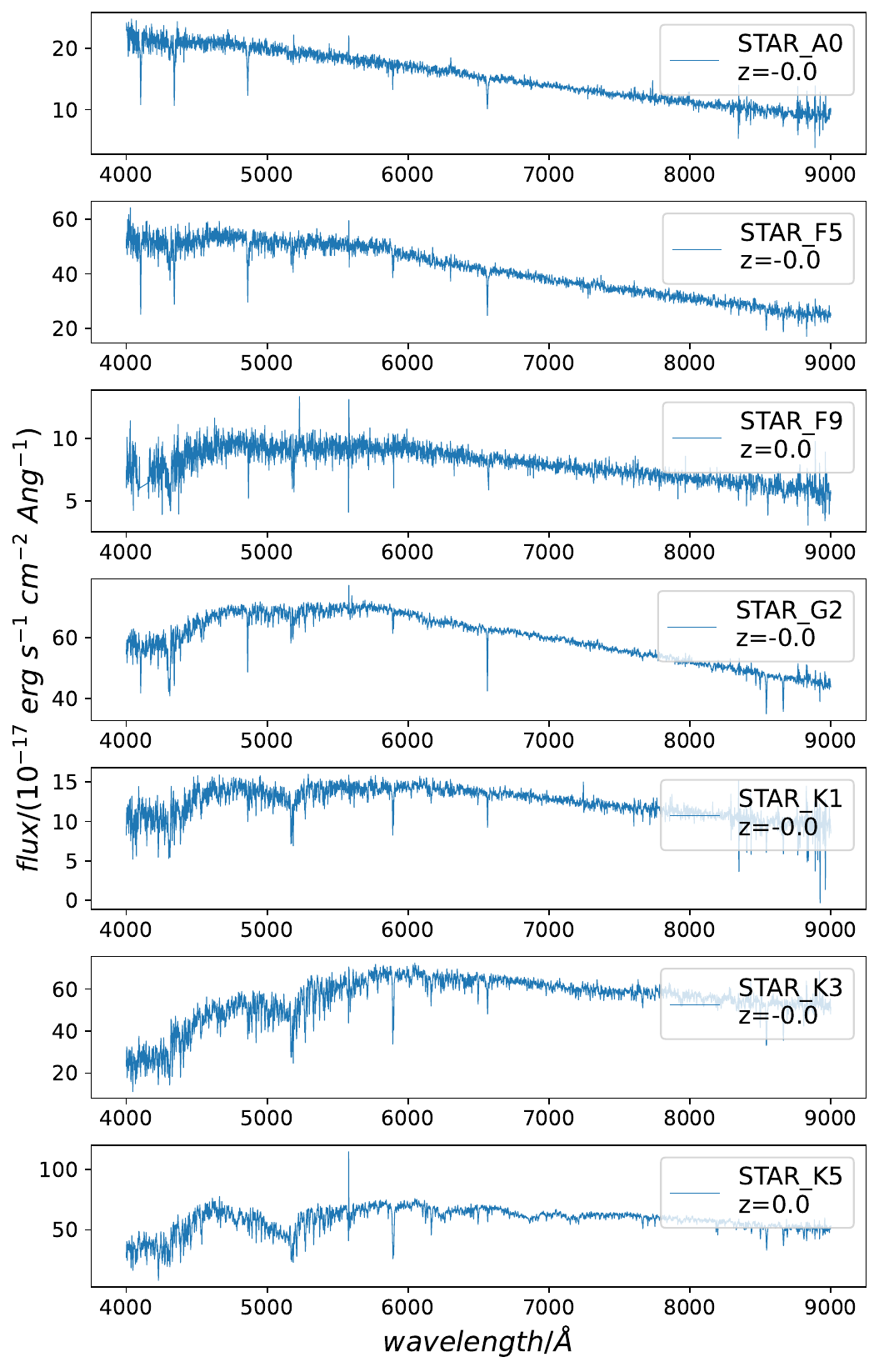}
    \caption{Example spectra of the 7 stellar sub-classes, corresponding to the first 7 of the 13 sub-classes constituting the SDSS sample listed in Table \ref{Table:1}. The A, F, G, and K stars with different subtypes are selected as the SDSS test samples to validate the ability of fine classification.}     
    \label{fig: spectra_1} 
\end{figure}

\begin{figure}
\centering
        \includegraphics[width=0.45\textwidth]{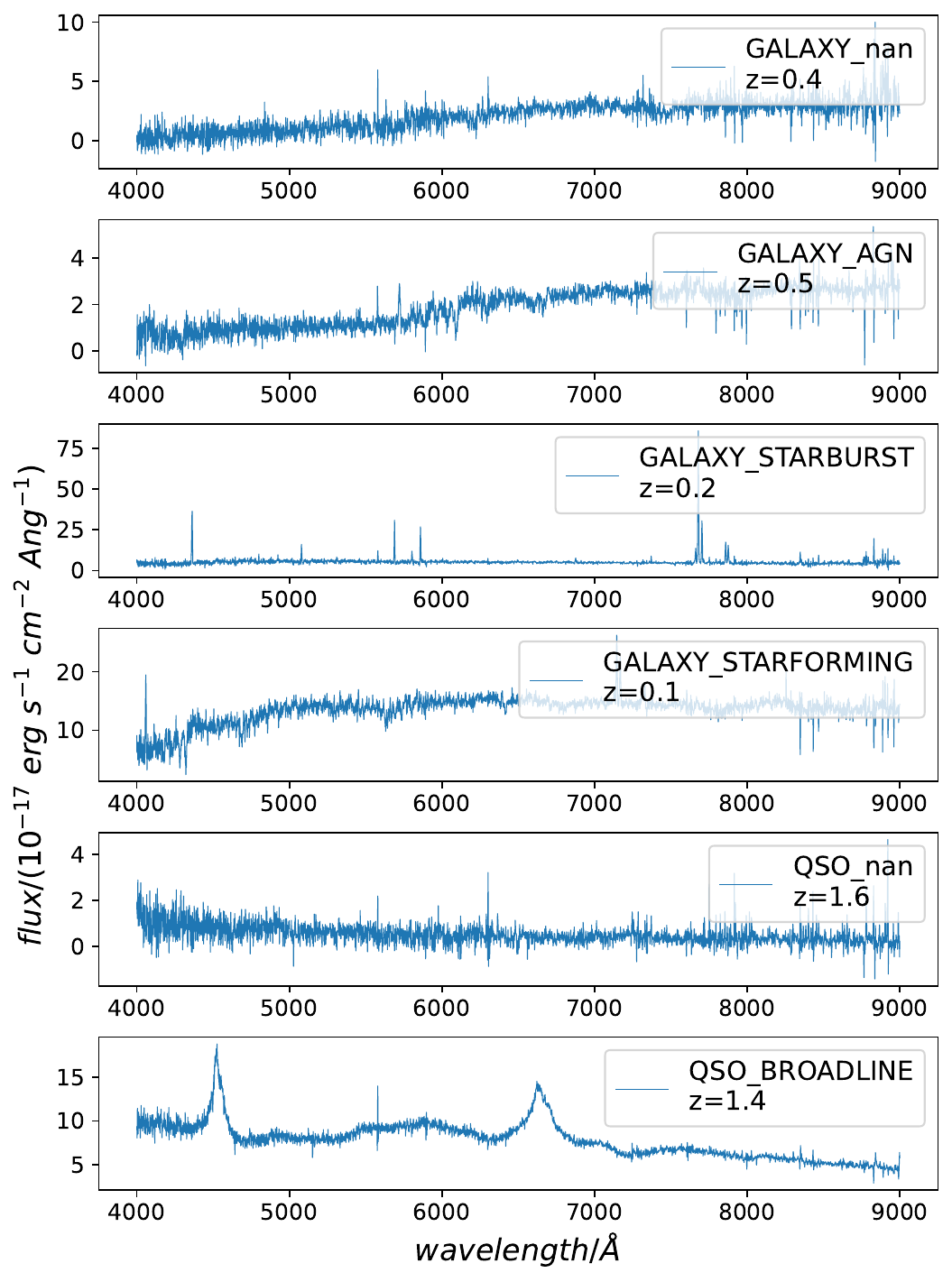}
    \caption{Example spectra of SDSS extragalactic sub-classes, 
    as listed in Table \ref{Table:1}. We can clearly see the different features characterizing the different classes. From top to bottom, in particular, we can notice the increasing importance of the emission lines that play an important role in redshift prediction. The `nan' type spectra generally lack such emission lines, although they might still contain some low-SNR ones, which are hard to see. This means that the `nan' sample might overlap with other emission line classes. QSOs also show a power-law continuum that does not carry any redshift information.}     
    \label{fig: spectra_2} 
\end{figure}

\subsection{Reference dataset: SDSS-DR16}
\label{sec:sdss_spectra}
SDSS-DR16 \citep{2020ApJS..249....3A}, contains around 0.44 million unique stars, 2.6 million galaxies, and 0.75 million quasars; 
all spectra are divided into three classes (star, galaxies, QSOs), each one having a different number of subclasses for a total of 181 subclasses. Most of the subclasses comprise a number of spectra smaller than a few hundred. The classification and redshift pipeline of SDSS is based on a $\chi^2$ minimization, by comparing each spectrum to the combination of basis templates, which are derived from rest-frame PCA of training samples \citep[][B+12 hereafter]{2012AJ....144..144B}. The number of labeled spectra is more than four million.\footnote{\href{https://www.sdss4.org/dr16/spectro/\#SDSSopticalMilkyWayobservingprograms}{DR16 Optical Spectra Overview}.} In Table \ref{Table:1}, we report the only 13 sub-classes that have more than 20,000 classified objects, as this is the minimal sample size we need for the best training of our tools.
Despite these representing a tiny fraction of the original class list (181), we stress that these 13 sub-classes are representative of the most common objects one would expect to classify in typical spectroscopic surveys, especially if we look at the extragalactic sample. Most of the excluded classes, though, consist of stellar types (e.g., O, B star, dwarf, special carbon star, etc.) that have small observational samples collected, due to their intrinsic rarity. 
\zhong{
Of course, this is a limitation if one wants to apply the current classifier to real data that we expect to solve in the future by collecting more complete samples to build a compelling training sample, e.g. using the early release of upcoming surveys (e.g. DESI and 4MOST). Also, the reduced number of sub-types adopted might not return the true final accuracy of the method, as we cannot predict if the classifier can perform closely to the average accuracy for all the missing classes. However, we believe that the number and variety of classes we have collected for this test, is already large enough to assess the potential of these (novel and unexplored) techniques.  
Indeed,}
since the main objective of this paper is to check if DL can efficiently and automatically classify spectra and measure redshifts of astronomical sources, the main conclusions we will draw will not be affected by the number of classes adopted, as long as the network can be trained for each class with a sufficiently large and representative knowledge base. Following this same line of argument, our results are also not affected by the accuracy of the classification performed in B+12, as long as all spectra are assigned to a given class following self-consistent criteria. In this respect, GaSNet-II would just replicate the same classification bias intrinsic to the SDSS-DR16 sample, if any. However, from the perspective of the application to upcoming surveys, the problem of cross-contamination among classes needs to be addressed to quantify how much this can impact the purity of classifications. Although this is not among the objectives of this paper, we briefly discuss this in Appendix \ref{sec:app_cross_cont}. 

Finally, for the 13 suitable classes from SDSS-DR16, we can randomly select 20,000 spectra from each of these classes to collect a total catalog of 260,000 spectra, constituting our primary dataset. 
Most of the classes do not overlap physically, except for the ``BROADLINE'' one, because if any galaxies or quasars have lines detected at the 10-sigma level with velocity dispersion $\sigma>$200 km/sec at the 5-sigma level, the label ``BROADLINE" is added to their subclass \footnote{\url{https://www.sdss3.org/dr9/spectro/catalogs.php}}.
The 20,000 spectra in each subclass are further split into random 70\%, 15\%, 15\% subsamples to be used in training, validation, and testing, respectively.

\begin{figure}
\centering
        \includegraphics[width=0.4\textwidth]{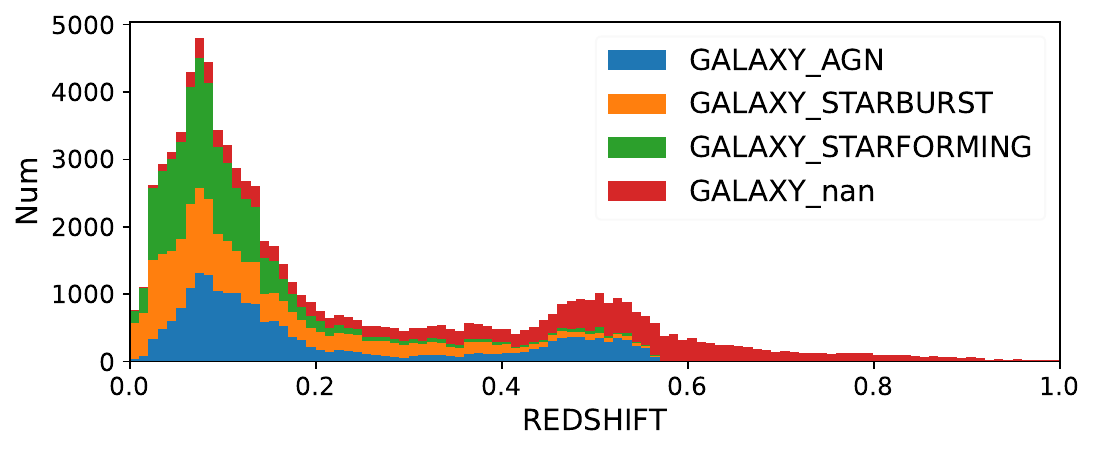}
        \includegraphics[width=0.4\textwidth]{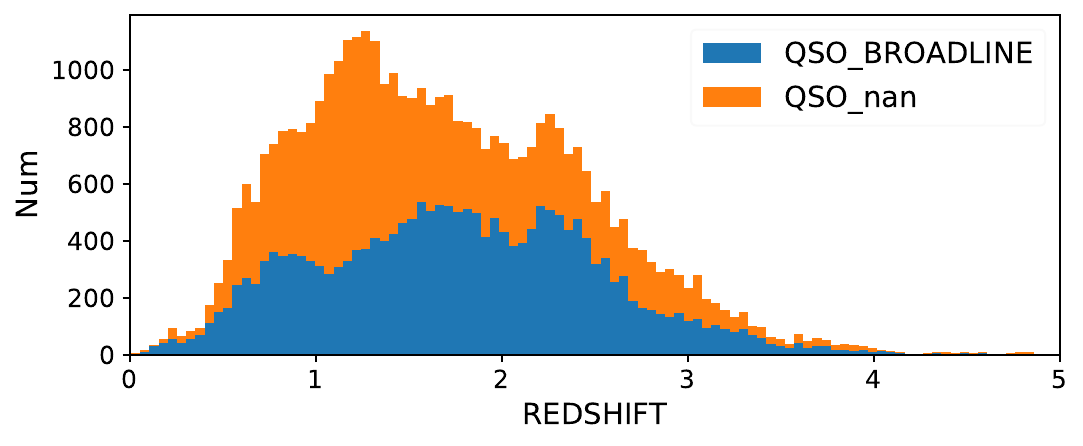}
    \caption{The redshift distribution of the SDSS-DR16 dataset (stacked histogram). The mean and range of redshift are already shown in Table \ref{Table:1}.}     
    \label{fig: redshift} 
\end{figure}

\begin{figure}
\centering
        \includegraphics[width=0.4\textwidth]{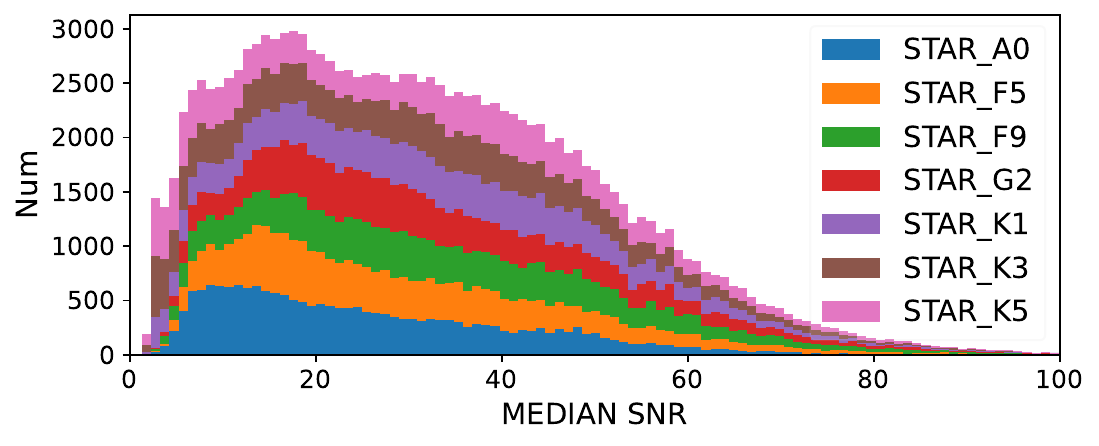}
        \includegraphics[width=0.4\textwidth]{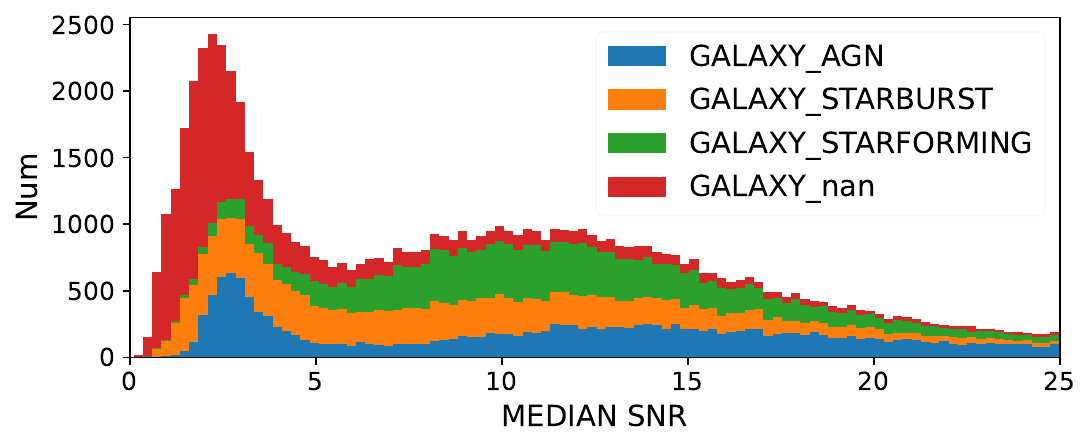}
        \includegraphics[width=0.4\textwidth]{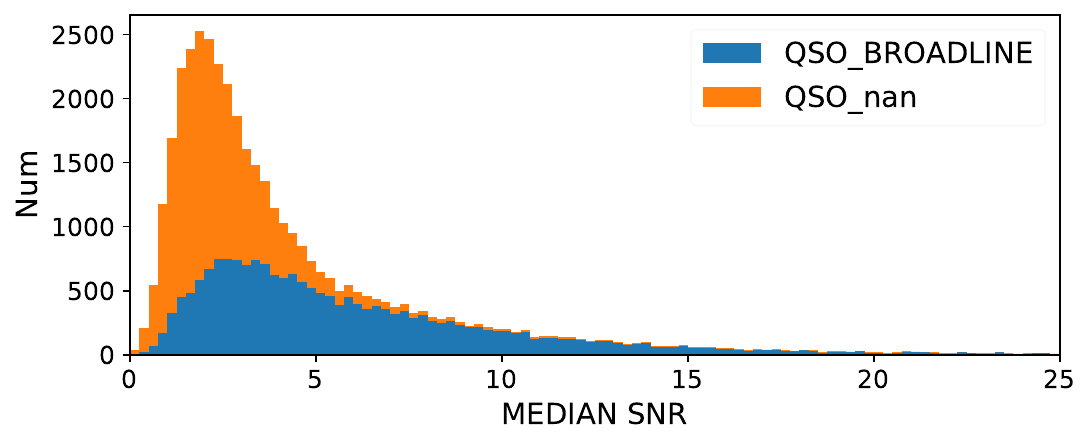}
    \caption{The SNR distribution of the SDSS-DR16 dataset (stacked histogram). Top: star classes; middle: galaxy classes; bottom: AGN classes. In general, the extragalactic objects are fainter than the star classes. The mean SNR is shown in Table \ref{Table:1}.}  
    \label{fig: SNR}
\end{figure}

Some typical spectra of different galactic (stars) and extragalactic (galaxies/AGN/QSOs) types are shown in Fig. \ref{fig: spectra_1} and Fig. \ref{fig: spectra_2}, respectively.  
The spectra are trimmed to stay within $4,000-9,000$\,\AA\ wavelength range, for uniformity, then re-sampled to cover 5,001 pixels, for a final effective binning of 1\AA/pixel. Besides this ``preprocessing'' step, producing a uniform binned spectrum with respect to the original one, no additional data manipulation has been applied to the data.
The range and mean redshift and SNR are listed in Table \ref{Table:1}. The distribution of redshift and SNR of different subclasses are shown in Figs. \ref{fig: redshift} and \ref{fig: SNR}. We stress that the high redshift end on the redshift distribution in Fig. \ref{fig: redshift} is populated by a few systems. This is important to keep in mind, as we expect that this under-sampling can impact the redshift predictions at the higher end of the class redshift distributions.
On the other hand, the SNR distribution covers quite a high range, except for the QSO, which also shows a significant under-sampling at SNR $>10$, and (counter-intuitively) causes worse predictions in this SNR range. 
Overall, to prevent such selection effects, one solution can be the use of simulated spectra, in order to collect a more balanced training dataset. Although useful to solve these ``completeness'' problems, this strategy has other limits which we will discuss in the next section, where we make use of 4MOST mock spectra, as an additional dataset to test.

\subsection{Other dataset: 4MOST mock spectra}
\label{sec: 4most data}

\begin{figure}
    \centering
        \includegraphics[width=0.45\textwidth]{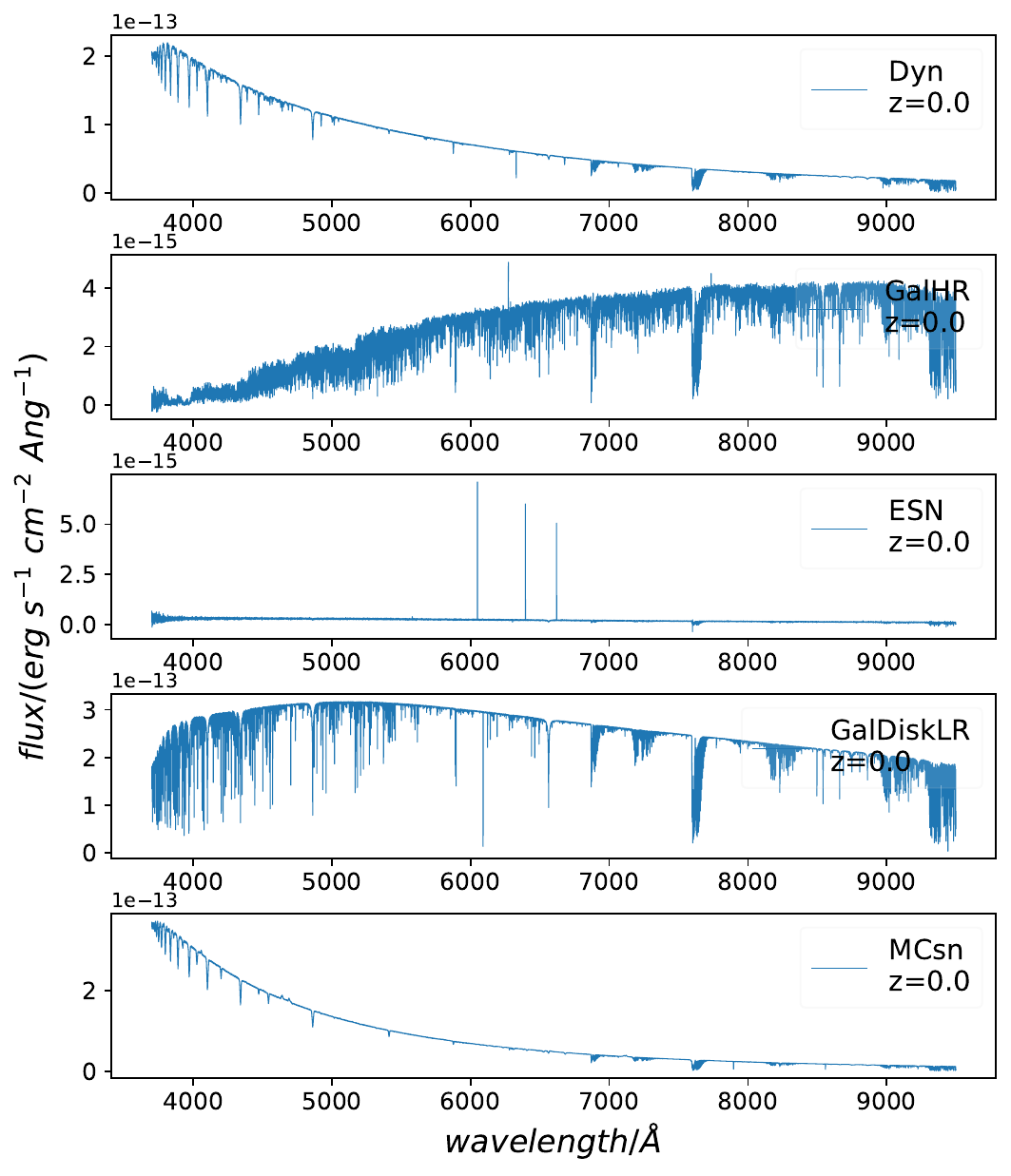}
    \caption{Example spectra of 5 galactic sub-classes of the 4MOST sample, as listed in Table \ref{Table:2}. From top to bottom, there are Dyn, GalHR, ESN, GalDiskLR, and MCsn.} 
    \label{fig: 4most spectra_1} 
\end{figure}

\begin{figure}
    \centering
        \includegraphics[width=0.45\textwidth]{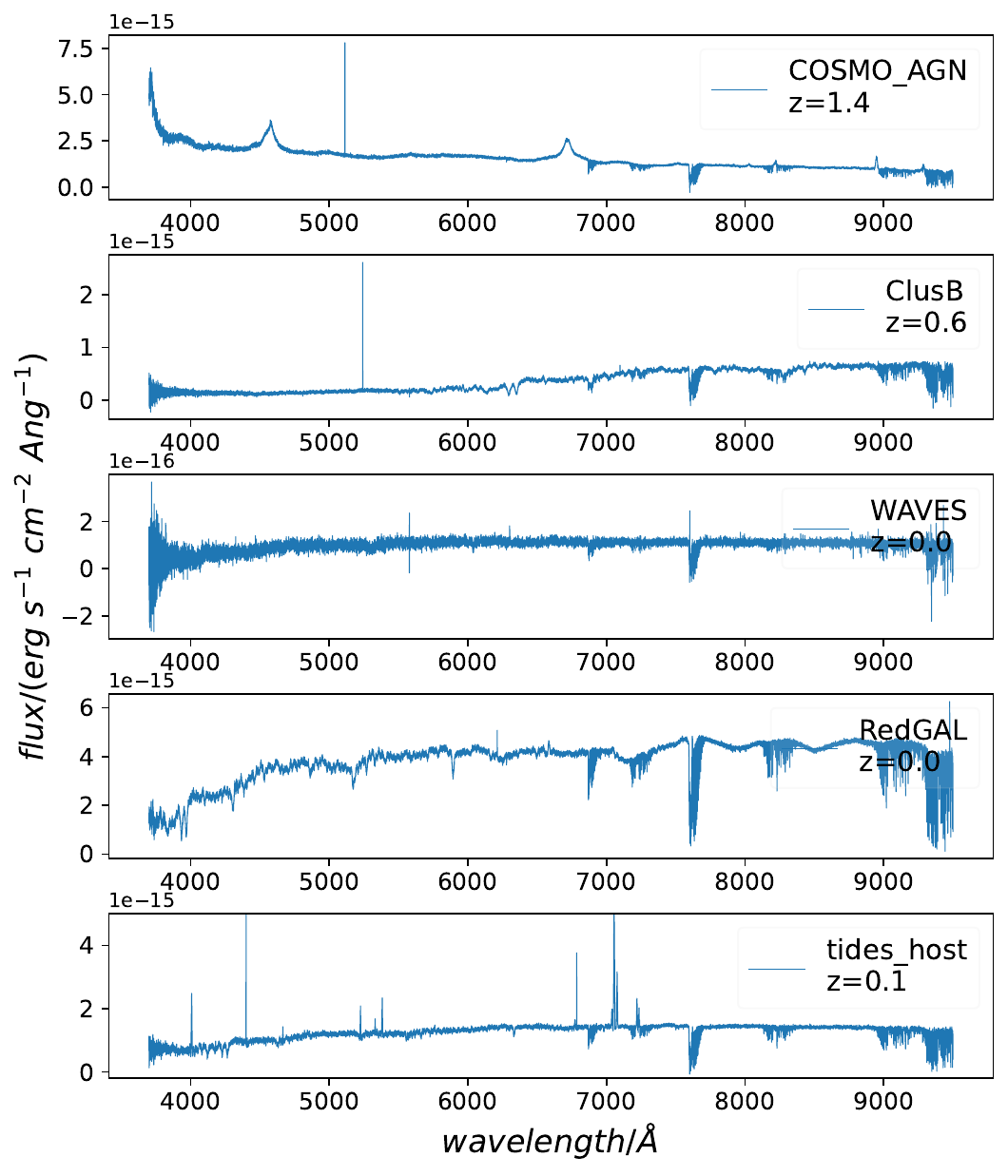}
    \caption{Example spectra of 5 extragalactic sub-classes of the 4MOST sample (simulated), as listed in Table \ref{Table:2}, From top to bottom, there are COSMO$\_$AGN, ClusB, WAVES, RedGAL, and tides$\_$host.} 
    \label{fig: 4most spectra_2} 
\end{figure}

\begin{figure}
        \centering
        \includegraphics[width=0.4\textwidth]{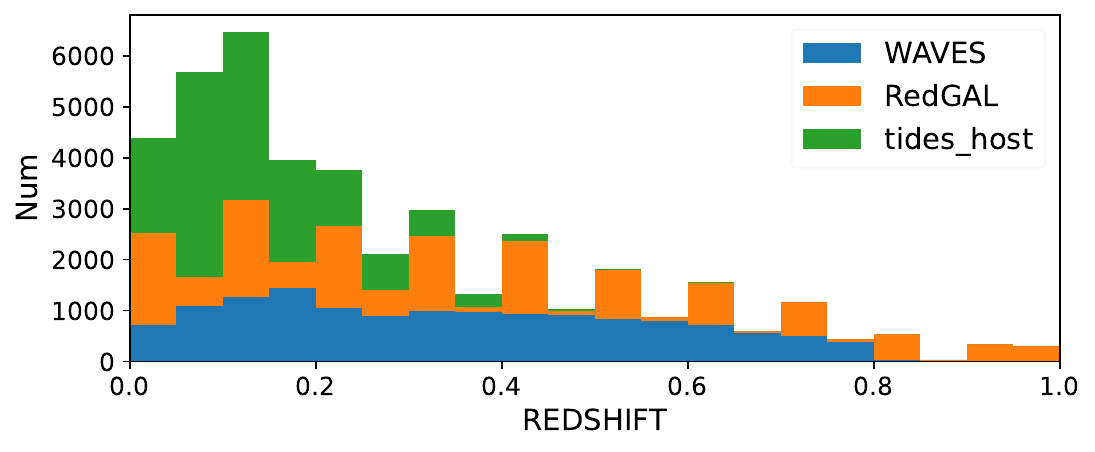}
        \includegraphics[width=0.4\textwidth]{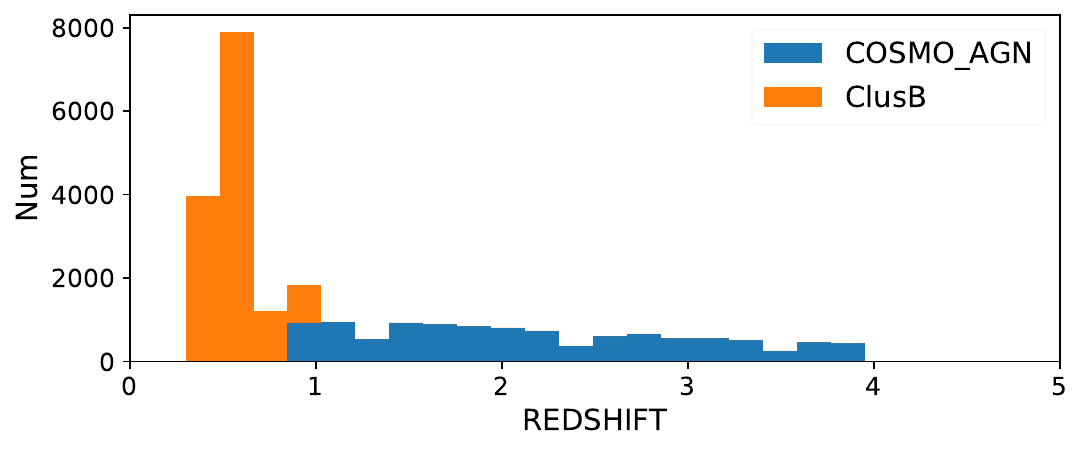}
    \caption{The redshift distribution of the 4MOST dataset. The mean and range of redshift are already shown in Table \ref{Table:2}.}     
    \label{fig: 4most redshift} 
\end{figure}

The dataset consists of approximately 200,000 mock spectra obtained to reproduce 4MOST observation conditions, which are categorized into 10 different sub-classes according to the adopted templates. 
{We make use of a mock catalog of spectra based on a customized software package\footnote{\url{https://escience.aip.de/readthedocs/OpSys/etc/master/index.html}} reproducing the Exposure Time Calculator prediction of observed spectra for 4MOST. The software makes use of a series of customized templates selected for the different surveys (see \S\ref{sec:intro}) to be tested within the Extragalactic Pipeline working group (IGW8) and the Classification working group (IGW9) of the 4MOST consortium.}
The spectral wavelength range is cut to between 4,000 and 9,000 \AA, and the number of pixels is interpolated to obtain 5,001 pixels. The simulated spectra are generated from the given SED templates for a given set of observation conditions and random noise (including cosmic rays and randomized Ly$\alpha$ forest)\footnote{\url{https://github.com/jkrogager/py4most}}. 
\begin{table}
    \caption{4MOST simulation dataset. Col 1: the name of the different subclass. Col 2: the label we used afterward. Col 3: the mean redshift of the subset. Col 4: the redshift range. Col 5: the mean median signal-to-noise, $\overline{SNR}$. 
    The first 5 subclasses are galactic objects and the last 5 are extragalactic objects.
    }
    \centering
    \begin{tabular}{|l||l|l|l|l|}
    \hline
     Col. 1 & 2 & 3 & 4 & 5 \\
    \hline \hline
     class\_subclass & label & $\bar z$ & $[z_{min},z_{max}]$ & $\overline{SNR}$   \\
    \hline \hline 
     Dyn            & 0 & -- & -- & 74.5 \\
    \hline 
     GalHR          & 1 & -- & -- & 39.9 \\
    \hline 
     ESN            & 2 & -- & -- & 12.8 \\
    \hline 
     GalDiskLR      & 3 & -- & -- & 140.4 \\
    \hline 
     MCsn           & 4 & -- & -- & 72.3 \\
    \hline 
     COSMO\_AGN     & 5 & 2.2 & [0.9, 4.0] & 6.3 \\
    \hline 
     ClusB          & 6 & 0.52 & [0.3, 1.0] & 5.8 \\
    \hline
     WAVES           & 7 & 0.32 & [0.0, 0.8] & 1.6 \\
    \hline 
     RedGAL         & 8 & 0.33 & [0.0, 1.1] & 8.7 \\
    \hline
     tides\_host     & 9 & 0.11 & [0.0, 0.6] & 19.7 \\
     \hline
    \end{tabular}
    \label{Table:2}
\end{table}
The spectral signal is obtained according to the exposure time and extinction: in particular, the exposure time is taken to be 1,200s for all spectra, and the extinction is determined by the average galactic reddening law parametrized by \citealt{2007ApJ...663..320F}. The final sample contains a total of 10 subclasses, 5 galactic and 5 extragalactic. The galactic objects are: metal-poor stars and other dynamics tracers (Dyn) of The Milky Way Halo Low/High-Resolution Survey \citep{2019Msngr.175...23H, 2019Msngr.175...26C}, Cepheids in Magellanic Cloud (GalHR) of 1001MC Survey \citep{2019Msngr.175...54C}, White Dwarf (ESN) of 1001MC Survey, Galactic disc stars (GalDiskLR) of 4MOST Surveys S1-S4 \citep{2019Msngr.175...23H, 2019Msngr.175...26C, 2019Msngr.175...30C, 2019Msngr.175...35B}, stars of Magellanic Cloud (MCsn) in 4MOST Survey S1 \citep{2019Msngr.175...23H, 2019Msngr.175...26C}. 
The extragalactic simulated sources are taken from mock catalogs and spectra provided by the 4MOST consortium extra-galactic surveys: S5 eROSITA Galaxy Cluster Redshift Survey \citep{2019Msngr.175...39F}, S6 Active Galactic Nuclei \citep{2019Msngr.175...42M}, S7 Wide-Area VISTA Extragalactic Survey (WAVES; \citealt{2019Msngr.175...46D, 2023MNRAS.tmp..715J}), S8 Cosmology Redshift Survey (CRS; \citealt{2019Msngr.175...50R}), S10 The Time-Domain Extragalactic Survey (TiDES; \citealt{2019Msngr.175...58S}). The respective contribution in simulated spectra of each survey is 2,099 (24,658, 6,056, 10,443, 13,386) for S5 (S6, S7, S8, S10). The templates used by S5, S6, S8 were obtained by stacking spectra with the method from \citet{2020A&A...636A..97C}\footnote{\url{ https://github.com/JohanComparat/qmost_templates}}. The stacked spectra were observed by SDSS within the eBOSS or the SPIDERS programs \citep{2023ApJS..267...44A} and have similar properties to the selected targets to be observed by 4MOST consortium surveys S5, S6, and S8.
As opposed to the SDSS-DR16, the classes available in the 4MOST sample are ``survey oriented''.
In fact, the templates simulated come from different methods, and they are not purely grouped by physical properties, e.g., star-forming vs. passive galaxies or AGN, but rather customized for the survey requirements, including the SNR. 
\footnote{The main reason for this particular choice is that at the moment we have finished this work there was not yet a uniform physically motivated set of templates available for galactic/extragalactic targets in 4MOST, although a list of FGK star targets (from the galactic working group, IWG3) and a catalog of stars with known labels for half a million stars from GALAH/APOGEE/RAVE/Gaia (from the ISSI team) will be available, and will be used for future GaSNet analyses. This does not represent a major issue for the purpose of this paper which aims to show the capabilities of the deep learning to perform classifications/regression tasks, regardless of the physics behind the spectra.}
This is evident, e.g., for the WAVES sample, which requires only redshift measurements of the targets, with the minimal exposure time and SNR needed to reach a reliable measurement. Table \ref{Table:2} shows the label of subclasses, SNR, and redshift distribution, while in Figs. \ref{fig: 4most spectra_1} and \ref{fig: 4most spectra_2} we show some typical spectra from each of the 10 classes. 
The galactic objects have a higher average median SNR than the extragalactic objects. 
In the 4MOST sample, galactic objects exhibit a higher $\overline{SNR}$ than those in the SDSS samples, whereas the extragalactic objects show a slightly lower $\overline{SNR}$.
The redshift distributions of the 5 extragalactic classes are shown in Fig. \ref{fig: 4most redshift}. The galaxy classes show a distribution that is similar to the one seen for the SDSS-DR16, while the quasars show a flatter distribution than the real data. As mentioned in Sect. \ref{sec:sdss_spectra}, this might help alleviate the bias associated with incompleteness. However, this also raises the question of how realistic the ``prior'' distribution adopted in simulation can be (e.g., see discussion in \citealt{2022_galnet}, for imaging mock data). We postpone this test until we can access deep 4MOST observations, fully accounting for selection effects.
Until then the 4MOST mock dataset provides us a unique opportunity to test GaSNet-II as {\it a general purpose ``survey-oriented'' classifier, based on a large variety of classes, at the same time.}
Each subclass consists of approximately 20,000 spectra, which are split into 70\%/15\%/15\% for training, validation, and testing, respectively.

\subsection{Other dataset: DESI spectra}
\label{sec:desi}
\begin{figure}
        \centering
        \includegraphics[width=0.45\textwidth]{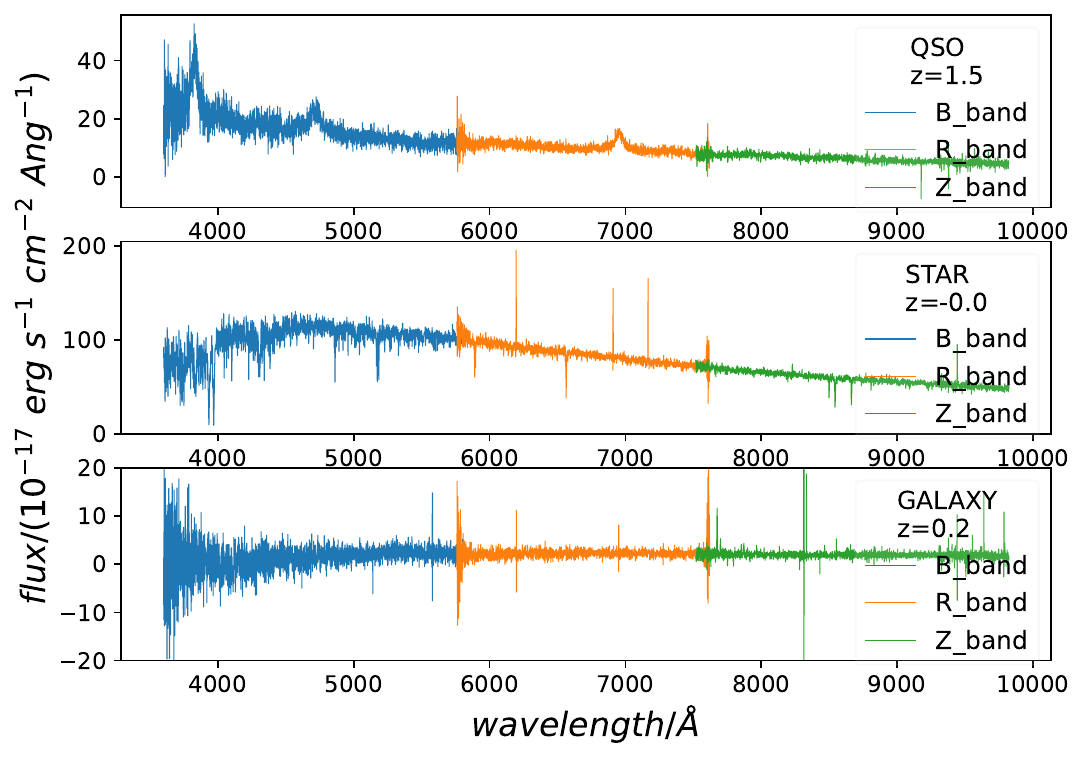}
    \caption{The typical DESI spectra of QSO, STAR, GALAXY classes.}
    \label{fig: DESI spectra}
\end{figure}

\begin{figure}
        \centering
        \includegraphics[width=0.4\textwidth]{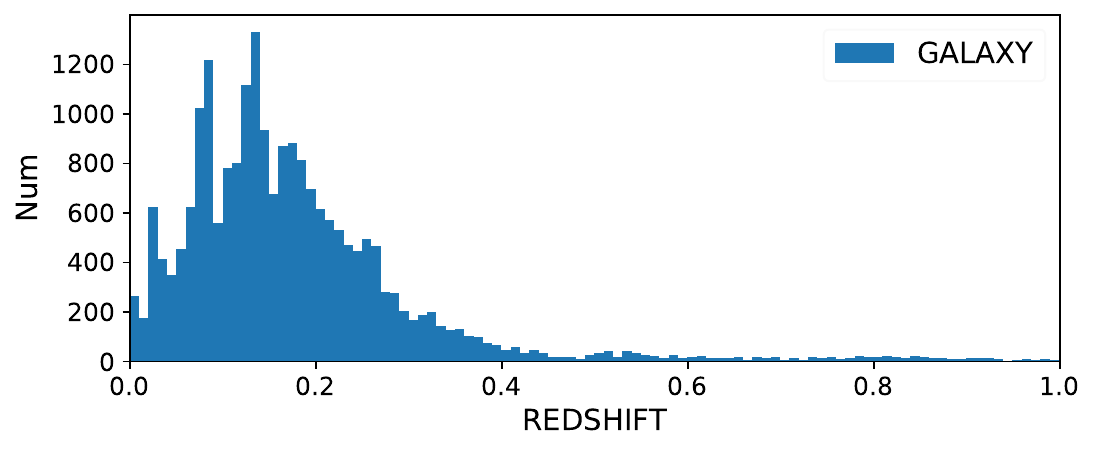}
        \includegraphics[width=0.4\textwidth]{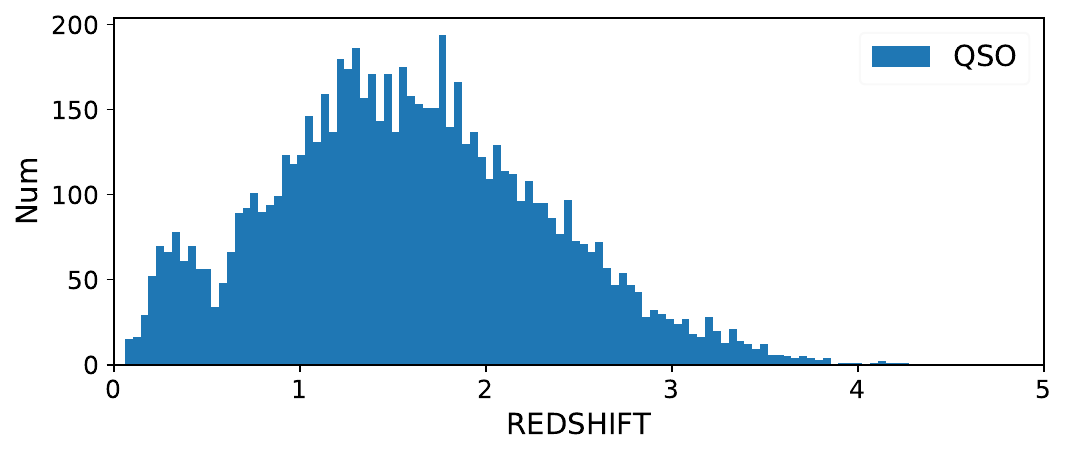}
    \caption{The redshift distribution of the DESI dataset. The mean and range of redshift are shown in Table \ref{Table:3}.}     
    \label{fig: DESI redshift} 
\end{figure}

The dataset is constituted of 21,000 randomly selected DESI spectra,\footnote{\url{https://data.desi.lbl.gov/public/edr/}} which are categorized into 3 classes, QSO, STAR, GALAXY. Each class consists of 7,000 spectra in the dataset. The DESI spectra are randomly selected from ``sv1" (``Target Selection Validation") samples and ``sv3" (One-Percent Survey) samples with SNR larger than 2 and $\it ZWARN$ flat equal 0. Spectra are split into 70\%/15\%/15\% for training, validation, and testing, respectively. In the early data release version, DESI only provides a separation of the observed object into QSO-STAR-GALAXY, with only stars possessing further subclasses (8 in total), but with too few spectra to be used for training here. Hence, the DESI dataset can be used to test GaSNet-II for a coarsely classified, poorly sampled dataset (e.g., to be compared to a similar test on SDSS-DR16 as in Appendix \ref{sec: The coarse classifier of SDSS}). 
The DESI classification and redshift prediction pipeline used {\it redrock}, a software package\footnote{\url{https://github.com/desihub/redrock/releases/tag/0.15.4}} based on fitting a set of PCA templates to every target at every redshift \citep{2023arXiv230606308D}. 
The DESI spectra consist of three bands (B, R, and Z band), with a wavelength range from 3,600 - 9,800 \AA. Once again, spectra are interpolated to cover 5,001 pixels in the wavelength range 4,000 - 9,000 \AA, which are then used for the training. More details of the dataset are shown in Table \ref{Table:3}. The samples have a similar level of $\overline{SNR}$ to the SDSS samples (Table \ref{Table:1}) after the selection conditions were imposed. In Fig. \ref{fig: DESI spectra}, we show some spectra from the 3 different classes. Here, we have also highlighted, in different colors according to the legend, the sub-spectra collected from the three DESI arms, that are combined in the final DESI full wavelength range spectra. Finally, the redshift distributions of galaxy and quasar samples are shown in Fig. \ref{fig: DESI redshift}.

\section{Pipeline description and training}
\label{sec: Method}
Thanks to their flexibility, efficiency, and accuracy, the multi-networks combination can be applied to the prediction of various astronomical parameters, and  possibly form a fully-automatic DL pipeline. The Convolution Neural Network \citep[CNN;][]{Krizhevsky2012ImageNetCW} and the Residual Connection \citep[$\it ResNet$;][]{2015arXiv151203385H} are two of the most widely tested DL architectures. CNN and $\it ResNet$ have been extensively applied to classification and regression problems in astronomy, such as the photometric strong lens detection \citep{2019MNRAS.482..807P, 2021ApJ...923...16L, 2019MNRAS.482..313L, Huang:2019lei}, galaxies morphology classification \citep{2004MNRAS.348.1038B, 2020A&A...638A.134D, 2022MNRAS.509.4024D}, star, galaxy, or quasars identification \citep{2017MNRAS.464.4463K, 2018MNRAS.476.1151P, 2018arXiv180809955B, 2019ApJ...879...72G}, photometric redshift predictions \citep{2016A&C....16...34H, 2019A&A...621A..26P, 2022A&A...666A..85L}, and stellar parametrization \citep{2018MNRAS.475.2978F, 2019MNRAS.483.3255L,2024A&A...682A...9G}.

\begin{table}
    \caption{DESI dataset. Col 1: the name of different classes. Col 1: the label. Col 3: the mean redshift of the subset. Col 4: the redshift range. Col 5: mean signal-to-noise, $\overline{SNR}$. 
    }
    \centering
    \begin{tabular}{|l||l|l|l|l|}
    \hline
     Col. 1 & 2 & 3 & 4 & 5 \\
    \hline \hline
     class & label & $\bar z$ & $[z_{min},z_{max}]$ & $\overline{SNR}$   \\
    \hline \hline 
     STAR           & 0 & -- & -- & 19.1 \\
    \hline 
     QSO            & 1 & 1.59 & [0.06, 4.27] & 6.54 \\
    \hline 
     GALAXY         & 2 & 0.196 & [0, 1.69] & 7.53 \\
     \hline
    \end{tabular}
    \label{Table:3}
\end{table}

In this paper, we construct a multi-network pipeline system, which is constituted by several, small, self-similar $\it ResNet$ network models. The pipeline intends to map the pixel-level 1D spectra to return a classification probability and redshift. The classifier first is able to distinguish between subclasses. For instance, in the case of SDSS-DR16 (see Table \ref{Table:1}), it separates the 7 subclasses of stars (A0, F5, FG, K1, K3, K5) that, being ``galactic'' objects, are assumed to have redshift $z=0$, and the 6 extragalactic objects, 4 of galaxies (nan, 
AGN, STARBURST, STARFORMING) and 2 of QSOs (nan, BROADLINE). In total, there are 13 different classes. Then, on these extragalactic classes, GaSNet-II performs the redshift predictions and error estimates. Similarly, for 4MOST (see Table \ref{Table:2}), the classifier separates the objects in the 5 star classes (Dyn, GalHR, ESN, GalDiskLR, MCsn) and extragalactic classes (COSMO\_AGN, ClusB, WAVES, RedGAL, tides\_host), then, for these latter, the GaSNet-II predicts the redshift and the errors. For DESI the classifier just separates into three coarse classes (Table \ref{Table:3}) and the redshift is measured for the galaxies and QSOs. 

In this section, we introduce the details of the GaSNet-II architecture, the strategy for network training, and error estimates. We start by discussing in detail the training of the pipeline using the reference dataset over which we want to test the capabilities of the pipeline, i.e., the SDSS-DR16 sample. 
The structure and training of the pipeline will be the same for 
the other two datasets, i.e., 4MOST and DESI, except that, due to the different numbers of labels (see \S\ref{sec: Data}), only the structure of the output will be different. For the latter datasets, we will discuss directly the performances on the test sample in \S\ref{sec: Results}.

\subsection{GaSNet-II: Philosophy and Architecture}
\label{sec:architecture}
The philosophy behind the GaSNet-II architecture is based on two principles: simplicity and efficiency. 
Simplicity, because we want to build a network made of ``lighter'', self-similar $\it ResNet$s. The reason is that, by controlling each small network performance, we can easily check and control the whole pipeline performance. Also, having several $\it ResNet$ blocks makes it easy to customize different sub-networks for different tasks.
Efficiency, because GaSNet-II is able to parallelize classification and redshift predictions, which generally are part of a serial two-step process in classical pipelines, as the redshift accuracy is class dependent. Indeed, it is more difficult to determine the redshift for specific classes. An obvious example is passive vs. active galaxies, as the former does not have as many high SNR features as the emission lines of the latter \citep{2006MNRAS.370..721M}.

\begin{figure*}
    {\begin{minipage}{0.65\linewidth}
           \includegraphics[width=1\textwidth]{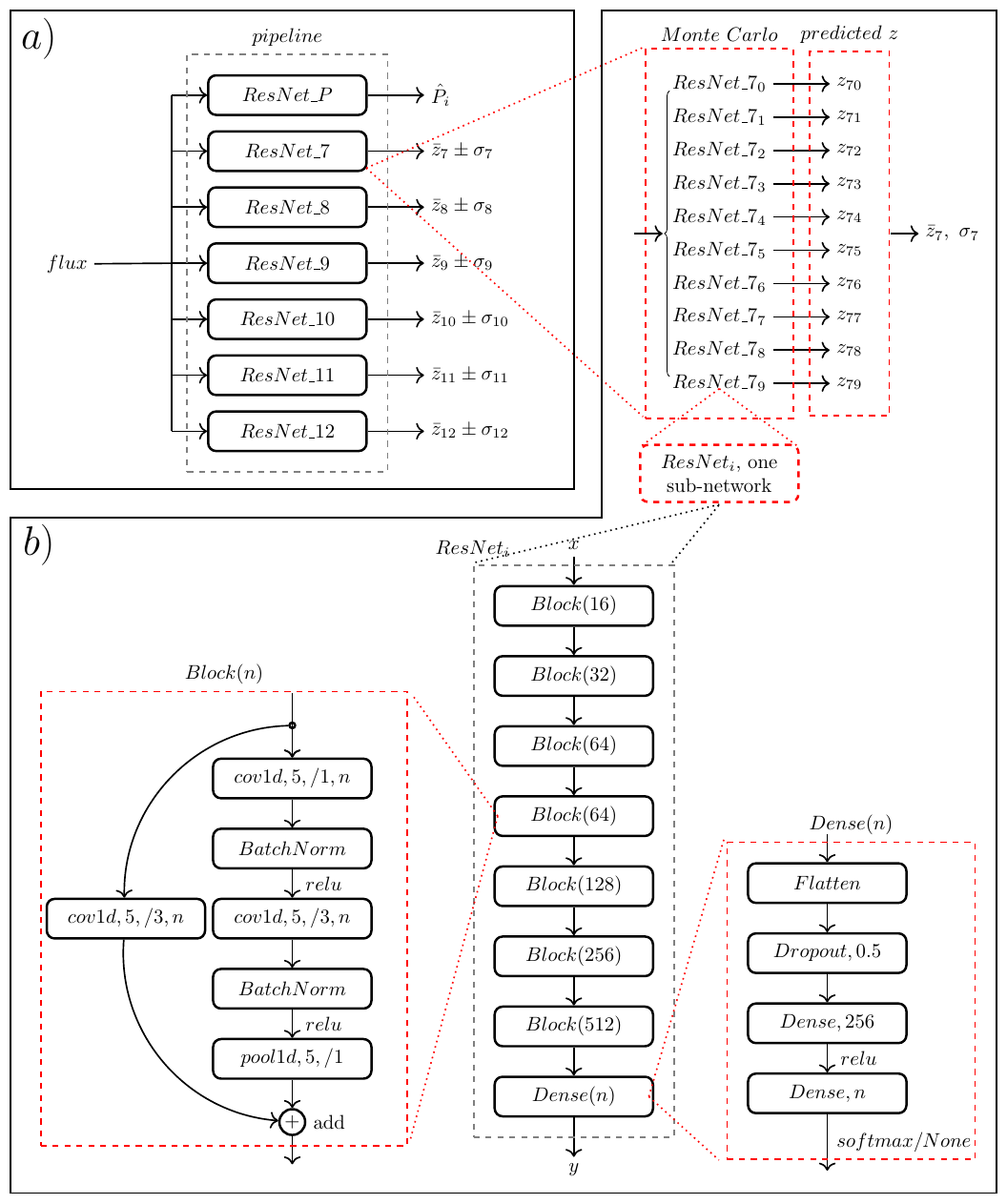}
    \end{minipage}}
\caption{{\it Panel a)}: the general structure of the multi-networks pipeline. $\it ResNet\_P$ is used as a classifier and  $\it ResNet\_7-12$ is used for redshift prediction of extragalactic targets (note that $\it ResNet\_0-6$ are missing because we do not need to predict the redshift of stars). One of the advantages of this structure is that it is simple and controllable, and can be trained and predicted in parallel. {\it Panel b)}: the detailed description of single sub-network $\it ResNet_i$ (bottom figures) architecture, made by small blocks. The input of the network is 5001-pixel spectrum flux, and the output is the probability or redshift. The difference between classification ($\it n=13, \ softmax$) and redshift prediction ($\it n=1, \ None$) is the output dimension and the activation in the last layer. A feature-extract block $\it Block(n)$ and a fully connected block $\it Dense(n)$ are shown. $\it cov1d$ is the 1-D convolution layer. In one $\it cov1d$ rectangle, $5$ is the kernel size; $/3$ is the stride size; $n$ is the number of channels. $\it relu$, $\it softmax$ are the activate function, $\it None$ represents no activate function here, that means liner. The left $\it cov1d$ in the $\it Block(n)$ shortcut is used to match the shape. $\it pool1d$ is a 1-D Maxpooling layer. As a schematic, the top right panel shows how to predict the redshift error of the label 7 (GALAXY\_nan) subclass in parallel. Though 10 (customized) same sub-networks, trained by the same data but with different initial weights, 10 different redshifts were obtained from a single spectrum input. The expectation and error can be calculated. Other redshift errors are obtained in the same way.} \label{fig: full pipeline} \label{fig: blocks} 
\end{figure*}

To achieve this second objective, for GaSNet-II we decided to use a particular architecture made of parallel sub-networks, each one specialized on a specific task. This is sketched in Fig. \ref{fig: full pipeline}$a$, where a sub-network is used to classify and give the probability to each object to belong to a series of predefined classes, while other parallel sub-networks, trained on each and only classes that need redshift estimates, are used to give the redshift predictions and error estimates. 
Obviously, the numbers of sub-networks are preassigned according to the number of those classes with redshift, i.e. the training sample. In fact, being GaSNet-II a supervised network, the classes and redshifts need to be known as labels of the training sample used to train the networks.

However, all sub-networks are almost the same, in terms of their internal structure. Specifically, the multi-network pipeline consists of 1 $\it ResNet\_P$ model to predict the probability, $\hat P$, of each subclass for classification, and 6 (identical) $\it ResNet\_i$ to predict the redshift, $z$, of different extragalactic objects, respectively. The index $i$ corresponds to the label in Table \ref{Table:1}. The input of all sub-networks are the 1-D spectra, in flux units. As we will detail later, in this latter phase, GaSNet-II performs a Monte Carlo (MC) test, that allows us to estimate the errors, $\sigma_z$, on the redshift predictions. 
Hence, the output of the GaSNet-II pipeline is a 13-dimensional array of terns ($\hat P,~z,~\sigma_z$). The final input/output can be schematically summarized as:
\begin{align}\label{eq: function}
    F(flux) =
    \begin{cases}
    (P_i, 0),   & i \in [0,6], \\
    (P_i, z_i, \sigma_{z,i}), & i \in [7,12]
    \end{cases},
\end{align}
where $\hat P_i$ are the probability from the $\it ResNet\_i$ classifier, $z_i$ are the redshift predictions and $\sigma_{z,i}$ are the redshift uncertainty, from the 6 $\it ResNet$ regression models. 

\zhong{
In terms of workflow, the classification is performed in parallel to (and hence independently from) the redshift prediction, hence this latter does not impact the classification. In principle, one can guess that this is a disadvantage as the knowledge of the redshift could improve the classification (for instance this is easy to understand for stars that have $z\sim 0$). However, the GaSNet-II seems to reach already very high classification performaces 
($\sim 99\%$, see Appendix \ref{sec: The coarse classifier of SDSS}, Fig. \ref{fig: coarse_classifier_SDSS}) without this information. On the other hand, there are advantages of this ``parallel'' approach: a) one can scale-up the network by adding training samples for more classes, 
making it easy to extend the classification to other objects or even other targets, such as stellar parameters; b) parallelization reduces the impact of correlations between different quantities; c) for this reason it is extremely flexible and can effectively applied to 
different SNRs and various surveys, as we will demonstrate later in this paper; d) it provides a reasonable uncertainty estimation, which is a robust starting point for subsequent Bayesian analyses; e) neural networks are powerful interpolators, thus also good at classifying spectra that lie within a learned multi-dimensional surface that cross-correlation won’t grasp.}

\subsection{GaSNet-II: pipeline description}
\label{sec:pipe_descr}
In this section, we describe in detail the full end-to-end pipeline, which we have broadly described in the previous section. In the following, for brevity, we define the input of the sub-networks, $x$, and the fitting labels of sub-networks, $y$, as:
\begin{align}
    & x    = {\it flux}/\sqrt{N}, \quad 
    N = \sum^{5001}_{j=1} {\it flux}_j^2  \label{eq:x} \\
    & \hat y  = {\it one-hot}(i), \quad i \in [0,12] \label{eq:y} \\
    & y_i  = z_i, \quad i \in [7,12], 
\end{align}
where the $j$ represents the pixel index, from 1 to 5001, and the {\it one-hot} encoder converts the categorical data into digits for example, {\it one-hot}(0)=001, {\it one-hot}(1)=010, {\it one-hot}(2)=100, etc. In Eq. \ref{eq:x}, the {\it flux} is normalized just like a vector. The fitting labels $\hat y$ are the labels converted by the {\it one-hot} encoder from Table \ref{Table:1}. The fitting parameters $y_i$ are the spectroscopic redshifts provided by the catalog. To prevent the prediction of very high-redshift values, where the currently available training samples are too poor to give accurate results, we limit them to the range $z \in [0,5]$. The loss functions used are
\begin{align}
    & {\it loss} = - \hat y_{t} \cdot \log(\hat y_{p}), \quad \text{(categorical cross-entropy)} \\
    & {\it loss}_i = \begin{cases}\frac{1}{2}(y_{it}-y_{ip})^2, & |y_{it}-y_{ip}| \leq \delta \\ \delta|y_{it}-y_{ip}|-\frac{1}{2} \delta^2, & |y_{it}-y_{ip}|>\delta\end{cases}, \quad \text{(Huber loss)}.
\end{align}
where $\hat y_{p}, y_{ip}$ are the prediction values and $\hat y_{t}, y_{tp}$ are the true values, and parameter $\delta = 0.1$. Huber loss combines the advantages of mean absolute error and mean square error, and alleviates the sensitivity to outliers.

As seen in the previous section, the GaSNet-II pipeline is constituted by 7 almost identical $\it ResNet$ sub-networks. 
This is shown now in more detail in Fig. \ref{fig: blocks}$b$, where we offer a complete schematic view of the full architecture, which we describe below. 
Starting from the general structure seen in Panel $a$, we see that the sub-network architecture consists of a series of $\it ResNet$ "blocks". One of the advantages of using the sub-network architecture, discussed in \S\ref{sec:architecture}, is that it is particularly convenient to perform MC tests, which are the foundation of the GaSNet-II error estimates, as shown by the ``zoom-in'' inset (top-right) in the same Fig. \ref{fig: blocks}$b$. 

The idea behind the MC run is to use the different (10) sub-networks\footnote{\zhong{The choice of 10 networks is primarily to optimize the computational resources, to make GaSNet-II usable in small medium scale servers with no much impact on the final results.
For instance, considering the convergence of uncertainty in high SNR, Fig. \ref{fig: Delta vs SNR} shows that 10 sub-networks are sufficient} \zhong{to robustly assess uncertainties and we do not expect to improve this result by increasing the number of sub-networks.}
} with the same data, e.g., a spectrum of an object of a given class, but with different initial network weights.
\zhong{In practice, the initial sub-network parameters are set by a random Gaussian distribution, which establishes a random initial condition for the entire process, thus mimicking a MC experiment. However, this can also be seen as an ensemble training/MC, which is a relatively common practice in deep learning \citep[e.g.,][]{2016arXiv161201474L, 2021arXiv210402395G}, and applied in the synthetic stellar spectra physical properties estimating\citep[e.g.,][]{2020MNRAS.498.3817B}.}
This allows us to evaluate the stability of the output, by changing the initial condition of the training process. For the robust data points, different sub-networks are expected to predict values that are close to the ground truth, like the best-fit values that find a global (or even a local) minimum in the $\chi^2$ topology. 

On the other hand, for the ``unstable'' points different sub-networks are expected to find different predictions, like happens in best-fitting if the $\chi^2$ has many local minima. In this way (despite the number of parallel experiments being only 10) we can separate the robust from unstable prediction targets. Hence, estimating the cumulative uncertainties on the final target estimates has two main objectives: 1) to associate a redshift and an error based on a probability distribution function (PDF) to every given target; 2) to test the robustness of the network, by quantifying the overall predictions scatter with respect to the ground truth.


Indeed, from the ``zoom-in'' inset of the MC test, in Fig. \ref{fig: blocks}$b$, we can see that the MC step provides a mean value, $\overline{z}$ and a variance, $\sigma$. 
\begin{table}
    \caption{Models detail. "pars" is the number of network parameters. "Num" is the number of training spectra used. "loss" is the minimal loss on the validation set. "acc" is the max accuracy on the validation set, and "MAE" is the minimum mean absolute error on the validation set. \protect\footnotemark}
    \centering
    \begin{tabular}{|l||l|l|l|l|l|}
    \hline
    Name &pars ($10^6$)&Num ($10^3$)&loss ($10^{-3}$)&acc/MAE\\
    \hline \hline 
    $\it ResNet\_P$        & 4.16 & 182 & 218   & 91.9\% (acc) \\
    \hline 
    $\it ResNet\_7$      & 4.16 & 14  & 0.868 & 0.011 \\
    \hline 
    $\it ResNet\_8$      & 4.16 & 14  & 0.152 & 0.003 \\
    \hline 
    $\it ResNet\_9$      & 4.16 & 14  & 0.066 & 0.001 \\
    \hline 
    $\it ResNet\_10$   & 4.16 & 14  & 0.112 & 0.002 \\
    \hline 
    $\it ResNet\_11$   & 4.16 & 14  & 10.2  & 0.107 \\
    \hline 
    $\it ResNet\_12$   & 4.16 & 14  & 2.32  & 0.027 \\
    \hline
    \end{tabular}
    \label{tabel: model detail}
\end{table}
\footnotetext{Both trained by an NVIDIA Tesla P40 GPU}

This is also done in parallel for the 6 extragalactic classes to obtain:
\begin{equation}
\bar z_i = \sum_{j=0}^{9} z_{ij} / 10 \\ \label{eq:mean}
    \end{equation}
\begin{equation}
    \sigma_i = \sqrt{\sum_{j=0}^{9} (z_{ij} - \bar z _i)^2} / 10, \label{eq:sigma}
\end{equation}
as shown in Fig. \ref{fig: blocks}$b$.
The predicted expectations and errors will be shown in section \ref{sec: Results}. In Table \ref{tabel: model detail} we show the number of parameters, and the number of spectra adopted for the training of the sub-networks.

To check the effectiveness of the use of the mean redshifts and their errors from Eqs. \ref{eq:mean} and \ref{eq:sigma}, we also provide the {\it point estimate} redshift for each target. These are based on a version of the network with only one $\it ResNet$ in the MC module in Fig. \ref{fig: full pipeline}$a$ for each of the classes, and compare these with the ones we obtain from the MC run. These point estimates are analogous to individual measurements from standard techniques, like cross-correlation ({\it redmonster}) or template fitting ({\it redrock}) and are meant to provide a realistic scatter of the estimates due to the combination of the data quality and the deep learning method. 

Finally, Fig. \ref{fig: blocks}$b$ (top right) shows the convenience of the sub-network architecture as the structure of each $\it ResNet$ is the same for each one of the sub-networks, regardless of whether it is used to classify (e.g., $\it ResNet\_P$) or to predict redshifts ($\it ResNet\_i \ _{0-9}$).

\begin{figure}
        \includegraphics[width=0.5\textwidth]{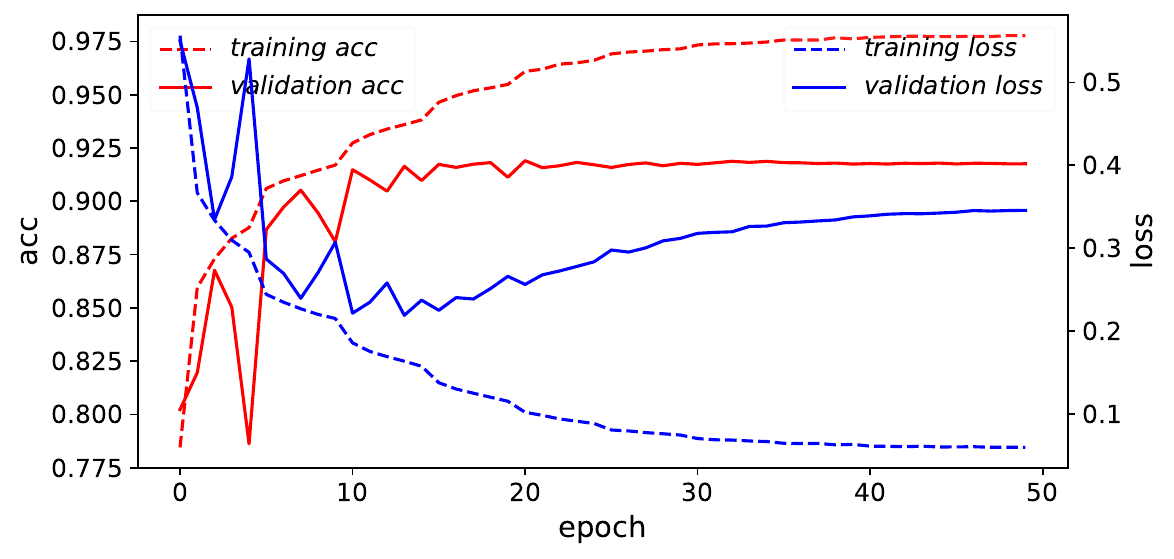}
        \hspace{-0.6cm}
        \includegraphics[width=0.48\textwidth]{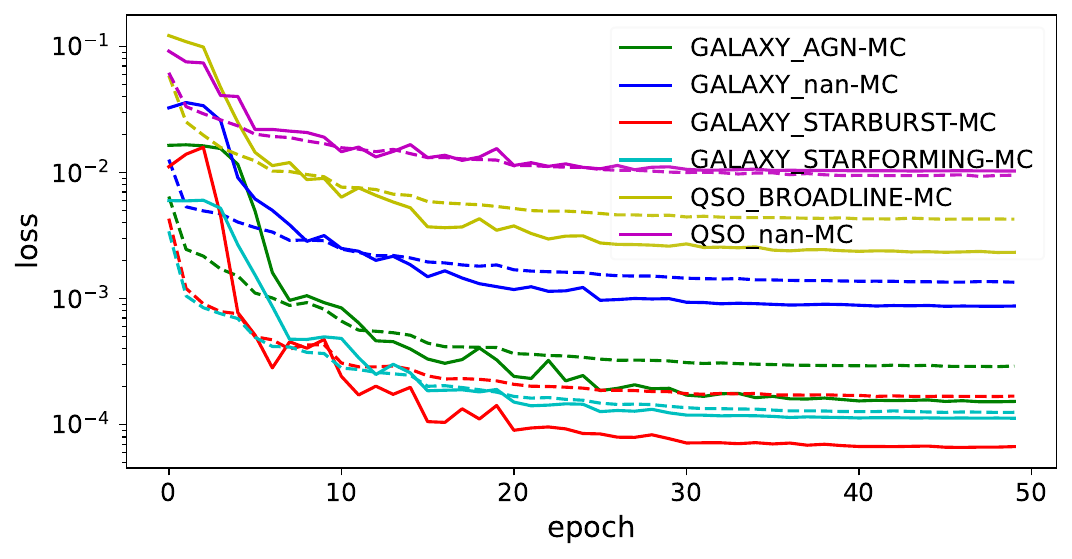}
        \hspace{-0.6cm}
        \includegraphics[width=0.48\textwidth]{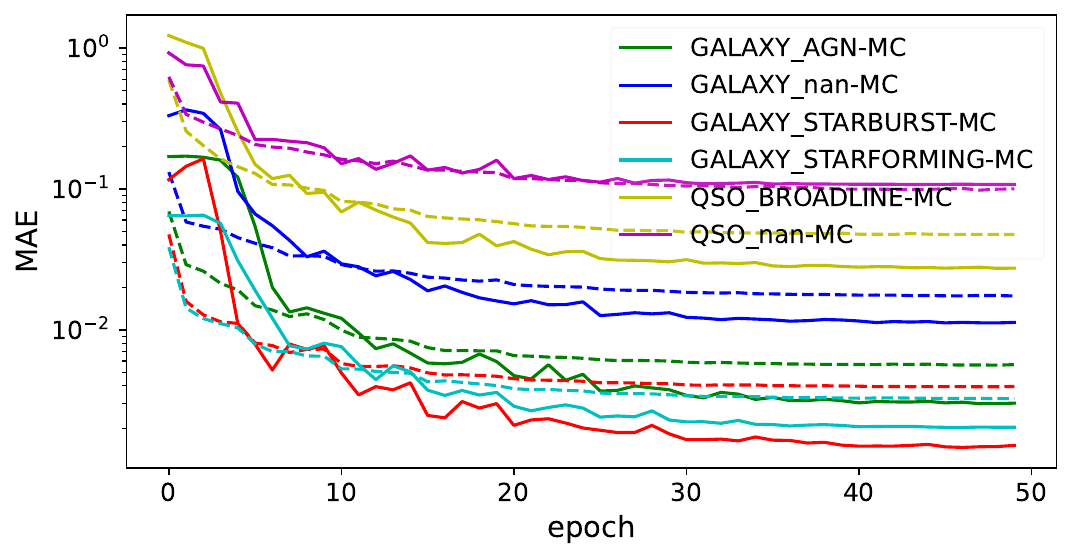}
    \caption{The training results for 50 epochs. 
    We adopted a dropout rate of 0.5 in the dense layer to prevent overfitting during training. The first panel is the loss and accuracy of $\it ResNet$, which is used to classify the spectra. The second and third panels are the loss and the MAE of $\it ResNet\_{\rm i}$, which are used to predict the redshift. The dashed lines are the results of the training set, and the solid lines are the results of the validation set.
    The significant fluctuation in the first 20 epochs is due to the significant varying of learning rates. The overall worse performance in the training set is because we only employed the dropout in the training processes.
    }
    \label{fig: training}
\end{figure}

\subsection{GaSNet-II training: SDSS-DR16}
\label{sec:training}
The training of GaSNet-II aims to minimize the loss function and maximize the accuracy (of the classification and predictions). As mentioned, all ``specialized'' sub-networks are trained in parallel. 

As a training set for the classifier network ($\it ResNet\_P$, in Fig. \ref{fig: full pipeline}$a$), we use a total of 182,000 SDSS-DR16 spectra, incorporating the 13 subclasses, each of them covered by 14,000 spectra for their training. By definition, each of the redshift prediction networks ($\it ResNet\_{\rm i}$ in Fig. \ref{fig: full pipeline}$a$), makes use of the same 14,000 used by the classifier for each subclass $i$, but with the purpose of mapping the input spectra to the labeled redshifts.

Under such partitioning of the training data, one can imagine that the classifier is set to search for the redshift in a larger parameter space, while the redshift ``regressor'' networks, $\it ResNet\_{\rm i}$, are set to search for the specific redshifts of each subclass of spectra.

The result of the training process over the validation set is shown in Fig. \ref{fig: training}, where a step learning rate is used. The learning rate starts at $10^{-3}$, then slowly decays to $10^{-6}$ at the end (halving every 5 epochs when the epoch < 50) during 50 training epochs. The loss curves in the upper panel of the figure might indicate some slight overfitting, while the accuracy curves show that it does not affect the performance. The accuracy curve remains flat as more training epochs are implemented, meaning that it has achieved its upper limit. We have used a 0.5 dropout rate in the final layer to mitigate potential overfitting in the training set. Overfitting could be further reduced by using fewer network parameters or increasing the size of the training data (e.g., through online additive-noise data augmentation), however, due to the small amount of overfitting to correct we decided to test these strategies in future analyses. The checkpoints with maximum accuracy or minimum Mean Absolute Error (MAE) are used as the model of the pipeline. Table \ref{tabel: model detail} shows an average classification accuracy of 91.9\% from $\it ResNet\_P$, as well as a range of MAE for redshift estimation across different subclasses, ranging from $0.001$ to $0.107$. The number of trainable parameters of the sub-network and the number of training samples are also provided.

\begin{figure}
    \hspace{-0.5cm}
        \includegraphics[width=0.5\textwidth]{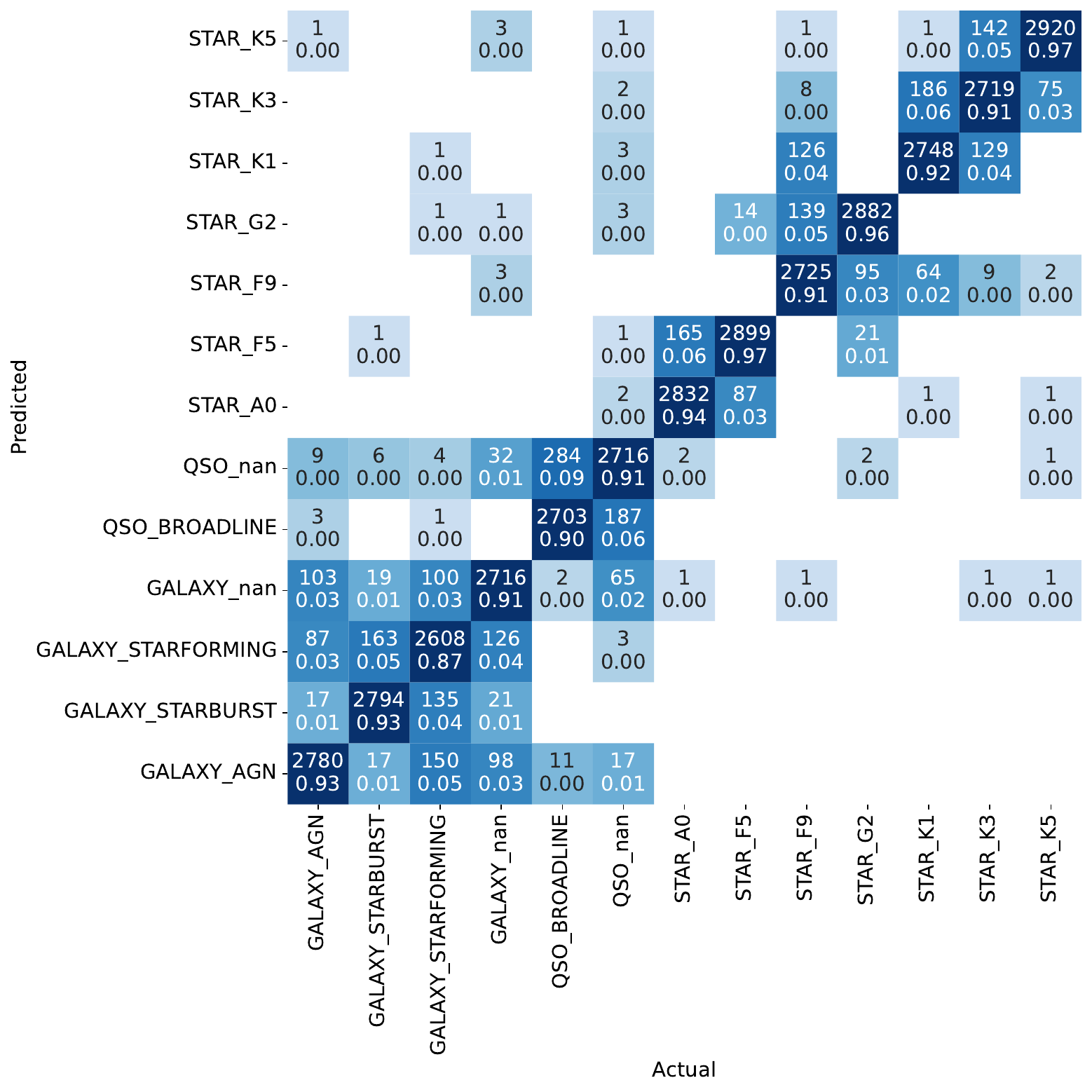}
    \caption{Confusion matrix results for the classification of the SDSS test set. The predicted and actual labels for each subclass (see Table \ref{Table:1}) are listed on the left and bottom sides, respectively. Each subclass has 3\,000 test samples. The average accuracy is 92.4\%, and most are larger than 90\% (except the GALAXY\_STARFORMING subclass). The matrix should be read along columns, that is the direction along which 100\% of the actual labels are distributed by the classifier.}
    \label{fig: confusion matrix}
\end{figure}


\begin{figure*}
        \centering
        \includegraphics[width=0.85\textwidth]{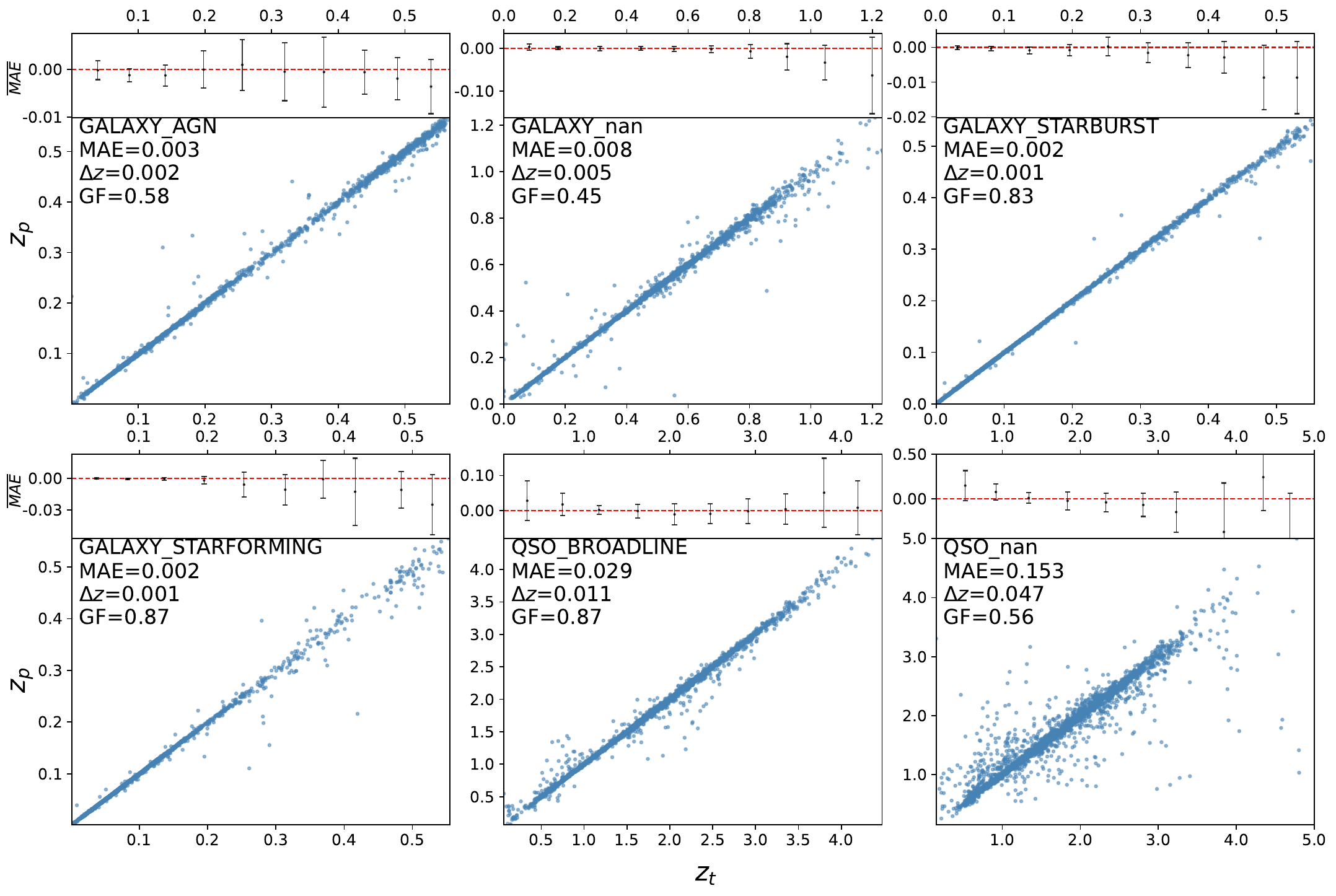}
    \caption{Redshift predictions of 6 extragalactic SDSS subclasses, each of which used one sub-network. The subclasses GALAXY\_STARBUSRST and GALAXY\_STARFORMING have the best redshift estimations, with an error of $\Delta z=0.001$. This can be attributed to the presence of significant emission lines in their spectra, as shown in Figure \ref{fig: spectra_2}. The subclass QSO\_nan has the worst estimation with an error of $\Delta z=0.047$. This subclass is characterized by the lowest signal-to-noise ratio (SNR), a high redshift range (Table \ref{Table:1}), and a weaker broad emission line signal in the spectrum (Figure \ref{fig: spectra_2}). Error bars on each redshift bin (10 bins) are plotted at the top of the panel. The $\it MAE$ for each bin is used as the error bar. The plot clearly indicates that errors become significant at the higher redshift end, which is attributed to the lack of training samples in that region.}
    \label{fig: redshift prediction}
\end{figure*}

\section{Results}
\label{sec: Results}
In this section, we show the results of the pipeline using the same SDSS DR16 dataset and test sample. However, in the second part of the section, we also show the results of GaSNet-II, customized for the 4MOST mock data and DESI early data release, to demonstrate the potential for future application on Stage-IV surveys. 

\subsection{Statistical parameters}
Before looking into the results, we introduce the statistical indicators to quantify the performance of GaSNet-II, specifically for the redshift accuracy. The first parameter is the ${\it Bias}$, defined as:
\begin{align}
    {\it Bias} = |\ln(\frac{1+z_t}{1+z_p})|,
\end{align}
where $z_t$ represents the real value and $z_p$ represents the prediction value. The {\it  Bias} measures the deviation of $z_p$ from $z_t$.
In particular, we can use it to define the fraction of the ``good'' estimates, {\it  Good\_Frac} (GF), as the fraction over the total number of spectra, $N$, of the redshift estimates for which the {\it  Bias} is smaller than the related threshold, $thr\_x$, that can differ for different classes ($\it x=gal,~qso$). Hence:
\begin{align}
   {\it  Good\_Frac\_x }= \frac{\it N(Bias < thr\_x)}{N}.
\end{align}
We set the threshold of the galaxy species (nan, star-forming, starbursts and AGN), {\it thr\_gal}$\rm =0.0015$, such that optimal predictions are defined as ${\it  Bias} < \rm 0.0015$, and the threshold of the QSO (nan and broadlines),  {\it thr\_qso\rm  =0.015}, which qualify as good the predictions with ${\it  Bias} < 0.015$. 

The second parameter is redshift relative bias $\Delta z$, defined as:
\begin{align}
    &\Delta z=|z_p-z_t|/|1+z_t|,
\end{align}
which is more intuitive than the {\it Bias} to interpret redshift discrepancies. 
In particular, this is closely related to the $MAE$, which is the mean of the $\Delta z$ numerator, i.e.:
\begin{align}
    \text{MAE}=\text{Mean}(|z_p-z_t|).
\end{align}
As a reference, for the SDSS and DESI pipelines, $\Delta z < 0.01$ was essentially used as the catastrophic prediction threshold \citep{2012AJ....144..144B,2016AJ....151...44D, 2023AJ....165..124A}, although it was less strict for high-velocity dispersion QSOs. 

\subsection{SDSS-DR16 spectra}
\subsubsection{Classification}
\label{sec:sdss_class}
As discussed in \S\ref{sec: Method}, the $\it ResNet\_P$ sub-network gives the classification probability ($P_i$) for each of the input spectra.
This is the fastest task performed by GaSNet-II; it can perform the classification prediction of the 39,000 spectra belonging to the test sample in about one minute (excluding read time). The corresponding confusion matrix is shown in Fig. \ref{fig: confusion matrix}. 
Here we see that most of the subclass accuracies are larger than 90\%, except for the subclasses GALAXY\_STARFORMING. The average accuracy of the 13 subclasses is 92.4\%. This average accuracy is certainly driven by the SNR of the spectra, as higher SNRs allow the network to better separate the spectra. This is shown in Appendix \ref{sec:average classification accuracy and SNR}, where we use the same GaSNet-II to classify increasingly higher SNR spectra and find that the average accuracy can reach a limit of $\sim 96\%$ for the highest SNRs. The star-forming galaxies are the class with lower accuracy, possibly due to a larger overlap (and thus a more uncertain classification) with other ``emission line'' classes, e.g. AGN and starburst, but also with normal galaxies (GALAXY\_nan class), possibly because low star-forming galaxies do not have strong enough emission lines to distinguish against non-star-forming systems. Some additional confusion can have a more physical origin, such as the smooth transition between AGN-dominated and host-galaxy-dominated signals.
Furthermore, the accuracy of QSO\_nan is also relatively low, but in this case, we track the reason for the typically low SNR, as seen in Fig. \ref{fig: SNR}. 
Despite QSO\_nan (GALAXY\_STARFORMING) performing  at 91\% (87\%) level, the missing sources are misclassified as QSO\_BROADLINE (GALAXY\_nan, GALAXY\_STARBURST, GALAXY\_AGN), which means that only the level of activity (intensity of the lines) moves some objects from one subclass to the other. If we also consider the arbitrariness in the separation of these subclasses in the SDSS classification, we believe that the accuracy reached by the GaSNet-II represents possibly a lower limit.

In Appendix \ref{sec: The coarse classifier of SDSS}, we have collapsed all the sub-classes on the three major classes of star/galaxy/QSO, which shows an average of 99\% accuracy.
This test is important to reproduce the ``primary'' coarse classification each of the forthcoming surveys will implement (see, e.g., DESI in \S\ref{sec:desi_class}, for comparison). The main result is that a higher accuracy can be achieved (99\% on average) with fewer classes, using the same training data and network architecture.

\subsubsection{Redshifts: point estimates}
\label{sec:point_SDSS}
As anticipated in \S\ref{sec:pipe_descr}, we want to first derive the redshift point estimate for a single measurement from the spectra. This has an intrinsic error, which is due to a series of factors that we simplify into two categories: 1) SNR of the spectra and 2) measurement method. The former is linked to the structure of the data and how the features used for the redshift estimates are detected and measured (emission/absorption lines, 4,000\AA\ break, etc.). The latter is linked to the accuracy of the method: for the DL tools, this lies in the impact of the weights and random seeds in the network. These two factors are not independent as, for instance, high SNR spectra make the impact of the weights minimal as the network tends to converge to a more robust estimate, and vice versa. Hence, the point estimate should reflect more the scatter due to these intrinsic sources of errors. 

Fig. \ref{fig: redshift prediction} shows the ``point estimate'' redshift predictions of the 6 extragalactic SDSS-DR16 subclasses, described in \S\ref{sec:pipe_descr}.
The overall impression is a rather good agreement between the predictions from GaSNet-II and the SDSS-DR16 redshifts, with rather small $\Delta z$ and MAE, and a minimal fraction of catastrophic estimates, except for the QSO\_nan subclass. The best accuracy is found for the GALAXY\_STARBUSRST and GALAXY\_STARFORMING , subclasses, $\Delta z=0.001$, while QSO\_nan shows the worst $\Delta z=0.047$. These accuracies are still about one order of magnitude larger than the ones required for redshift catalogs (see e.g. \citealt{2012AJ....144..144B}), but this is not a major concern for this analysis that is not meant to optimize the redshift accuracies\footnote{Preliminary tests considering anomaly detection show that we can achieve $\Delta z\sim10^{-4}$. This will be discussed in upcoming analyses.}. The GF is generally larger than $\sim50\%$ but reaches $80\%$ relevant fractions only for three classes.  
We see an increasing scatter of the predictions at higher redshifts in almost all categories, mainly driven by the poor coverage from the training samples of high redshifts. As we will see, training on mock spectra can strongly alleviate this problem.
The relatively poor performance of the QSO\_nan sample, as we mentioned above, is additionally driven by the low SNR of the spectra. As we will discuss later, the SNR has a large impact on the accuracy of the predictions. 
\begin{figure}
\centering
\vspace{-0.2cm}
        \includegraphics[width=0.44\textwidth]{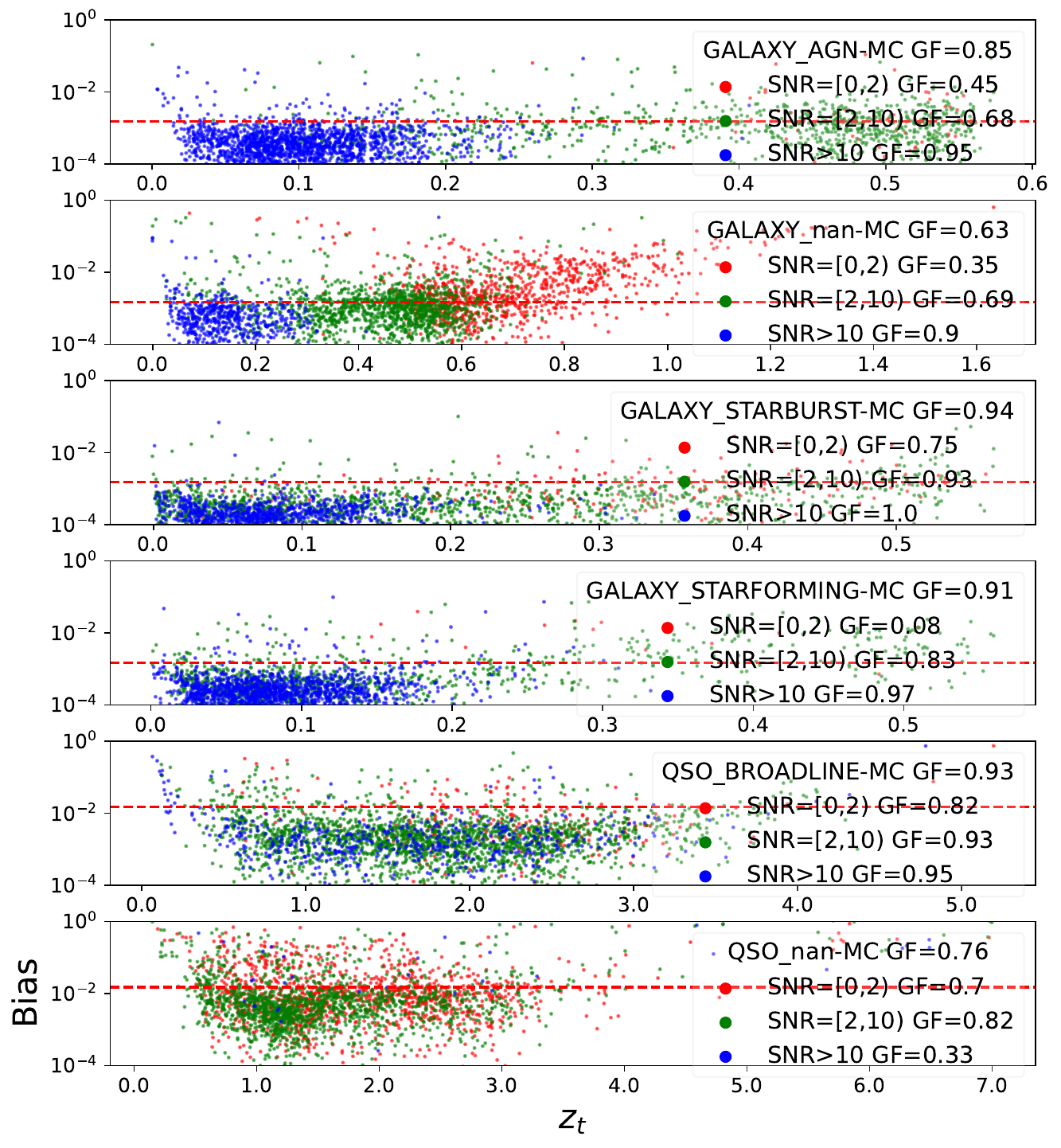}
    \caption{{\it} Bias as a function of the redshift for different extragalactic SDSS classes as indicated by the legend on the right. 
    The spectra are divided into low (red), medium (green), and high (blue) SNR to show the performance at different noise levels. The GF within each SNR bin is reported in the legend. The plot shows clearly that estimate deviations exhibit more scatter as the SNR decreases, implying larger statistical errors. The errors increase at the high redshift end, where the SNR is typically lower. Another source of scatter is that as the redshift increases, the training samples become smaller. See also \S\ref{sec:sdss_spectra}.}
    \label{fig: Good vs SNR}
\end{figure}

\begin{figure*}
        \centering
        \includegraphics[width=0.85\textwidth]{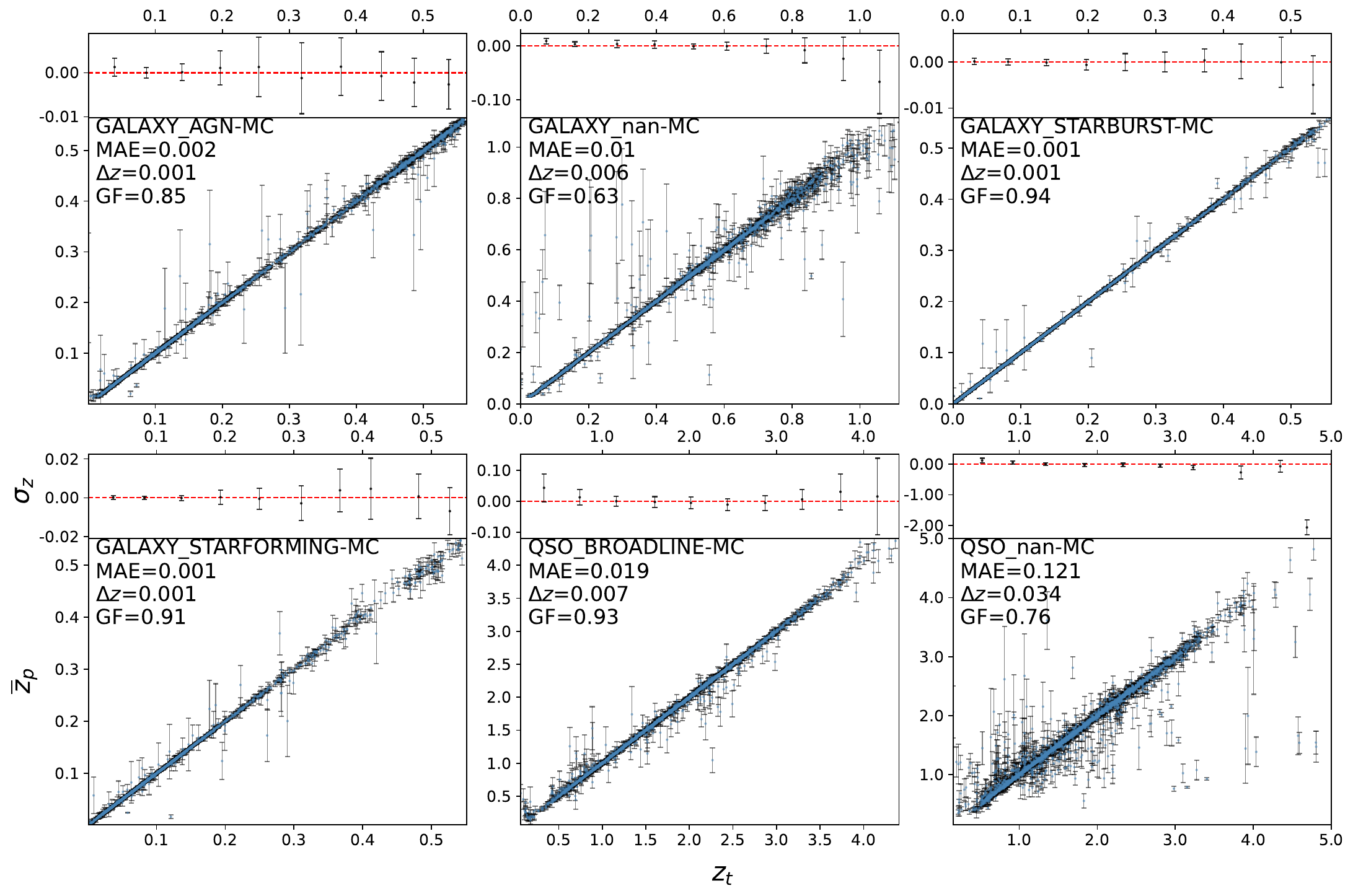}
    \caption{The mean redshift predictions and errors of the 6 extragalactic SDSS subclasses. The error bar of each sample point represents the standard deviation obtained from the MC estimation of 10 sub-networks. In the top left of each main panel the subclass name, MAE, $\Delta z$, and the GF are displayed. The points in the top panels display the mean of the distribution of the $\overline{z}_p$ residuals ($\overline{z}_p-z_t$) with respect to the true values ($z_t$) in each bin, and error bars corresponding mean $\sigma_z$ values (see text).
    }
    \label{fig: redshift error}
\end{figure*}

The values of $\it Bias$ of all subclasses are shown in Fig. \ref{fig: Good vs SNR}.
In this figure, we present the {\it Bias} values as a function of the redshift and color-coded by their SNR. The GF is reported in the legend for each SNR bin. It is evident that the number of ``good'' predictions increases with SNR, which also correlates with redshifts; the lower SNR spectra generally correspond to the higher redshift ones. This also explains why even classes with lower GF, like the GALAXY\_nan (GF=0.63), reach a rather large GF$\sim90\%$, for SNR$>$10 spectra. 
If we exclude the QSO\_nan, which has too few SNR$>$10 spectra to have reliable statistics (see \S\ref{sec:sdss_spectra}), all classes have GF going between 63\% and 94\%, while the average GF is larger than 90\% for starburst, starforming and broadline QSO, clearly because of their well detectable emission lines. On the other hand, the lower accuracy of the normal galaxies (GALXY\_nan) is due to the fact that GaSNet-II learns the redshift mainly from the continuum shape and possibly the absorption lines, whereby the spectra have lower SNR for key features compared to the emission line galaxies; this can limit the performance of the former subclass.

\begin{figure}
        \includegraphics[width=0.44\textwidth]{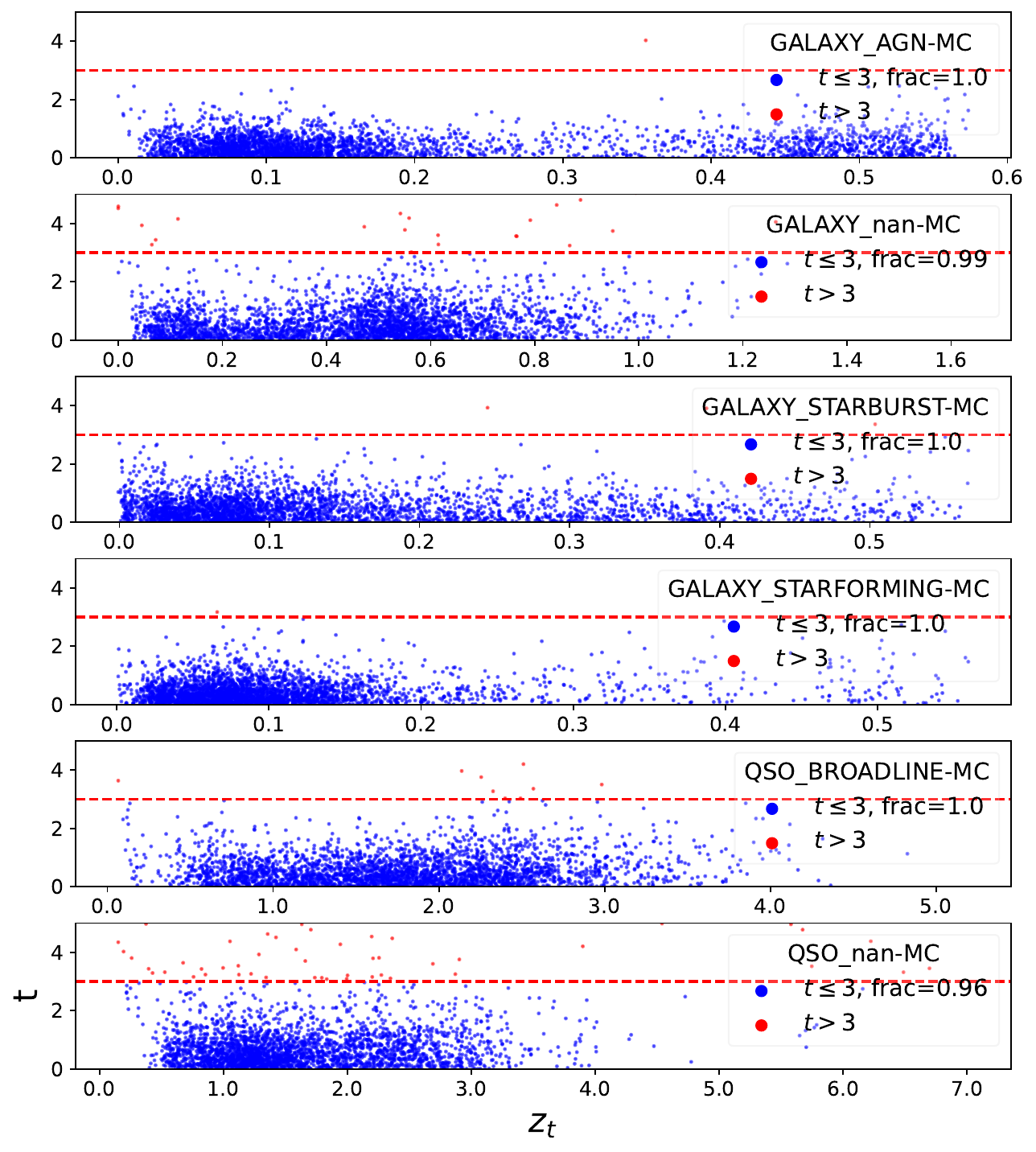}
    \caption{
    The distribution of $t$ vs redshift, where $t$ is defined as $t = |z_t-\bar z_p|/\sigma_z$.
    In the legend, `frac' denotes the proportion of the sample with $t \le 3$.
    }
    \label{fig: z_zm_sigma}
\end{figure}

\subsubsection{Redshifts: MC estimates}
\label{sec:montecarlo}
We finally discuss the redshifts and errors of the 6 extragalactic subclasses predicted by the MC test discussed in \S\ref{sec:pipe_descr}, which are shown in Fig. \ref{fig: redshift error}. 
The main evidence emerging from a quick view of the predicted values is that the accuracy is comparable to the point estimates, as measured by MAE and $\Delta z$, which are very close, or even identical to the ones 
shown in Fig. \ref{fig: redshift prediction}. Looking at the errors, they are extremely small for the predicted values that distribute along the 1-to-1 relation and become bigger for the (few) highly scattered predictions.

As discussed in the previous section, QSO\_nan is the most problematic subclass, showing a larger scatter, and, consequently, larger errors.
Looking at the high-$z$ end in all classes, we see the effect again of the sparse training samples which contribute to the larger errors, which are mirrored by the increased scatter in the estimates already noticed in \S\ref{sec:point_SDSS}. 
This is quantified in Fig. \ref{fig: redshift error}, where the upper panels show the mean $\sigma_z$ of the redshift estimates in different redshift bins.
Here, we can clearly see that the mean errors increase with increasing redshift in almost all classes, except the GALAXY\_AGN. 
Some points' errors are underestimated, particularly at the high redshift end. This is due to a lack of training samples in those regions, which results in lower accuracy in this region.
The bottom line is that the estimated errors are indeed a measure of the reliability of the GaSNet-II predictions, as large error bars emerge either because the estimated values are far from their true value, or because the predicted value is poor due to the poor knowledge base. In particular, we can use the estimated error, $\sigma_z$, to determine whether an estimate is ``robust'' or ``unstable'', using the MAE (listed in Table \ref{tabel: model detail}) as a lower limit for an estimate to be unstable.

Before we discuss the predicted errors, we want to see whether the mean redshift estimates behave similarly to the point estimates, or, in other words, whether the point estimates are drawn by the redshift probability distribution function (PDF) derived by the MC run. This is needed to check if the point estimates are ``unbiased'' predictions of the ``ground truth''. To do that, in Fig. \ref{fig: z_zm_sigma}, we plot the relative scatter normalized to the errors, $t=|z_t-\overline{z}_p|/\sigma_z$, for the different test sets, which should be enclosed in the range [0, 3] for a Gaussian distribution. Here we see that the great majority of the point estimates are within the 3$\sigma_z$ distribution with fractions of the order of 0.96 or higher. This is not fully compatible with a pure Gaussian distribution (expected to be $\sim 0.99$), but rather shows some excess outliers, which we can roughly estimate to be no more than 5\%. Also, we see that some subclasses are more prone to systematics than others, like the `GALAXY\_AGN' and `GALAXY\_STARBURST', that have a tendency to provide overestimated ``point'' redshifts. We stress here that the point estimates are obtained by a separate, independent pipeline, trained to optimize the redshift estimate on a single run, so they cannot be considered a random sample of the MC run, which is trained to optimize the mean $\overline{z}$. We take this into consideration in the discussion below.

\begin{figure}
        \includegraphics[width=0.44\textwidth]{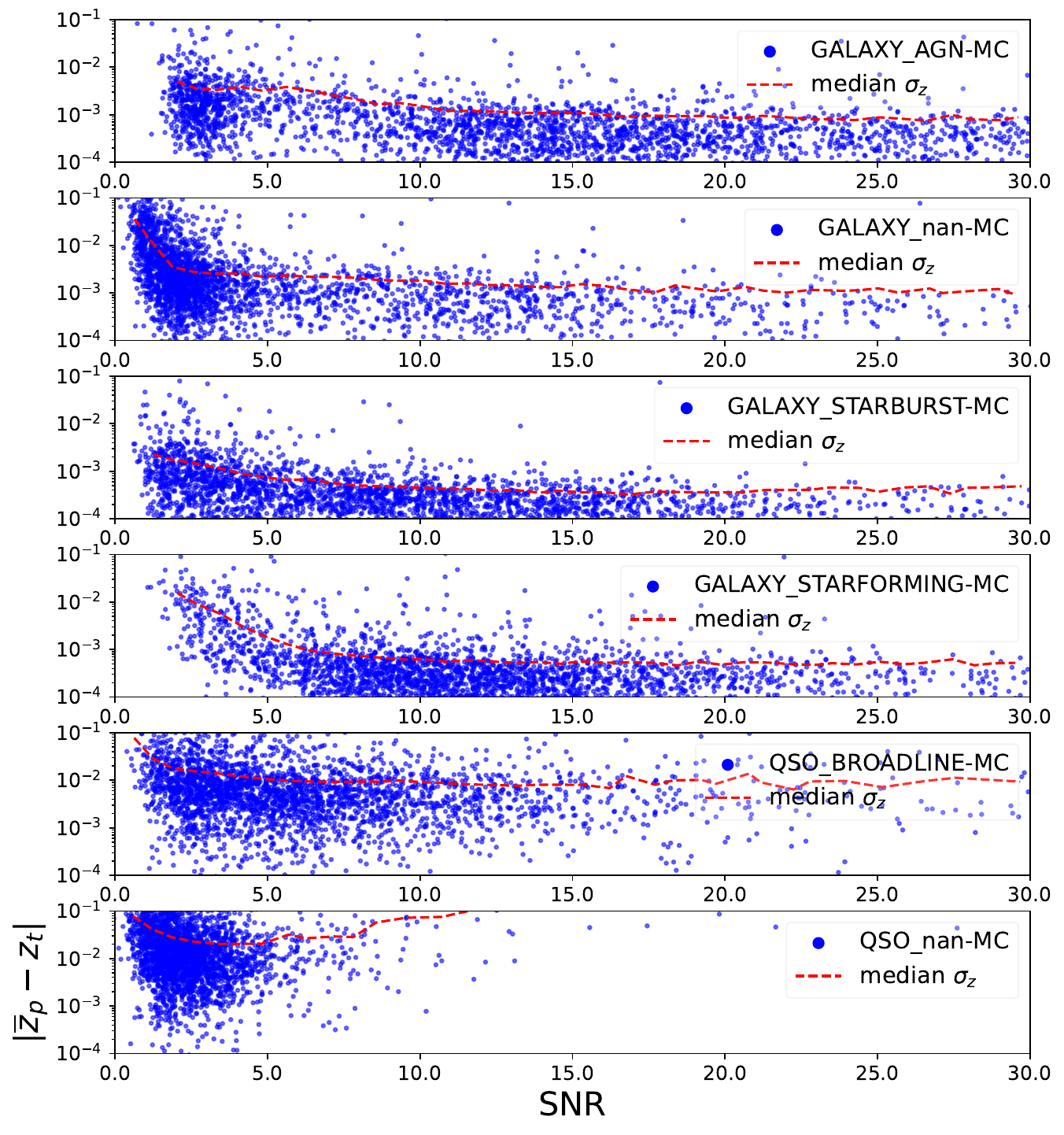}
    \caption{The distribution of $|\overline{z}_p-z_t|$ vs. SNR for the SDSS test data, tracking the performance of error estimations in different noise levels. Median $\sigma_z$ is indicated by a dashed red line. It demonstrates the MC method can reflect the uncertainty realistically.
    In low SNR regions, the value of median $\sigma_z$ is larger compared to the high SNR regions, as expected.
    } 
    \label{fig: Delta vs SNR}
\end{figure}

\begin{figure}
        \includegraphics[width=0.44\textwidth]{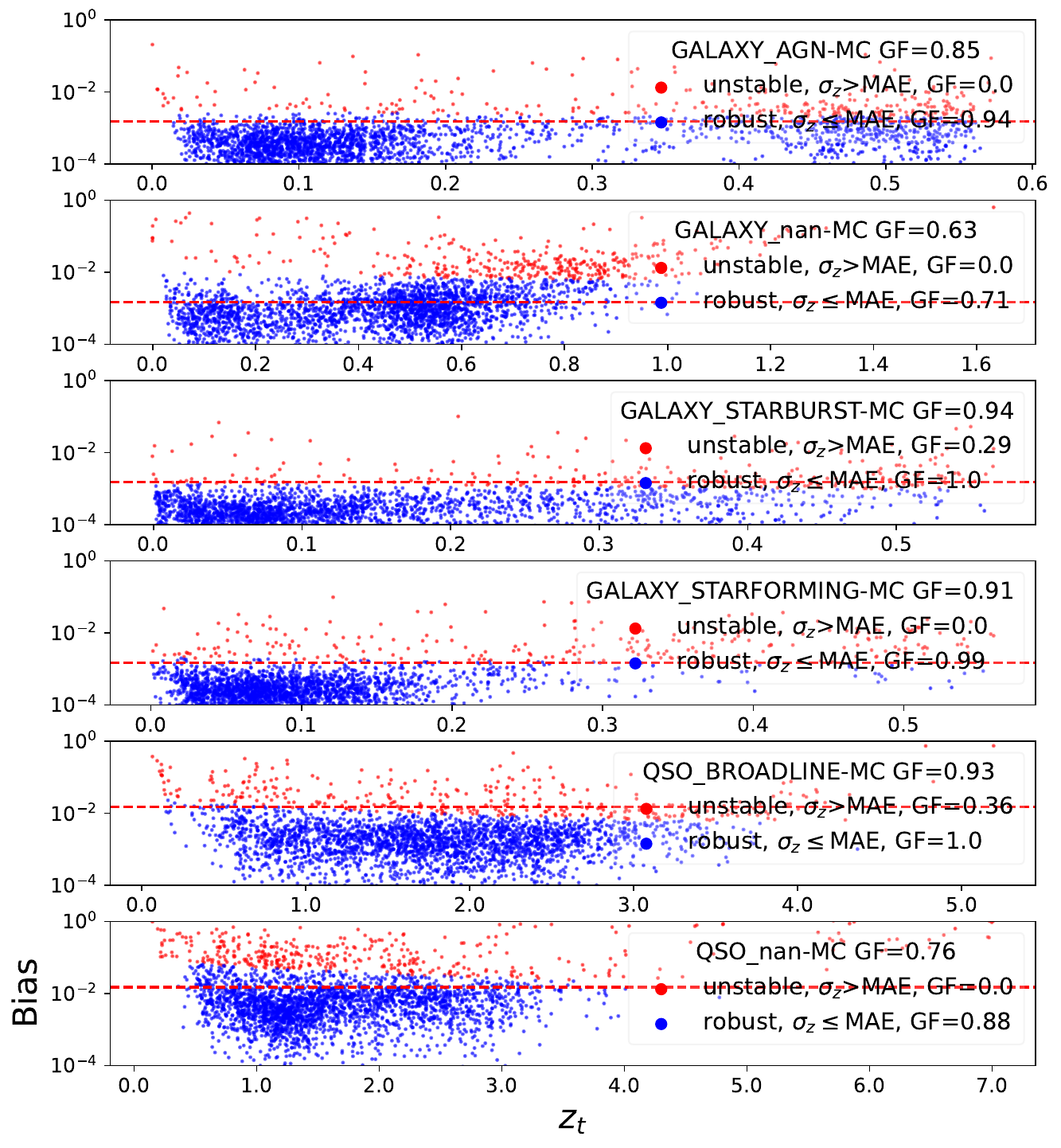}
    \caption{Bias of the 10 sub-networks used. The x-axis is the real redshift and the y-axis is the Bias. The MAE is listed in Table \ref{tabel: model detail}. "robust" is defined as $\sigma_z \leq $ MAE, where MAE is the mean absolute error in the validation set. This demonstrates that unstable points (larger deviation points) can be automatically found without knowing the ground truth.}
    \label{fig: Good vs SNR_mean2}
\end{figure}

Moving to the error estimate, we start by connecting these errors with the data structure. If the errors are artificially produced by internal network errors, due to the stochasticity of some processes, then these should not have any correlation with the spectra uncertainties. To show that, in Fig. \ref{fig: Delta vs SNR} we compare the $\sigma_z$ and SNR of the spectra, where we see a correlation between the error size and the SNR, as quantified by the median values (dashed line), showing that the lower the SNR the larger the $\sigma_z$ tends to be. This is proof that the errors are driven by the data noise, which was assumed without proof so far in this section, and is consistent with the impact of the SNR in classification, discussed in \S\ref{sec:sdss_class} and Appendix \ref{sec:average classification accuracy and SNR}. 
However, at any fixed SNR value, we also see the scatter of the $\sigma_z$ from class to class, with the QSO generally showing larger errors. If we exclude the regions with sparse sampling (see e.g., SNR$\sim5$ for `GALAXY\_STARFORMING', or SNR$\sim 6-8$ for `GALAXY\_AGN'), where the larger scatter of the errors might reflect lower precisions due to a poor training sample, the reason of the $\sigma_z$ variation from class to class should reside in the type of features that GaSNet-II used for the predictions. For instance, in the case of the `QSO\_BROADLINE' (and perhaps also partially true for `QSO\_nan') it is the line broadening that leads to more insecure estimates, especially at lower SNR. Interestingly this is not seen for `GALAXY\_nan', which lets us speculate that for these systems the absorption lines are not driving the redshift estimates, but rather the full spectrum and there is a smooth and regular degradation of the errors for smaller and smaller SNRs, similar to what is seen for GALAXY\_AGN. Direct analysis of the impact of the spectral features on the accuracy is beyond the purpose of this paper and would require more sophisticated techniques like self-attention methods of anomaly detection, which we will address in forthcoming analyses. However, to give a preliminary insight into the importance of the spectral features in classification and redshift predictions, in Appendix \ref{sec: gradients}, we show the gradients of the classification probability and the output redshift with respect to input flux, which allows us to visualize the impact of spectral features in the GaSNet predictions, although they cannot give a real measure of the impact of the continuum.

On the other hand, GALAXY\_STARFORMING seems to be insensitive to SNR until they reach SNR$\sim7$, below which prominent emission lines start to blend into the noise, and then the continuum takes over dominating the larger errors, similar to GALAXY\_nan. We also notice different behavior between GALAXY\_STARFORMING and STARBURST, as, for the latter, $\sigma_z$ is increasingly noisier toward low SNR. As the most important features for these two classes are the emission lines, one would expect a similar behavior for $\sigma_z$. There are two reasons for this: 1) emission lines in starburst galaxies dominate the spectra and GaSNet-II does not learn much from the continuum for star-forming systems. Thus the redshifts are fully determined by the ability of the {\it ResNets} to cross-correlate emission lines over a large wavelength range; 2) {\it ResNets} is perhaps not the ideal tool for this emission line redshift estimation task, which is typically well-handled by other deep learning structures, like ``self-attention'' networks (e.g., \citealt{2020arXiv201212556H}). We will discuss this in detail in \S\ref{sec: Discussion}.
Finally, another source of uncertainty in both redshift and classification can be 
the velocity dispersion, as this can produce a different broadening of the line that might reduce the accuracy of both tasks. In Appendix \ref{sec: correlated with velocity dispersion}, we demonstrate that both $\sigma_z$ and classification accuracy show almost no correlation with the velocity dispersion, inside the different classes. 

The bottom line is that the estimated error sizes as a function of SNR and redshift seem to be mainly driven by the data quality and data features as one should expect from standard analysis methods, rather than the stochasticity of the deep learning network.
As a consequence, we are motivated to use $\sigma_z$ as a proxy of the `robustness'' of the redshift estimates, as we now can interpret $\sigma_z$ as the cumulative effect of the variance of the weights of the network (see \S\ref{sec:pipe_descr}) and the data noise. Also, we can expect that the estimates with smaller $\sigma_z$ are more tightly distributed around the true value. 
In Fig. \ref{fig: Good vs SNR_mean2}, we show again the Bias vs. $z$, which is split into ``robust'' or ``unstable'' categories based on whether their $\sigma_z \leq $ MAE or $>$ MAE, respectively, where MAE is the mean absolute error in the validation set (Table \ref{tabel: model detail}). The robust limit is very close to the GF limit, and only in the `GALAXY\_nan' or `QSO\_nan' subclasses it is significantly larger. Thus, the robust estimates have a fraction over the total samples that are larger than the GF defined by the Bias threshold. This result is particularly relevant for 
practical applications, as for new spectra with no {\it a priori} information on the redshift, the use of the redshift errors proposed here allows us to discard unstable estimates (larger deviation points) without knowing the ground truth. 



\subsection{4MOST mock spectra}

Next, we analyze the 4MOST dataset introduced in \S\ref{sec: 4most data}. The main reason to use this dataset is to test GaSNet-II with spectra close to expected data from major Stage-IV upcoming spectroscopic surveys, but classified on the basis of the survey requirements, thus providing a different classification approach, more survey-oriented. Overall, this would allow us to test the versatility of the pipeline, to respond to different requirements, both in classification and in redshift predictions. 

The training of GaSNet-II with the 4MOST spectra follows the same procedure discussed for SDSS-DR16 in \S\ref{sec:training}. As anticipated, the size of the sample for each class (total of 10 classes) is the same as SDSS-DR16 (20,000) and we use the same training, validation, and test division (70\%, 15\%, 15\%).

In the 4MOST observation phase, the labeled training data rely on the classification of the first months of 4MOST observations to develop a customized training sample based on data collected from the different survey teams. Alternative approaches might rely on the use of mock data, or using visually classified data. 

\subsubsection{Classification}
\label{sec:4most_class}
\begin{figure}
    \centering
    \includegraphics[width=0.45\textwidth]{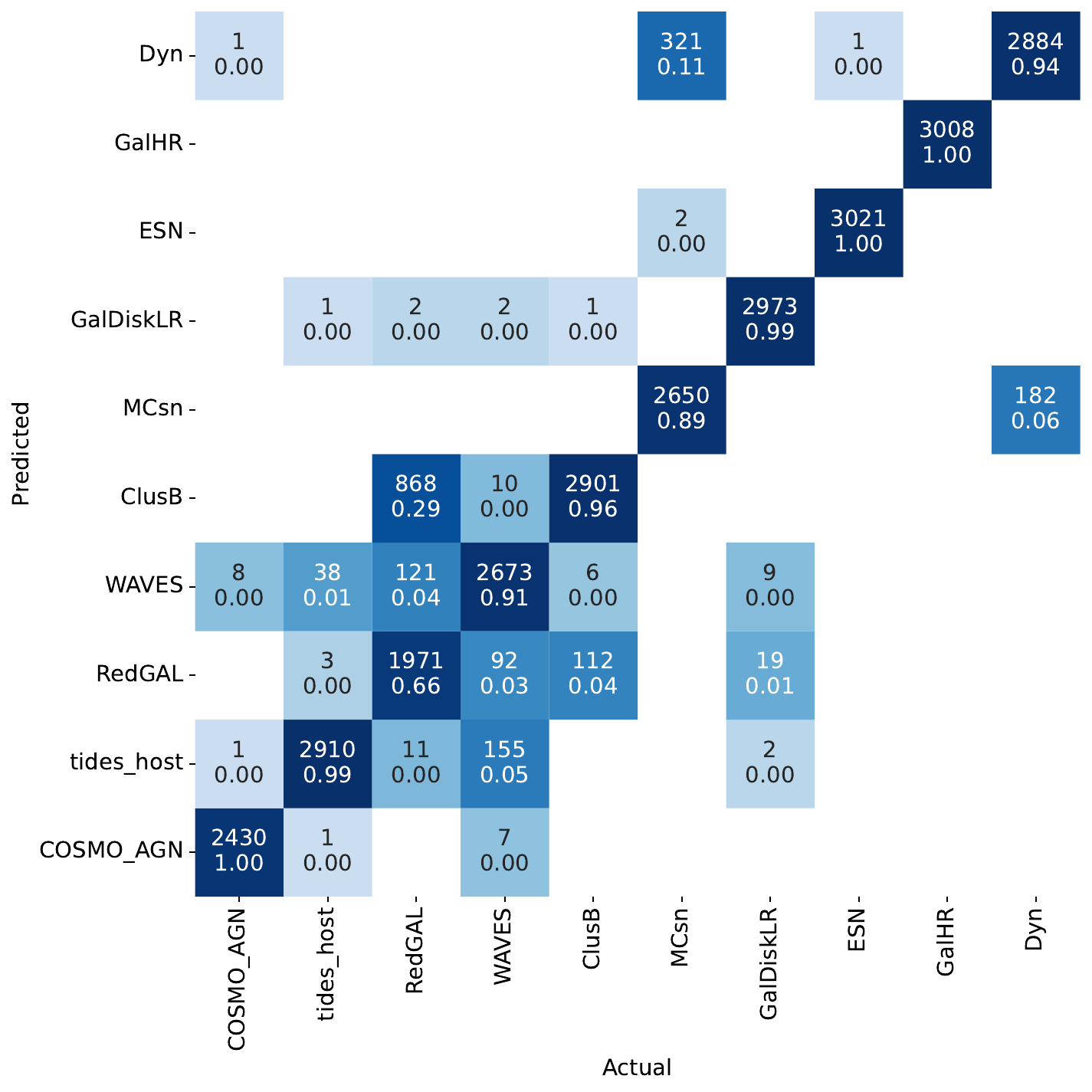}
    \caption{The figure displays the classification results of the 4MOST model on the testing set. It presents a confusion matrix where the legends are the same as Fig. \ref{fig: confusion matrix}. This figure indicates an average accuracy of 93.4\%. The worst performance is observed in the subclass RedGAL, which has an accuracy of only 66\%. 29\% of the spectra in RedGAL are misclassified as ClusB. Note that the matrix has to be read along columns, that is the direction along which the 100\% of the true labels are distributed by the classifier.}
    \label{fig: 4most classification}
\end{figure}

Starting with the classification, in Fig. \ref{fig: 4most classification} we show the confusion matrix obtained over the test samples. GaSNet-II achieves an accuracy beyond 90.0\% for the majority of subclasses, and an average overall accuracy of 93.4\%, which is slightly better than the one found for SDSS-DR16 (92.4\%). One reason can be the absence of contamination discussed above, which we will address at the end of this section; another reason is likely to be the even stronger disparity in SNR between subclasses. Before we check that, we first discuss some other relevant features from the confusion matrix. 

In particular, we notice a striking 100\% score by the COSMO\_AGN class that is superior to the 91\% scored by the GaSNet-II on the SDSS AGN sample. 
Since the mean SNR of the two datasets is very close in galaxy and AGN, (see Tables \ref{Table:1} and \ref{Table:2}), we identify the reason for this overperformance on the COSMOS\_AGN sample to the different redshift distributions, whereby the 4MOST sample lies at a much higher average redshift compared to the SDSS AGN.
This makes it easier for GaSNet-II to unequivocally distinguish the brightest AGN features from, e.g., starburst/starforming galaxy emission lines, for faraway systems than for closer ones. However, another factor that might help this outperformance is the limited chance of cross-contamination among the training/testing classes, which have been constructed here on distinct templates to obtain the mock spectra
(see also below).

\begin{figure}
        \centering
        \includegraphics[width=0.45\textwidth]{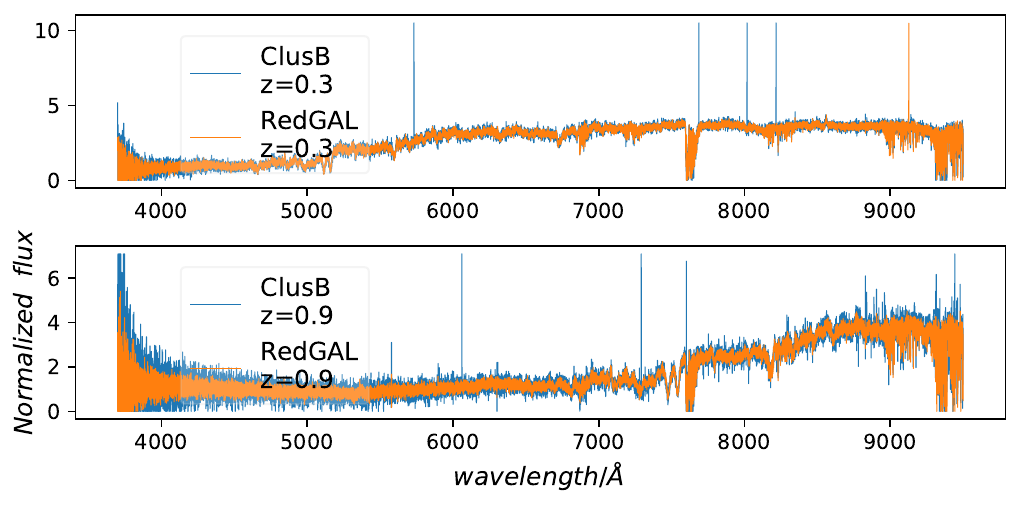}
    \caption{We randomly pick 4 spectra. The upper panel shows ClusB and RedGAL spectra with a redshift of $0.3$. The bottom panel shows the spectra with a redshift of $0.9$.}
    \label{fig: ClusB and RedGAL}
\end{figure}

The only clear case of such contamination is the mixing between subclass `ClusB' (label 6, corresponding to bright cluster galaxies) and `RedGAL' (label 8, i.e., red galaxies). 
ClusB likely systems are a peculiar sub-sample of the RedGAL systems, at least at low redshift, as bright central cluster galaxies are generally old, red galaxies, particularly in their centers (see e.g. \citealt{2007AJ....133.1741B}), which is where 4MOST fibers would be placed.
Fig. \ref{fig: ClusB and RedGAL} shows the templates of two ClusB spectra at redshift 0.3 and 0.9, respectively vs. two redGAL templates at the same redshifts, normalized to the same flux at 6000\AA\ at each redshift. 
We are asking the classifier to separate spectra that are nearly indistinguishable at the same redshift.
Surprisingly, in Fig. \ref{fig: 4most classification}, we see that GaSNet-II can correctly predict the clusB galaxies, while it confuses the RedGAL for ClusB in $\sim29\%$ of the cases. We can possibly explain this with the fact that ClusB galaxies often systematically show emission lines in their spectra, while the RedGAL mostly do not (see again Fig. \ref{fig: ClusB and RedGAL}), hence we argue that the emission lines are features that GaSNet-II associates to ClusB galaxies and not RedGAL, where they are not dominant. This means that RedGAL spectra with emission lines have a larger chance of being classified as ClusB.
To conclude this section, we refer the reader to Appendix \ref{sec: The coarse classifier of SDSS} where, as for SDSS, we have performed the classification of the spectra by grouping the different star, galaxy, and AGN classes to emulate a coarse STAR-GALAXY-AGN classification, to be compared with a similar one from SDSS and DESI. We stress here that this experiment, besides putting the performances on 4MOST templates in the context of other reference surveys, provides us also a test on a more physically-oriented sample, rather than a survey-oriented classification discussed so far. This is closer to what GaSNet will be required to perform in the early stage of 4MOST operations. In this case, we can see that the coarse classifier can reach an even higher mean accuracy of 98\%, comparable with what we have seen for SDSS.


\subsubsection{Redshifts}
\label{sec:4most_redshifts}
\begin{figure*}
        \includegraphics[width=0.9\textwidth]{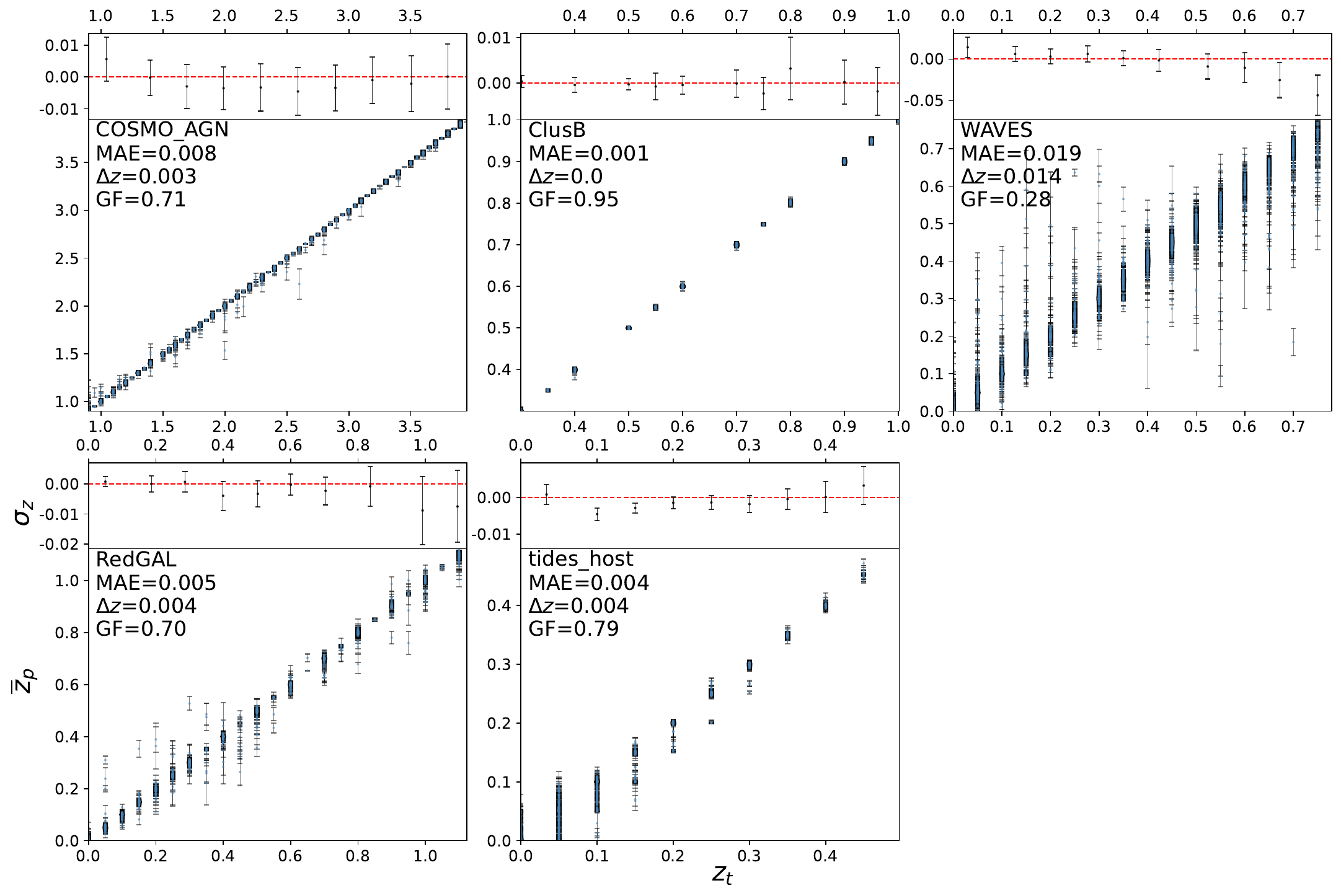}
    \caption{Redshift predictions for the five extragalactic 4MOST mock subclasses. It is worth noting that the simulated spectra are produced on a coarse grid of redshifts, hence the quantization. Legends are identical to Fig. \ref{fig: redshift error}. 
    }
    \label{fig: 4most redshift prediction}
\end{figure*}

We finally show the results for the redshift predictions, limiting ourselves to the MC estimates with errors. In Fig. \ref{fig: 4most redshift prediction}, we show the predicted redshifts for all the 4MOST extragalactic subclasses. 
The figure indicates an average $\Delta z$ of 0.0055 for galaxy types (ClusB, WAVES, RedGAL, and tides\_host), while it becomes 0.003 for AGN. The average GF for the galaxy is 0.68, while for AGN is 0.71.
These latter are the class for which GaSNet-II also provides the most accurate classification, meaning that the combination of good SNR and emission lines, permits high performances for both tasks.  
Among the galaxy types the average error is dominated by the WAVES class which has the largest errors, possibly due to the low average SNR (see Table \ref{Table:2}). The same WAVES class also shows the highest relative scatter $\Delta_z=0.014$ compared to $\Delta_z \sim 0.004$ shown by the majority of the other subclasses. 
Overall the $\Delta_z$ found for the 4MOST mock sample seems slightly worse than the one measured for SDSS ($\Delta_z\sim0.003$), although a direct comparison is not appropriate, with the two samples having different observational constraints, especially in terms of SNR, for instance, 4MOST AGNs and galaxies have a lower SNR except the ``tide\_host" subclass (e.g., comparing Tables \ref{Table:1} and \ref{Table:2}).
The 4MOST redshifts also show a GF on average slightly lower than the one of SDSS as reported by the mean good fractions in the legends of Fig. \ref{fig: 4most redshift prediction}, against the GFs reported in Fig. \ref{fig: redshift error}, for SDSS. Once again the WAVES spectra are the ones with the worst GF, which are a consequence of the systematically larger errors, ultimately driven by the low SNR.

As for the comparison with standard methods, here a full detailed check of the relative performance of GaSNet-II with respect to tools like {\it redrock} and {\it redmonster} is beyond the scope of this paper. However, to put the GaSNet-II performances into perspective, on a series of benchmarking tests on simulated 4MOST consortium datasets, we have found GaSNet-II GF to be $\sim$20\% worse than {\it redrock} and {\it redmonster}, although, for some classes, like AGN/QSO, GaSNet-II shows a GF even better than classical tools. For instance, {\it redmonster} shows an average GF of 0.71 (GF for AGN/QSO is 0.43), Mean Absolute Deviation (MAD) of 0.00042, and Time (in the unit of seconds per spectrum per core, sec/spec/core) of 1.02; {\it Redrock} shows an average GF of 0.48 (GF for AGN/QSO is 0.23), MAD of 0.051, and untested Time; while for GaSNet-II we find an average GF of 0.40 (GF for AGN/QSO is 0.70), MAD of 0.0086, and Time of 0.00089 on the AGN/QSO/GALAXY redshift testsets. This indicates that there is still room for GaSNet improvements, which can be consolidated with final, more sophisticated, mock data, and eventually with the first 4MOST observations.

\subsection{DESI spectra}
We finally apply GaSNet-II to the early release DESI data. As seen in \S\ref{sec:desi}, the DESI classification taxonomy is less complex only a very broad classification (i.e. star, galaxy, and quasars), and their numbers are less abundant, as we could test our tools over $\sim$1050 classified spectra for each class. This allows us, besides testing GaSNet-II on a further dataset, with a different observation set-up and size, to perform a basic analysis 
over a `coarse' classification which is similar to what we expect to implement for 4MOST earlier data releases (see also Appendix \ref{sec: The coarse classifier of SDSS}).
The classification and redshift estimates are quickly discussed below.


\subsubsection{Classification}
\label{sec:desi_class}
\begin{figure}
    \centering
    \includegraphics[width=0.3\textwidth]{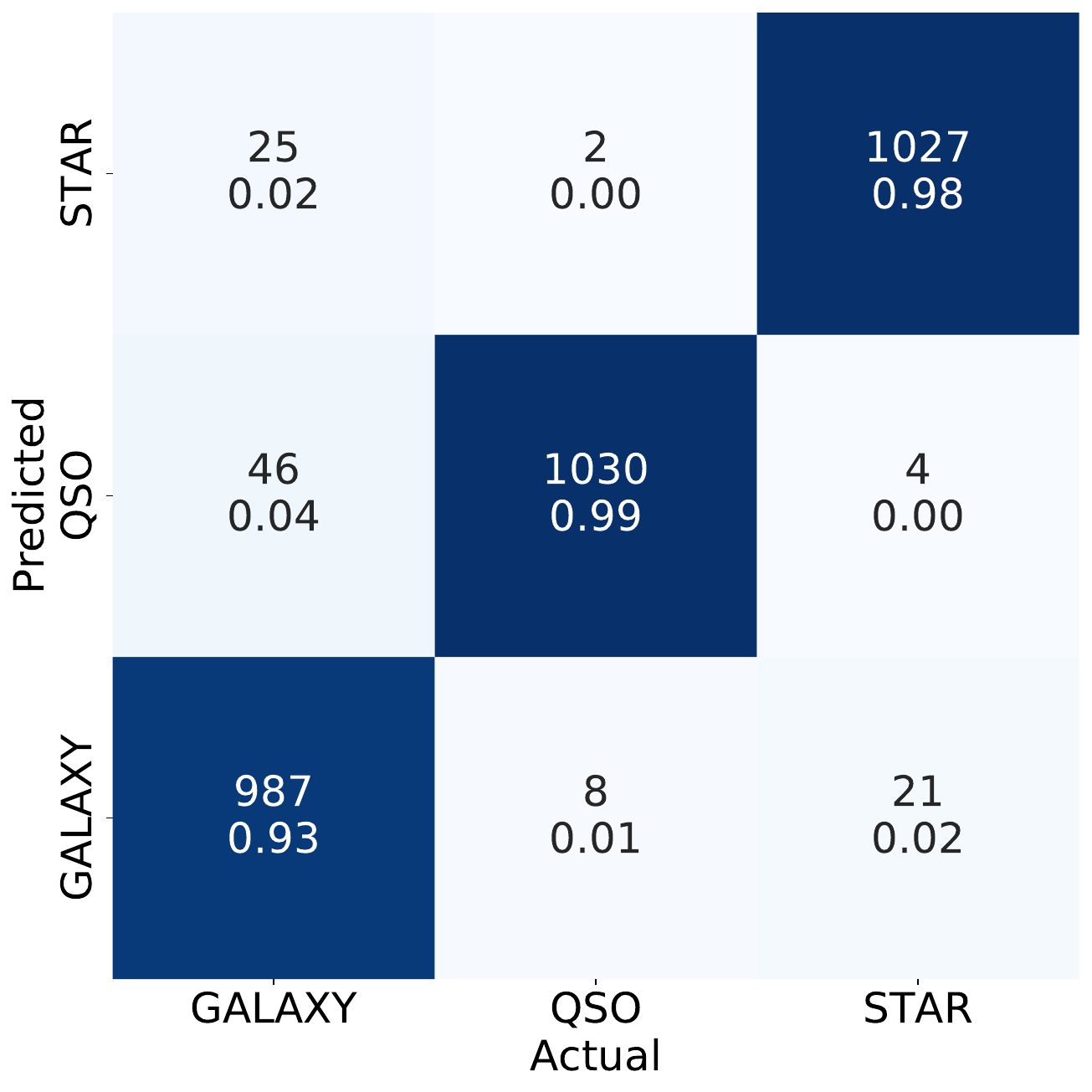}
    \caption{The DESI classification on the test set. Legends are identical to Fig. \ref{fig: confusion matrix}.
    As before, the matrix should be read along columns, that is the direction along which the 100\% of the true labels are distributed by the classifier.}
    \label{fig: DESI classification}
\end{figure}

The separation of the test sample on the 3 DESI classes is shown in Fig. \ref{fig: DESI classification}, where the confusion matrix indicates the accuracy of each of the 3 classes is larger than 93\%, and the average accuracy is 96\%. 
The high accuracy is obviously highly dominated by the small number of classes, however, this also shows an almost absent ambiguity of the classification for classes notoriously prone to confusion, e.g., stars and galaxies. This is likely due to the ability of GaSNet-II to guess the redshift and (eventually) the shapes of the spectral features. We expect though that with a larger training sample the accuracy will be 
further increased. To put these results in perspective with other datasets, in Appendix \ref{sec: The coarse classifier of SDSS} we have performed a similar analysis for SDSS-DR16, by collapsing all spectra sub-classes into three broad classes as for the DESI dataset. We anticipate that, using the same number of SDSS training samples, we find a 99\% accuracy for such a coarse classification, that seems rather higher than the one obtained for DESI. This implies that the quality of the spectra, rather than the number of training samples, is the major factor contributing to the accuracy. We expect to return to such a test in upcoming DESI releases to confirm this result.

\subsubsection{Redshifts}
\begin{figure}
    \centering
        \includegraphics[width=0.5\textwidth]{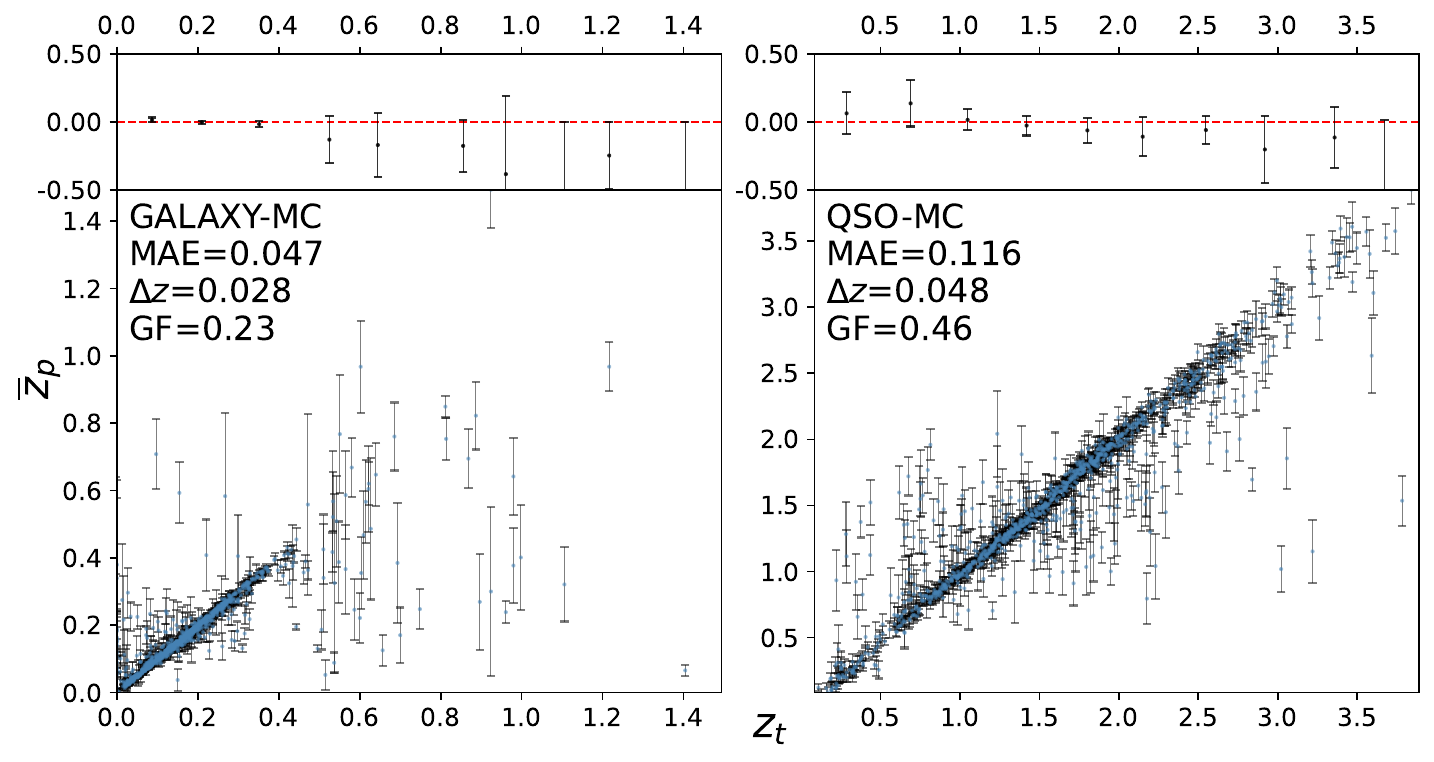}
    \caption{Redshift predictions of two DESI classes (GALAXY and QSO). Legends are identical to Fig.\ref{fig: redshift error}. 
    }
    \label{fig: DESI redshift prediction}
\end{figure}

Finally, we show the MC predictions of the redshifts and their errors for the DESI GALAXY and QSO objects. In Fig. \ref{fig: DESI redshift prediction}, we can see a good agreement between the predictions with the ground truth and an average redshift error ($\Delta z$) of the two classes of 2.8\% for galaxies and 4.8\% for QSOs.
These errors are larger than the ones obtained for former datasets for two main reasons.
The first is the unbalanced redshift distribution, especially in the high redshift part (i.e., $z>0.4$ for galaxies and $z>2.5$ for QSO), where there are fewer systems, especially for the galaxies. The second is the overall smaller training samples available for these early-release data from DESI (about 1/10 of the former datasets), resulting in typically larger errors on the individual spectra.
Once we can include more DESI training samples, and use customized sub-networks for the special subclasses, we expect the accuracy will rise to the level found for SDSS and 4MOST.

\section{Discussion}
\label{sec: Discussion}

In this section, we will discuss the potential strategies for improvements in performance and further developments.

\begin{figure}
    \centering
        \includegraphics[width=0.5\textwidth]{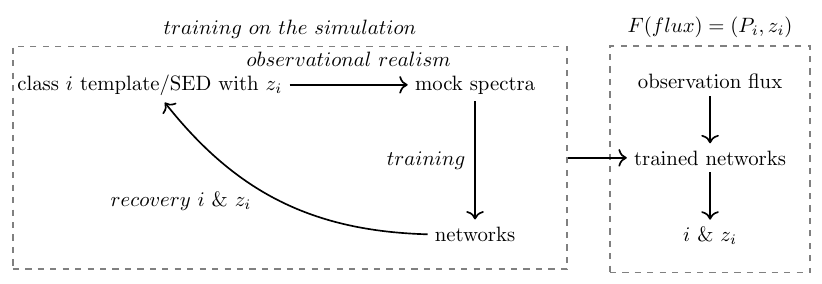}
    \caption{The general process of networks trained by simulation involves training on mock samples, finding the mapping $F$, and predicting real data. The training data is generated with specific parameters ($i, z_i$) and observational realism. The networks are trained to recover the labels $i, z_i$, and ultimately, the well-trained networks are used to fit the parameters $i, z_i$ based on observational data input. It is a first-principles-based method rather than an empirical-based one.
    } \label{fig: train on the mock}
\end{figure}

As far as classification is concerned, a key problem is how to improve the ``absolute'' accuracy of the classification method.
So far, we have benchmarked GaSNet-II with respect to the labels assigned from the different datasets (relative performances). For the SDSS and DESI datasets, the labels are deduced from the PCA fitting, and this can bring some systematics. In fact, when using a classification based on real spectra as labels for the training of the DL tools, the upper limit of the ``absolute'' accuracy of the trained networks is decided by the accuracy of the training set, which in turn is set by the accuracy of the ``traditional'' pipeline used for labeling it. 
A viable alternative is to incorporate human-labeled data, like, e.g., SDSS-DR12 superset \citep{2017A&A...597A..79P, 2018A&A...613A..51P}.
However, this approach is not bias-free either, introducing a different form of bias: human judgment. 
Another physically motivated alternative is to utilize mock data, based on theoretical templates, e.g., similar to those used for the 4MOST sample in \S\ref{sec:4most_class}. 
In Fig. \ref{fig: train on the mock}, we describe a general procedure for training on simulated data. Here, the function $F$ represents: 
\begin{align}
    F \ (flux) =
        \begin{cases}
        (P_i, 0),   & i \in {\rm galatic}\\
        (P_i, z_i), & i \in {\rm extragalactic.}
        \end{cases}
\end{align}
The networks shown in the figure serve as a powerful fitting tool that minimizes the need for manual adjustments. The mock data, produced under specific physical conditions ($i, z_i$), are used as training data for the networks. Subsequently, well-trained networks are set up by optimizing the prediction accuracy of the parameters ($i, z_i$). If the training sample is complete and accurate, these well-trained networks can be considered, by construction, as the optimal tools maximizing the ``absolute'' accuracy of the predicting parameters ($i, z_i$) when applied to real observational data. In practice, this is possibly true only if: 1) the theoretical models are correct, 2) one introduces into the process all the observational conditions to maximize the fidelity between mock train/test sets and observations, including Poissonian noise, realistic distributions of SNR, seeing, intrinsic broadening of the features (e.g., galaxy kinematics), artifacts, etc. (see e.g., Fig. \ref{fig: train on the mock}). 
The former condition is generally satisfied for most of the objects one expects to classify in galactic and extragalactic surveys as there are rather robust theoretical stellar (e.g., \citealt{2014MNRAS.440.1027C}) and galaxy/QSO templates (e.g., \citealt{2001ApJ...556..121K}). However, there might still be remaining systematics due to specific model shortcomings or even ``unknown'' phenomena that are not fully accounted for by standard theories or empirical models. In principle, these latter systems would possibly appear as ``anomalies'' in theoretical-based classifications that can be studied separately either to improve models or explore new phenomena.
With regard to ``observational realism'', the inclusion of more observational conditions is something that is currently under development (in the case of imaging data, see, e.g. \citealt{2022PASP..134d4502Y}).
Despite these difficulties, which we aim to address in future analyses, the main advantage of using mock datasets is the freedom to choose the hyperparameters that one is expected to predict with spectra, and then optimize the training sample accordingly (a kind of active learning loop), e.g., using theoretical-based simulation spectra covering a wide and physical range of these hyperparameters.
Another advantage of using the mock spectra is that they do not suffer the poor sampling problem, which plagues empirical datasets (e.g., rare events, like strong gravitational lenses, or high-redshift galaxy samples, etc.). 
As a result, they eliminate the biases introduced by incomplete or poor sampling. 

Regardless of the philosophy behind the training sets, there might be further strategies that can help improve the classification. One is the hierarchy. Classifications can be done in one step (as we have proposed in \S\ref{sec:sdss_class}, \ref{sec:4most_class}, \ref{sec:desi_class}) or multiple steps. Spectra can be roughly classified in the first step, followed by a more sophisticated sub-classification in subsequent steps (see, e.g., \citealt{2021AJ....161..141S}). This decision-tree-like classification can allow us to have a more fine-grained and detailed classification process. The architecture of multiple identical sub-networks, similar to what we currently use, can be easily rearranged into a decision-tree-like hierarchical structure to realize a multi-ML model combination "tree" structure, with more branches and deeper layers.

Moving to the redshift estimates, we foresee that relevant improvements can be obtained using ``self-attention'' \citep{2017arXiv170603762V}, which is becoming popular as the state-of-the-art model in DL applications. 
For instance, Fig. \ref{fig: ClusB and RedGAL} is an example where the classifier based on the ResNet struggles to effectively recognize the slight difference in the spectrum when there is a mix of features like the spectrum continuum and emission lines.
``self-attention'' has shown to be superior in recognizing the global features and ``long-range correlation" compared to CNNs \citep{2020arXiv201212556H} with the net effect that both classification and redshift estimates can highly be improved (see also \S\ref{sec:montecarlo}). 
We plan to implement these alternative approaches in future work by replacing the convolutions with "self-attention" in the small blocks of our network.

Finally, alternative methods of estimating the redshift error exist. 
In standard networks, keeping the inputs the same leads to the same outputs, which is stable but does not allow us to generalize the error estimates. Apart from introducing multiple sub-networks to estimate the errors, as we have already experimented with in this paper, there are other approaches to introduce uncertainty, such as MC drop-out techniques \citep{2022A&C....4000615P} or Bayesian neural networks \citep{2017ApJ...850L...7P, 2022RAA....22k5017Z, 2023MNRAS.522.5442G}. 
We stress though that we expect that these methods
are unlikely to yield significant differences with respect to our approach as these methods obtain the error by repeating predictions. We aim to test these different techniques in future analyses.


\section{Conclusion}
\label{sec: Conclusion}

We have developed new tools for spectroscopy classification and redshift prediction using deep learning techniques and constructed a pipeline that we have tested on SDSS, 4MOST, and DESI datasets. The performance of our pipeline on these three different datasets can be summarised as follows: on SDSS, the classifier achieves an average accuracy of 92.4\% for a 13-subclass classification task (with most types exceeding 90\%), and redshift prediction accuracy around 0.23\% for galaxy and 2.1\% for QSO subclasses. On 4MOST, the classifier achieves an average accuracy of 93.4\% for a 10-subclass classification task and redshift prediction accuracy of around 0.55\% for galaxy and 0.3\% for AGN. On DESI, the classifier achieves an average accuracy of 96\% for a 3-class classification task and redshift prediction accuracy of around 2.8\% for galaxy and 4.8\% for AGN. The accuracy of classifiers is strikingly consistent. However, the aspect of redshift prediction is clearly dependent on various factors such as the types of subclasses/classes, 
the average spectral element signal-to-noise ratio,
and the sample size of the training data. For example, the poor SNR of subclass WAVES results in the highest error on the 4MOST dataset, while the relatively sparse training data for DESI contributes to a larger redshift error compared to SDSS and 4MOST.

GaSNet-II's efficiency and accuracy make this tool suitable for real-time analyses of nightly observations.
The predictions for 39,000 spectra can be completed in less than one minute. 
Among the data products, GaSNet-II can provide realistic redshift errors from a built-in sud-network architecture simulating an MC test.
As seen in the discussion of the SDSS-DR16 results. The redshift error of each data point can be also used to assess the robustness of the predicted redshifts. 


 
In summary, deep learning methods offer significant advantages for Stage-IV spectroscopic infrastructures like DESI, 4MOST, and MOONS in various aspects, such as efficiency, ``data-driven", better performance in low SNR, better consistency and systematics, and so on. Although the current redshift accuracy leaves room for improvement, deep learning, as a new tool, holds huge potential for further development. Many aspects of improvement can be done with the future 4MOST simulations. 
Further datasets such as theoretical spectra and improvements such as a `self-attention' structure will be applied to GaSNet-II in the future to improve the ``absolute'' accuracy of classification and redshift estimates, respectively.

\section*{Acknowledgements}
We thank D. De Martino and P. Szkody for their useful comments and suggestions. 
NRN and FZ acknowledge that part of this work was supported by the National Science Foundation of China, the Research Fund for Excellent International Scholars (grant n. 12150710511), and the research grant from China Manned Space Project n. CMS-CSST-2021-A01. 
CH's work is funded by the Volkswagen Foundation. CH acknowledges additional support from the Deutsche Forschungsgemeinschaft (DFG, German Research Foundation) under Germany’s Excellence Strategy EXC 2181/1 - 390900948 (the Heidelberg STRUCTURES Excellence Cluster).
This work was funded by ANID - Millennium Science Initiative Program - ICN12\_009 (FEB), CATA-BASAL - FB210003 (FEB, BLR), and FONDECYT Regular - 1200495 (FEB, BLR). 
Boudewijn F. Roukema acknowledges the Polish LSST/Rubin grant from the Ministry of Science and Higher Education (MNiSW) DIR/WK/2018/12.
Y.-L.K. has received funding from the Science and Technology Facilities Council [grant number ST/V000713/1].
RJA was supported by FONDECYT grant number 123171 and by the ANID BASAL project FB210003. 
LJMD acknowledges funding by the Australian Research Council (ARC) Future Fellowship scheme (FT200100055). 
G.G. acknowledges support by Deutsche Forschungsgemeinschaft (DFG, German Research Foundation) – project-IDs: eBer-22-59652 (GU 2240/1-1). This project has also received additional funding from the European Research Council (ERC)under the European Union's Horizon 2020 research and innovation programme (Grant agreement No. 949173). 
E.T. acknowledges the ETAg grant PRG1006 and the CoE project TK202 funded by the HTM.

\section*{Data Availability}
The code and data (SDSS) are available in the GitHub link: \url{https://github.com/Fucheng-Zhong/GaSNet-II}.



\bibliographystyle{mnras}
\bibliography{references}

\appendix

\section{cross-contamination on dataset}
\label{sec:app_cross_cont}
We have anticipated in \S\ref{sec:sdss_spectra} that the empirical classification of the SDSS-DR16 cannot guarantee full accuracy, and we cannot exclude cross-contamination among the different classes. This might have an impact both on the classification
and the accuracy of the redshifts. As discussed in \S\ref{sec: Discussion}, a possible workaround is to train on a purer sample of mock spectra based on well-established theoretical or observational templates \citep{2020MNRAS.495..905R, 2020MNRAS.498.5581B, 2021MNRAS.505..540T}. An example of how this might lead to higher performances has been offered by the 4MOST sample, where for some classes we have reached 100\% accuracy (e.g. COSMOS\_AGN, ESN, and GAL\_HR) for a combination of clean templates and rich training sample, although the 4MOST training sample is not exactly built over physically motivated templates, but, rather, specific survey targets, that might have very specific properties, including high SNR, that make the spectra easier to classify (e.g., tides\_host).  

\begin{figure}
    \includegraphics[width=0.21\textwidth]{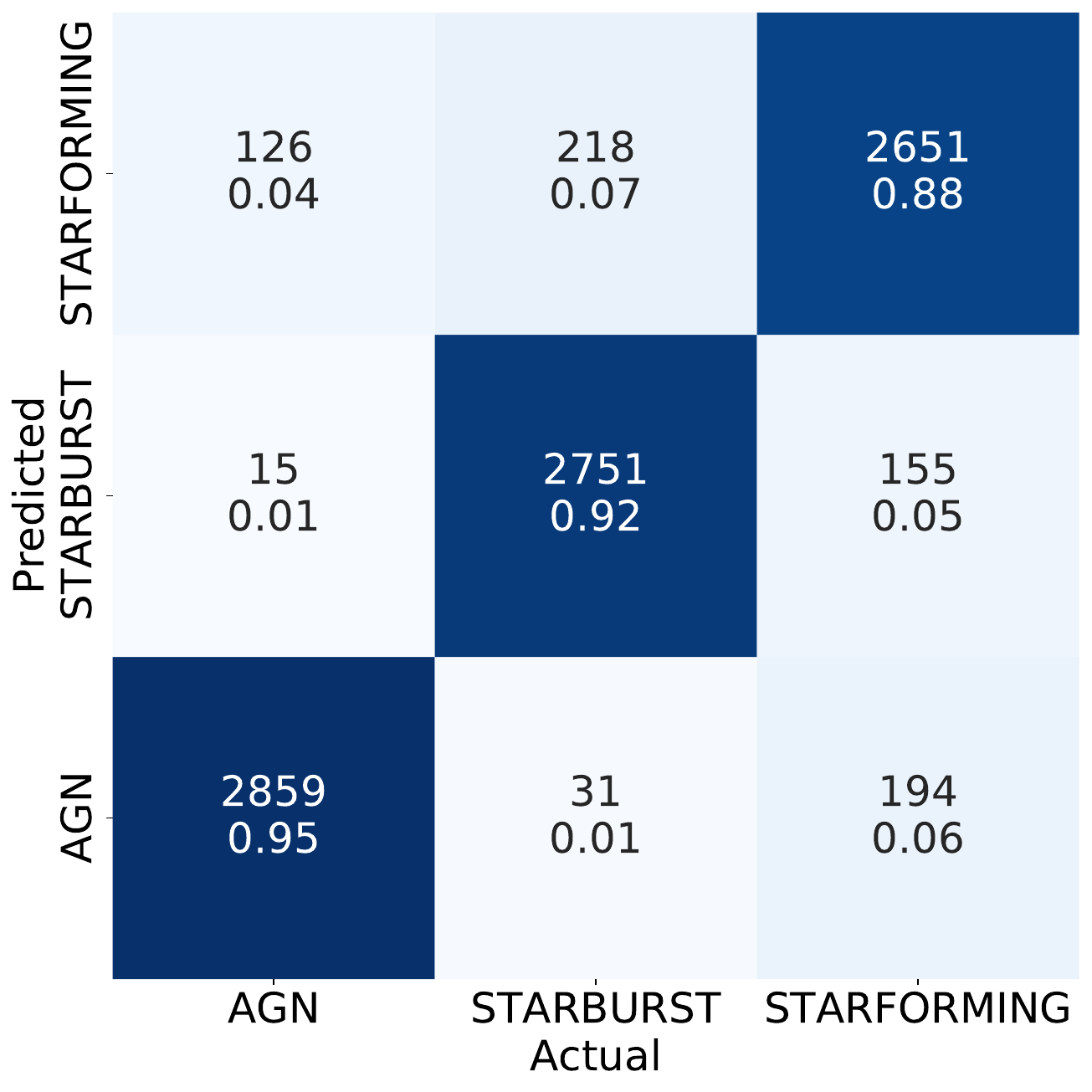}
    \includegraphics[width=0.21\textwidth]{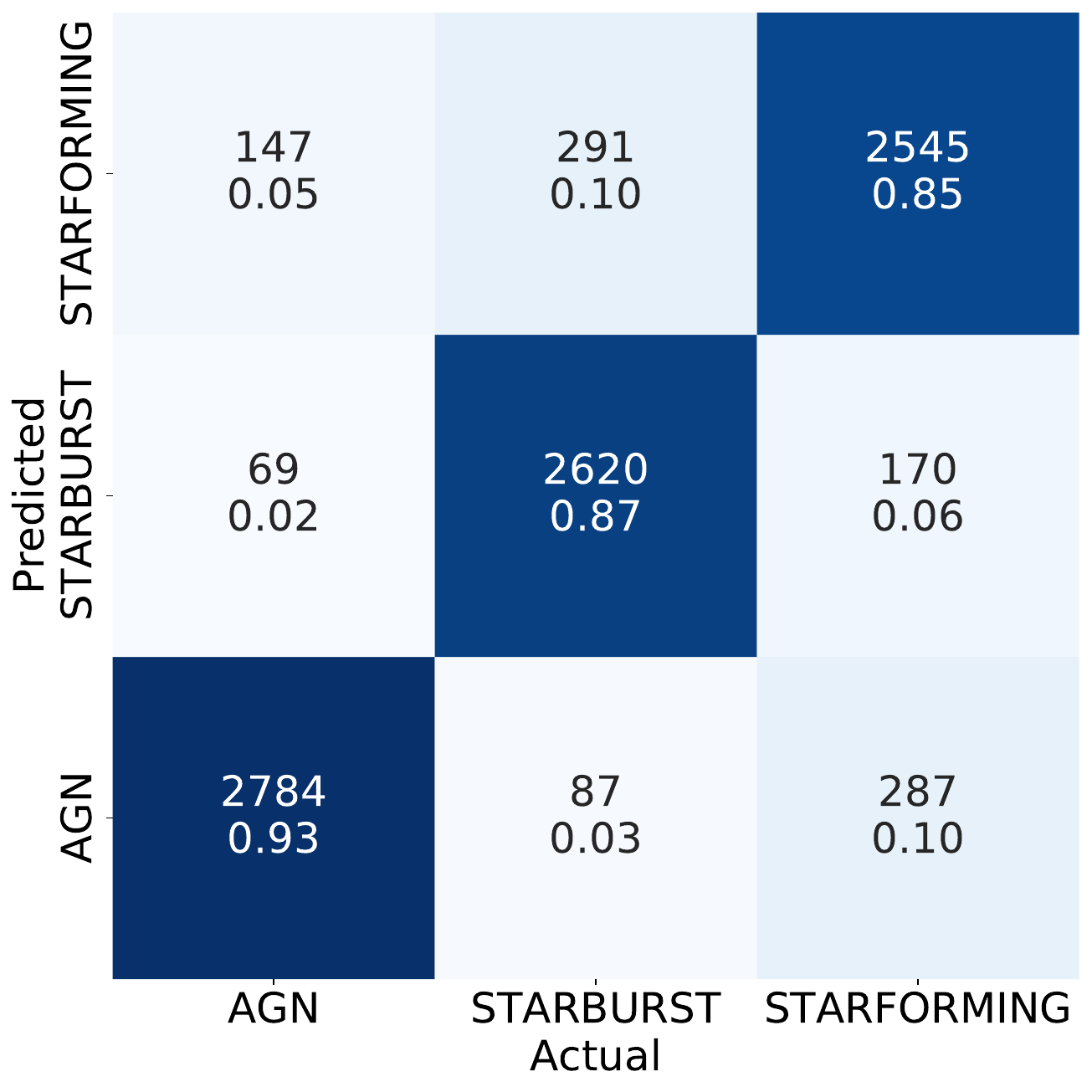}
    \includegraphics[width=0.21\textwidth]{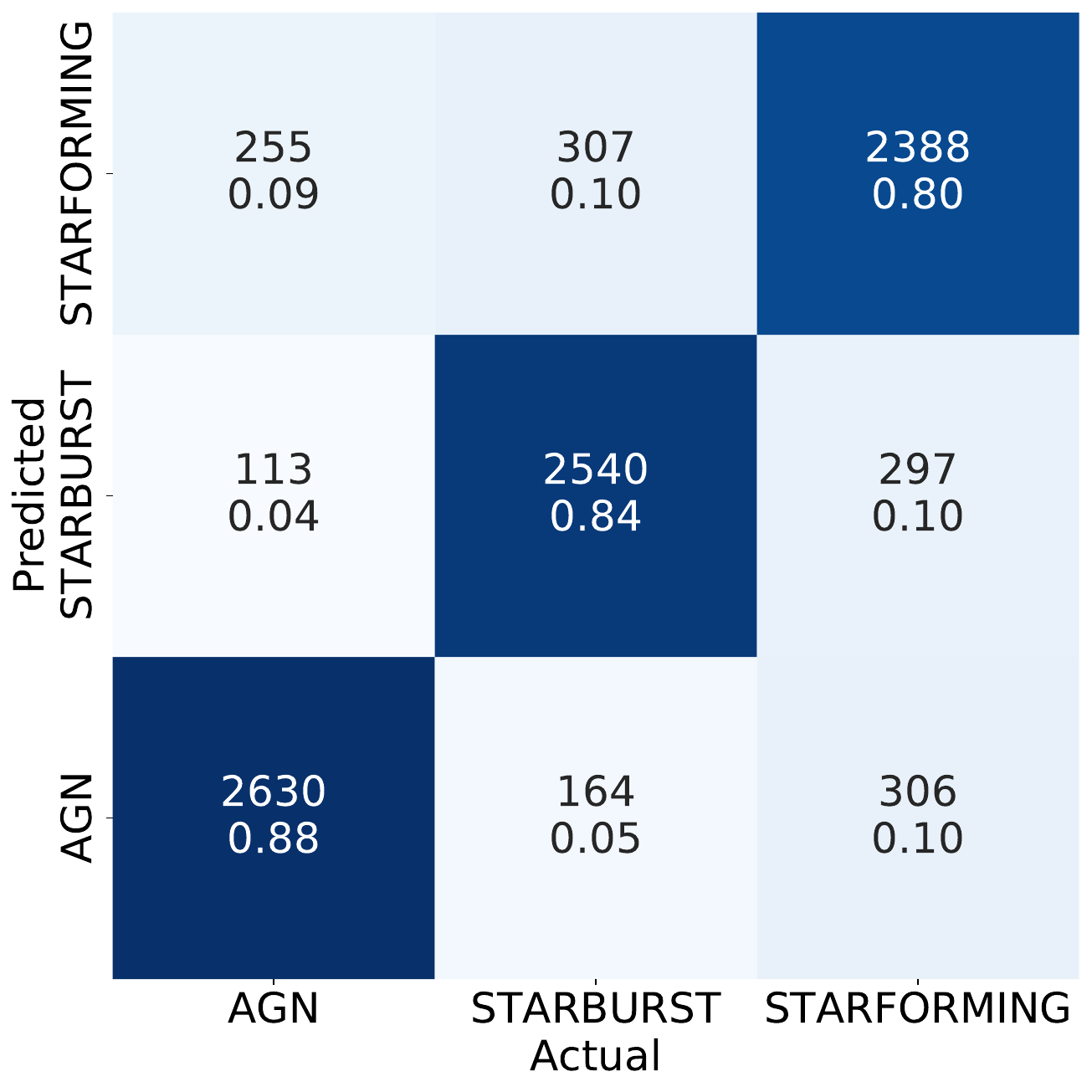}
    \caption{Confusion matrix of a 3-class classification (AGN, STARBURST, STARFORMING), showing how this changes with increasing random contamination (0\% - 5\% - 10 \%). The contamination fraction refers to randomly selected and shuffled labels in the dataset. As expected, as contamination increases, the accuracy decreases. Roughly speaking, with respect to the original sample (0\% artificial contamination), we observe an average decrease in accuracy by 3.3\% for 5\% contaminants and 7.7\% for 10\% contaminants.}
    \label{fig: SDSS contaminations}
\end{figure}

Here we intend to check, more quantitatively, the possible impact of the misclassified spectra in a given class. We use the SDSS-DR16 dataset as a reference for this test and add, to each class, 5\% or 10\% contamination from the relatively similar classes seen from the confusion matrix. 
For the sake of brevity and clarity, we use only three extragalactic classes: AGN, STARBURST, and STARFORMING. The result is shown in the confusion matrix in Fig. \ref{fig: SDSS contaminations}. All of those 3 classes belong to the subclass of GALAXY and show some degree of mixing with each other in the SDSS sample. 

\begin{figure}
    \includegraphics[width=0.21\textwidth]{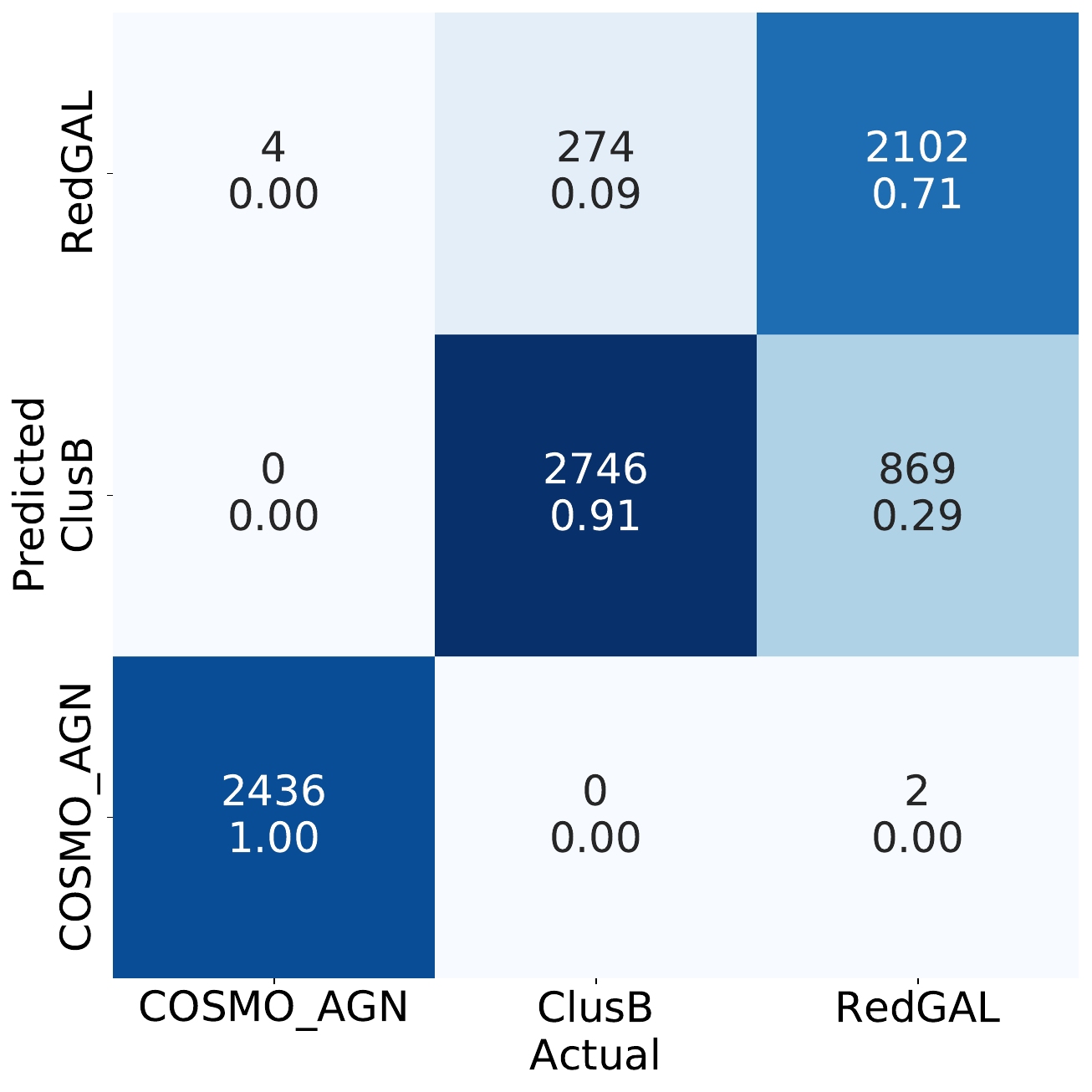}
    \includegraphics[width=0.21\textwidth]{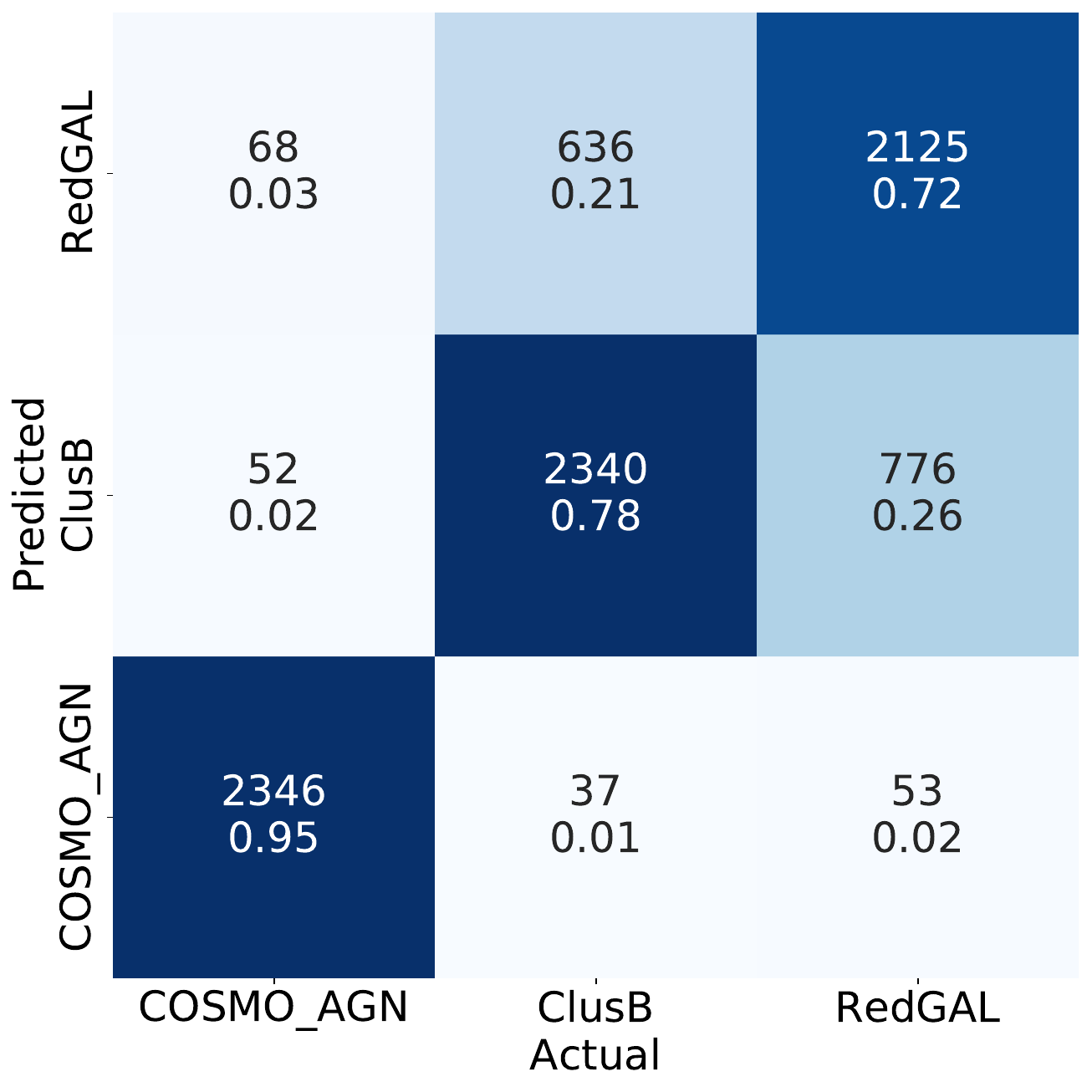}
    \includegraphics[width=0.21\textwidth]{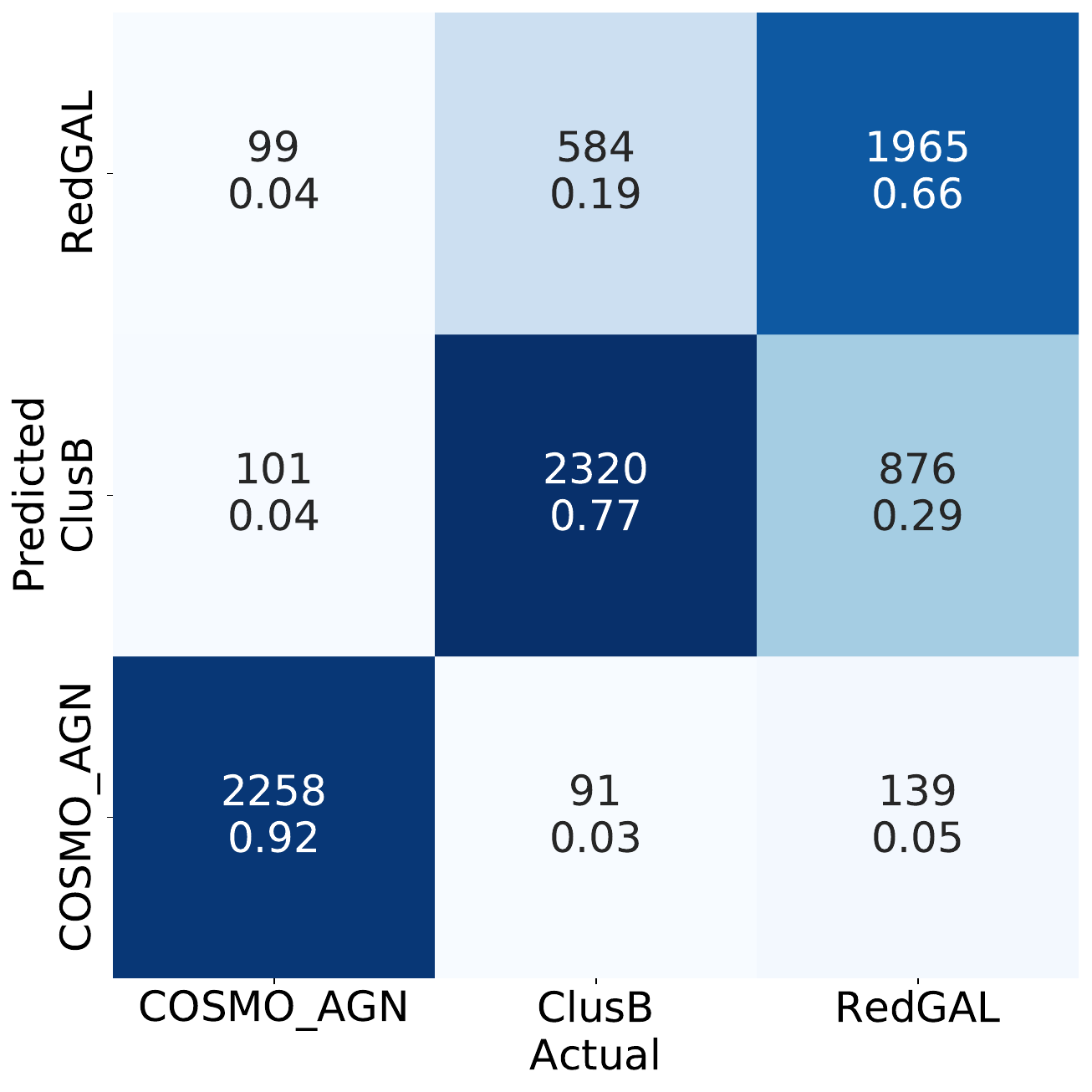}
    \caption{Confusion matrix of a 3-class classification (COMOS\_AGN, ClusB, RedGAL), showing how this changes with increasing random contamination (0\% - 5\% - 10 \%). The contamination fraction refers to randomly selected and shuffled labels in the dataset. As contamination increases, the accuracy of both the COMOS\_AGN and ClusB decreases. Roughly speaking, with respect to the original sample (0\% artificial contamination), we observe an average decrease in accuracy by 5.7\% for 5\% contamination and 9\% for 10\% contamination.}
    \label{fig: 4MOST contaminations}
\end{figure}

As additional testing, the results of cross-contamination on the 4MOST dataset (RedGAL, clusB, COSMO\_AGN) are also shown in Fig. \ref{fig: 4MOST contaminations}. 
In \S\ref{sec:4most_class}, the confusion matrix shows some strong degeneracy between the class RedGAL and clusB, so we test the cross-contamination on those two classes and COSMO\_AGN. The additional class COSMO\_AGN was used to reflect the upper limit of classification (the cross-contamination rate). We find an average decrease in accuracy by 5.7\% for 5\% contamination and 9\% for 10\% contamination. 

\begin{figure*}
    \centering
        \includegraphics[width=0.75\textwidth]{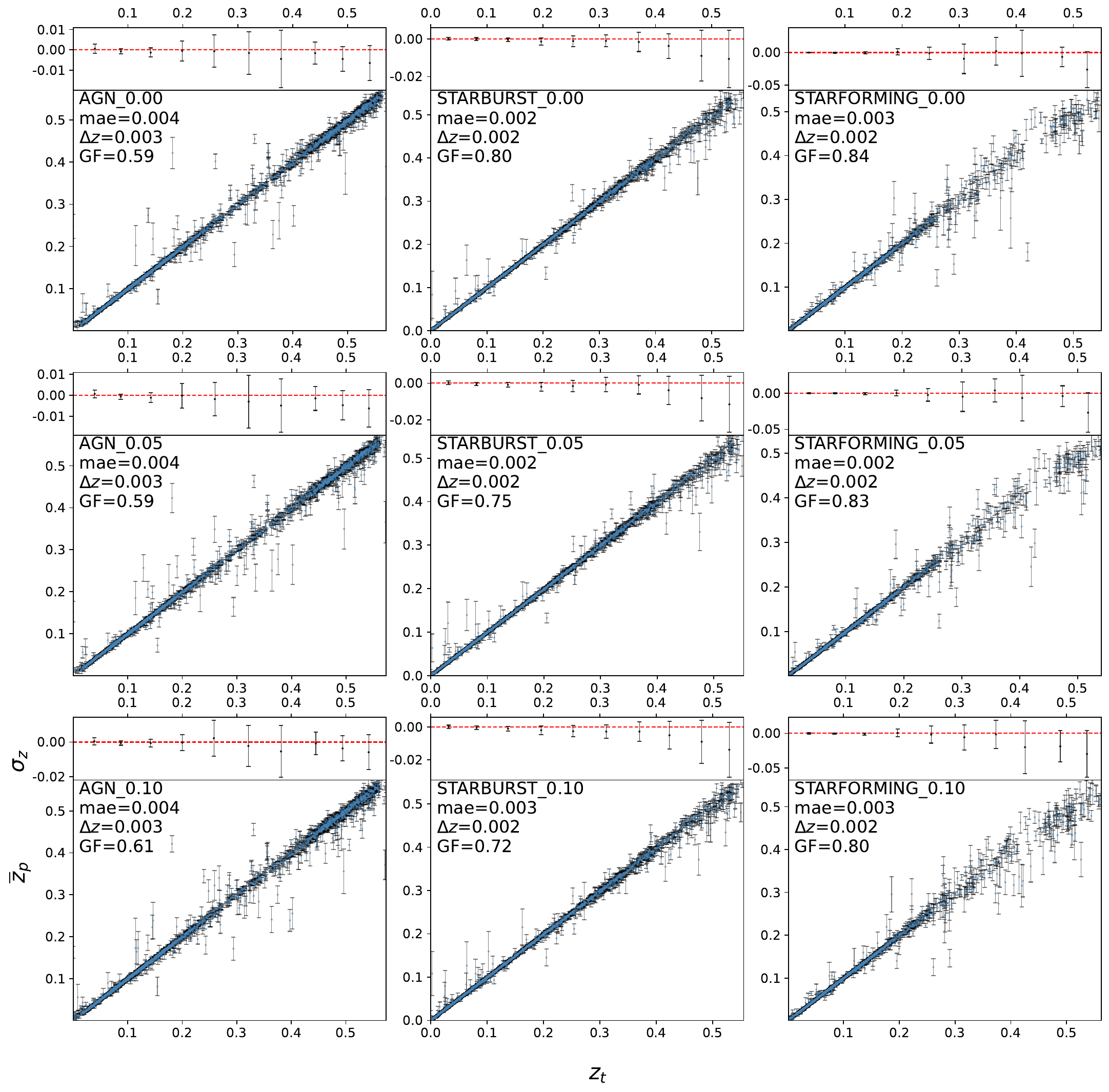}
    \caption{Redshift prediction results of 3 subclass (AGN, STARBURST, STARFORMINGL) in the 3 different cross-contamination levels (0\% - 5\% - 10 \%). The first row represents 0\% contamination. The second row represents 5\% contamination. The third row represents 10 \% contamination. The figure indicates that the small contamination on subclasses does not significantly decrease the accuracy of redshift prediction, but we still observe an average decrease of GF by 2\% for 5\% contamination and 3.3\% for 10\% contamination.}
    \label{fig: SDSS_contamination_one2one} 
\end{figure*}

We end this section by showing the redshift predictions for the samples with contamination, discussed at the end of the \S\ref{sec:4most_class}. Fig. \ref{fig: SDSS_contamination_one2one} shows the redshift prediction results for 3 subclasses (AGN, STARBURST, STARFORMING) in the 3 different cross-contamination levels (0\% - 5\% - 10 \%). The figure shows that the small contamination among subclasses does not significantly decrease the accuracy of redshift prediction, but it still causes performance degradation, such as the average decrease in GF by 2\% for 5\% contamination and 3.3\% for 10\% contamination.

\section{The coarse classifier of SDSS-DR16 and 4MOST}
\label{sec: The coarse classifier of SDSS}
\begin{figure}
    \centering
        \includegraphics[width=0.21\textwidth]{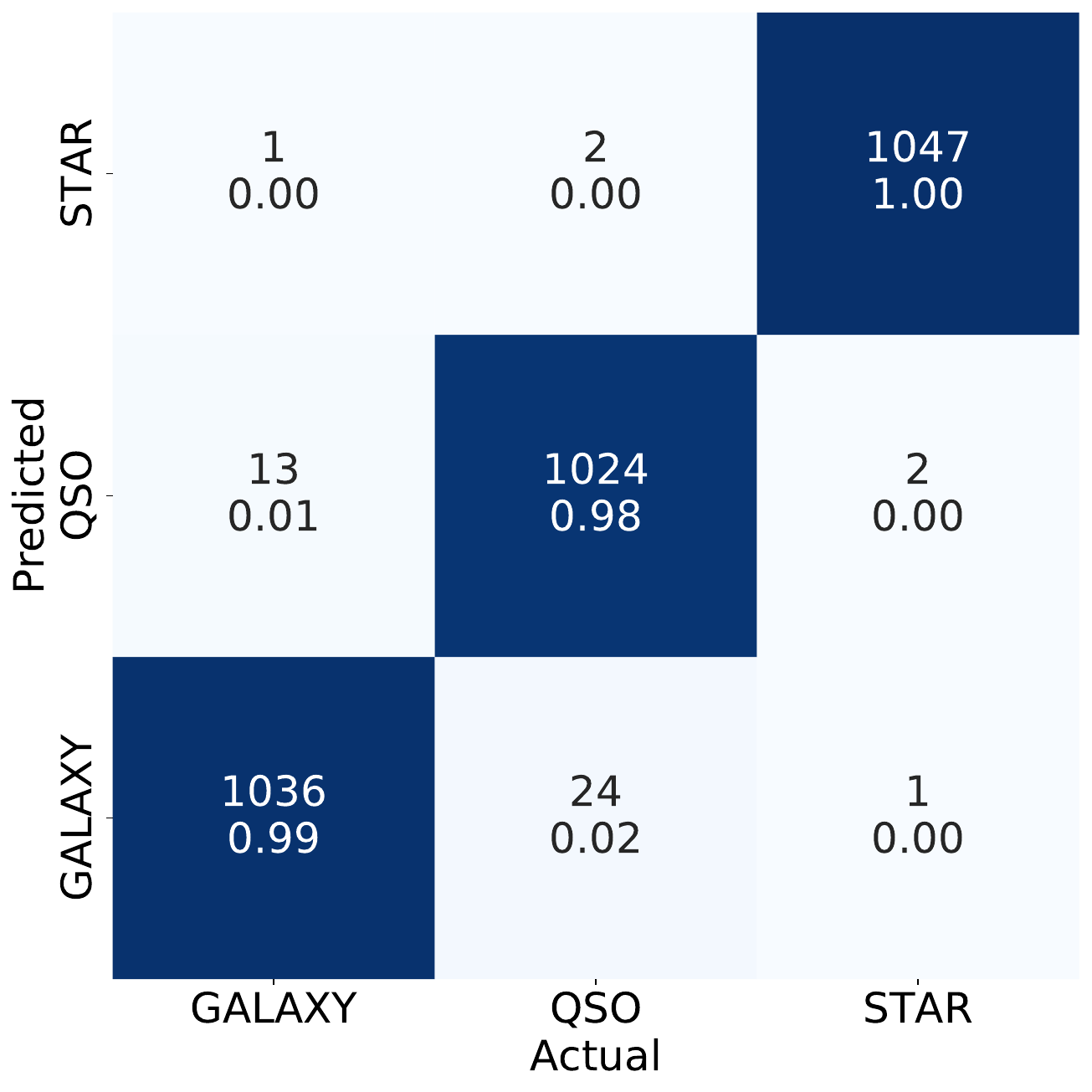}
        \includegraphics[width=0.21\textwidth]{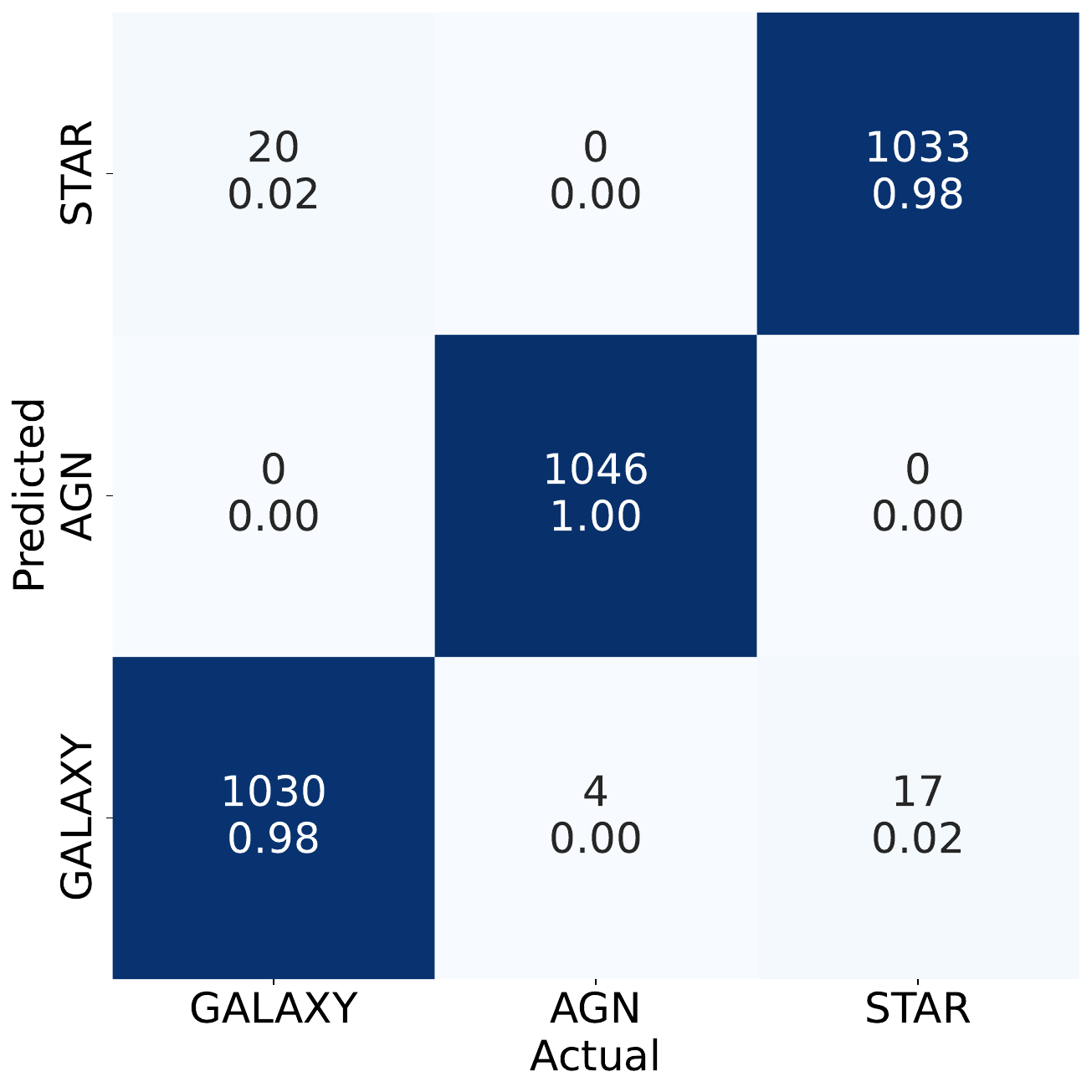}
    \caption{Results of the `coarse' classification of the SDSS (left) and 4MOST (right) dataset. The spectra are categorized into 3 classes: GALAXY, QSO (AGN), and STAR. GaSNet-II achieved an average accuracy of 99\%. The STAR class nearly achieved 100\% accuracy. The class with the lowest accuracy is QSO, but it still achieved an impressive 98\% accuracy.}
    \label{fig: coarse_classifier_SDSS} 
\end{figure}

In this appendix, we briefly describe a more homogeneous comparative check of the performances of the GaSNet-II on the three datasets discussed in the paper, by emulating the situation where we have the same data size and number of classes. We use the DESI sample as a reference, as it contains the smaller dataset (21\,000 entries) and coarser classification (GALAXY, QSO/AGN, and STAR).
To do that, we have regrouped the spectra belonging to these three broader classes for SDSS (STAR: raw 1-7; GALAXY: raw 8-11; QSO: raw 12-13, in Table \ref{Table:1}) and 4MOST (STAR: raw 1-5; AGN: raw 6; GALAXY: raw 7-10, in Table \ref{Table:2}) respectively. To be uniform with the DESI case, we have also randomly extracted 7\,000 spectra from these re-grouped classes, to train and test the GaSNet-II, using the same set-up of DESI training/testing.
Fig. \ref{fig: coarse_classifier_SDSS} shows the results of this `coarse' classification of SDSS and 4MOST datasets, to be compared with the same for DESI in Fig. \ref{fig: DESI classification}. 
The figure indicates that GaSNet-II can achieve an average accuracy of 99\% for classification. The STAR class nearly achieved 100\% accuracy. The class with the lowest accuracy is QSO, but it still achieved an impressive 98\% accuracy. {Compared to the DESI classification, it exhibits a higher accuracy for the same coarse classes and the same amount of training data, with an improvement of about $3\%$, which can be attributed to the qualities of the SDSS spectrum. For instance, the mean SNR of stars and galaxies in SDSS spectra is higher than that of DESI. This can be seen by comparing Tables \ref{Table:1} and \ref{Table:3}.}

\section{Relationship between average classification accuracy and SNR}
\label{sec:average classification accuracy and SNR}
\begin{figure}
    \centering
        \includegraphics[width=0.45\textwidth]{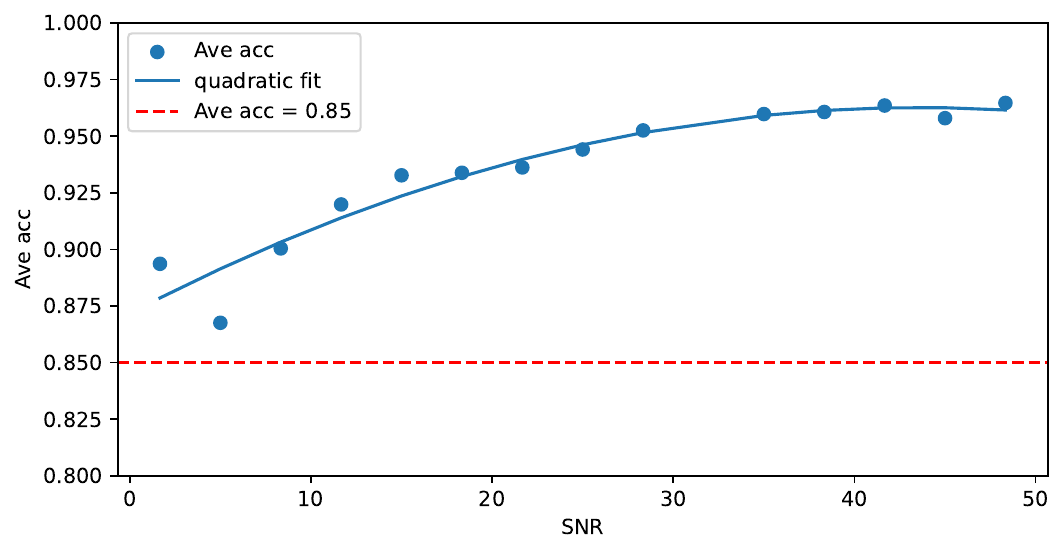}
    \caption{The average classification accuracy of the SDSS 13-subclasses classification with respect to the SNR. We only consider the SNR range of 0-50.}
    \label{fig:SDSS_Ave_acc-vs-SNR}
\end{figure}

Here, we want to test the dependence of the classification accuracy on the SNR of the spectra (see also \S\ref{sec:montecarlo} for the redshift estimates). In Fig. \ref{fig:SDSS_Ave_acc-vs-SNR}, we show the average classification accuracy over the SDSS 13-subclasses with respect to the SNR. We consider 14 bins in the SNR range of 0-50. The figure shows that as the SNR increases, the accuracy systematically increases and finally reaches an upper limit of an average classification accuracy of $\sim 96\%$.

\section{visualization, the gradients of output}
\label{sec: gradients}
\begin{figure}
    \hspace{-0.5cm}
        \includegraphics[width=0.45\textwidth]{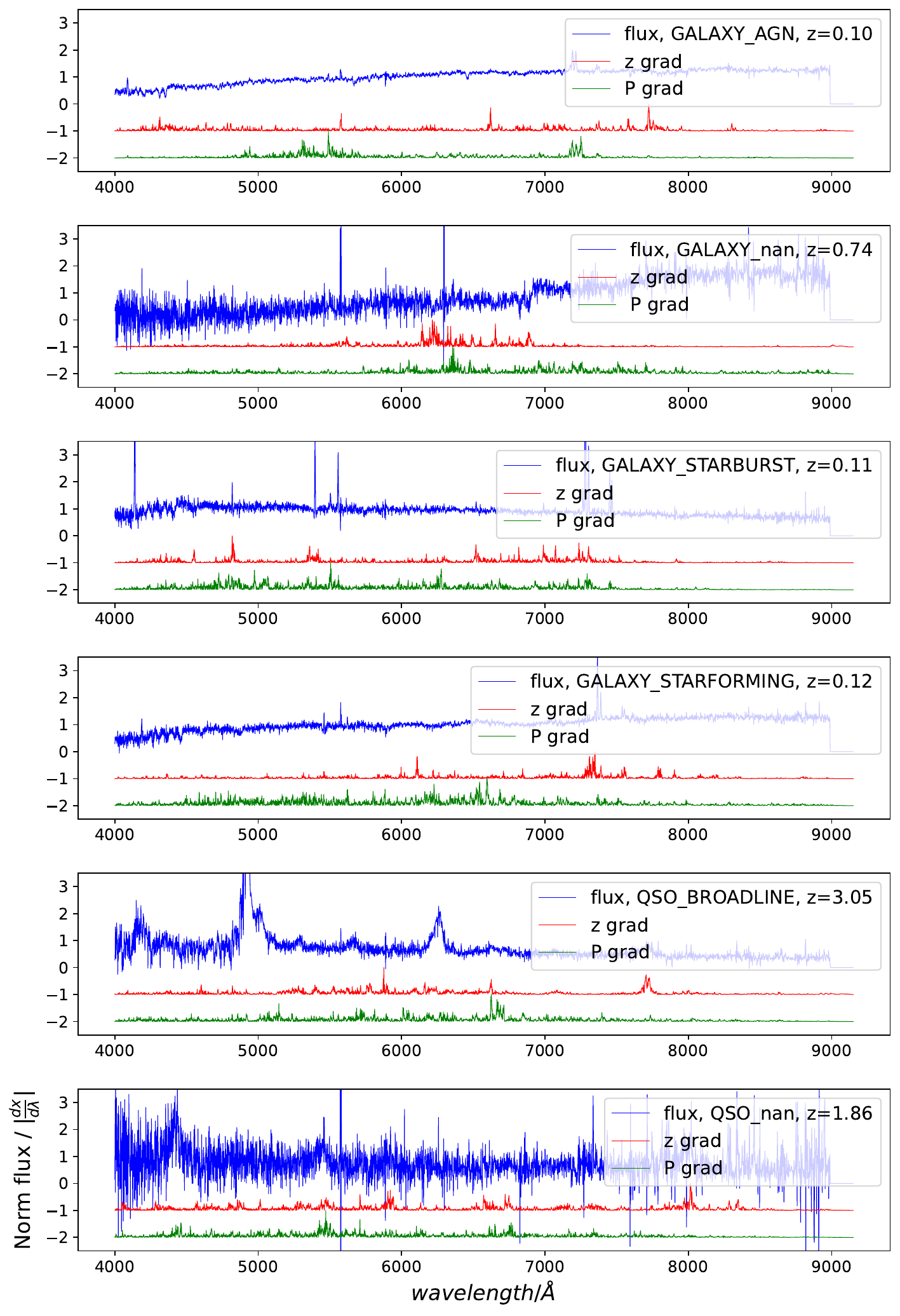}
    \caption{The normalized flux and gradients of 6 SDSS extragalactic spectra. The red line (z\_grad) represents the absolute redshift gradients, $|\frac{dz}{d\lambda}|$, which is shifted by -1; the green line (P\_grad) represents the absolute probability gradients, $|\frac{dP}{d\lambda}|$, which is shifted by -2.}
    \label{fig:SDSS_Gradients}
\end{figure}

As discussed in \citealt{2023A&A...671A..61N}, target (i.e. output label) gradients as a function of input neuron (or wavelength), in the form of partial derivatives of the output with respect to $\lambda$ can give information about the sensitivity of output labels to each of the input fluxes. This allows us to visualize whether the CNN
is learning from the spectral features.
In Fig \ref{fig:SDSS_Gradients} we have selected 6 SDSS extragalactic random spectra including objects from different classes. We have paid attention to avoiding too low SNR to avoid the gradient being dominated by noise rather than the impact of the spectral features.  
As it can be seen, the gradients of both the classification probability, $|\frac{dP}{d\lambda}|$, and the redshift predictions, $|\frac{dz}{d\lambda}|$, show strong increases around the most prominent features (e.g. emission lines in star-forming and starburst galaxies) and possibly some absorption lines from normal galaxies. Interestingly, they seem to be less sensitive to the very broad lines from quasars, meaning that these are too smoothly varying, maybe looking more like a continuum feature. Also interesting is the fact that the gradients show a burst around the ``redshifted'' 4000\AA\ break for the GALAXY\_nan spectrum (at $\sim 7000$\AA), implying that this is a feature that can be seen by the CNN.

\section{classification accuracy, redshift uncertainty, and velocity dispersion}
\label{sec: correlated with velocity dispersion}
\begin{figure}
    \centering
        \includegraphics[width=0.45\textwidth]{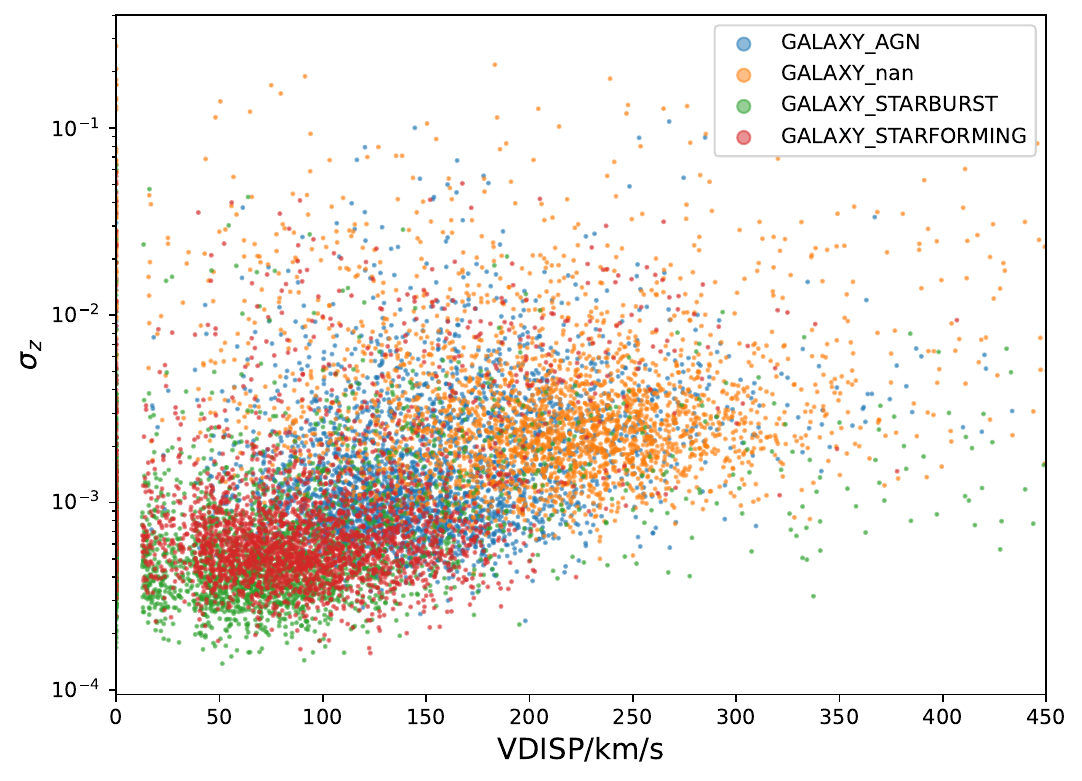}
        \includegraphics[width=0.45\textwidth]{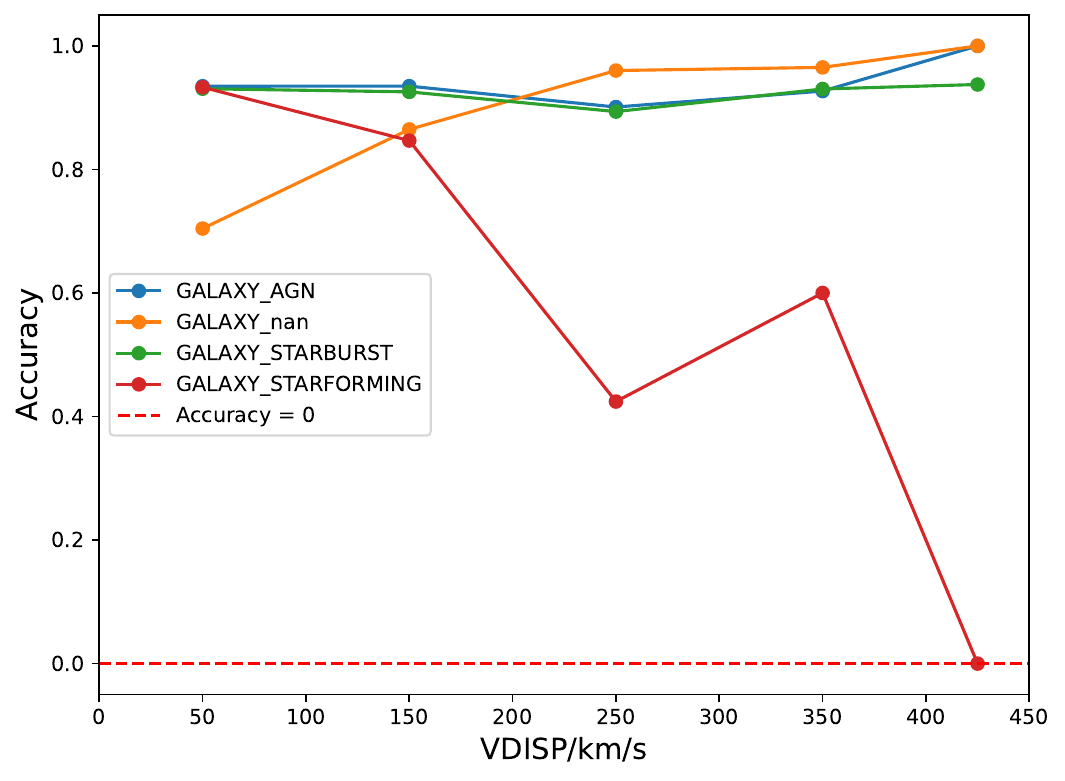}
    \caption{
    The predicted $\sigma_z$ values by the MC and the velocity dispersion of four SDSS subclasses of galaxies are plotted. The x-axis represents the velocity dispersion, limited to values up to 450 km/s.}
    \label{fig: sigmaz velocity dispersion} 
\end{figure}

In this appendix, we test the impact of the velocity dispersion on the spectra classification and redshift estimates. The line broadening caused by the velocity dispersion might enlarge the width of emission or absorption lines, affecting the accuracy of redshift prediction. In Fig \ref{fig: sigmaz velocity dispersion} (left panel), we demonstrate that the predicted $\sigma_z$ of the SDSS dataset is slightly correlated with the velocity dispersion of galaxies, as the larger the velocity dispersion of the galaxy, the larger the predicted uncertainty. However, in the same figure, we also show the $\sigma_z$ separated in the different subclasses and we see that, for each subclass, the $\sigma_z$ is almost independent of the velocity dispersion (VDISP). This is mirrored by the classification accuracy (bottom panel) where we see that, except for `GALAXY\_STARFORMING' which has a sparser sampling, the accuracy also stays almost constant with the velocity dispersion. Hence, we conclude that the accuracy of classifications and redshift estimates are mainly driven by the class type (meaning spectral features) and SNR (see \S\ref{sec:montecarlo} and Appendix \ref{sec:average classification accuracy and SNR}), rather than the velocity dispersion.

\label{lastpage}
\end{document}